\documentclass{nature3}
\usepackage{graphicx}
\usepackage{xcolor}
\usepackage{float}
\usepackage{longtable}
\usepackage{tabularx}
\usepackage{url}
\usepackage{amsmath}
\usepackage{subcaption}
\usepackage{courier}
\usepackage[labelfont=bf, labelsep=none]{caption}

\usepackage[super]{natbib}
\usepackage[colorlinks=true,citecolor=blue,urlcolor=cyan]{hyperref}
\usepackage{astjnlabbrev}

\usepackage[affil-it]{authblk}

\newcommand{\sep}{$\,\vert\,$}
\newcommand{\eureka}{\texttt{Eureka!}}
\newcommand{\kzz}{$K_{\rm zz}$}
\newcommand{\waspstar}{WASP-18}
\makeatletter
\def\@maketitle{%
  \newpage
  \begin{center}%
  \let \footnote \thanks
    {\Large \@title \par}%
    \vskip -3em
    {\large
      \begin{tabular}[t]{c}%
        \@author
      \end{tabular}\par}%
  \end{center}%
  \par
  \vskip 0.5em}
\makeatother

\title{A broadband thermal emission spectrum of the ultra-hot Jupiter WASP-18b}

\author[1]{Louis-Philippe Coulombe\footnote{Corresponding authors' emails: louis-philippe.coulombe@umontreal.ca, bjorn.benneke@umontreal.ca}} 
\newcounter{ind}

\stepcounter{ind}\edef\UdeM{\arabic{ind}}
\affil[\arabic{ind}]{Department of Physics and Trottier Institute for Research on Exoplanets, Universit\'{e} de Montr\'{e}al, Montr\'{e}al, QC, Canada}

\author[\UdeM]{Bj\"{o}rn Benneke$^*$}

\stepcounter{ind}\edef\UofM{\arabic{ind}}
\affil[\arabic{ind}]{Department of Astronomy, University of Michigan, Ann Arbor, MI, USA}
\author[\UofM]{Ryan Challener}

\stepcounter{ind}\edef\CarnegieEPL{\arabic{ind}}
\affil[\arabic{ind}]{Earth and Planets Laboratory, Carnegie Institution for Science, Washington, DC, USA}
\author[\CarnegieEPL]{Anjali A. A. Piette}

\stepcounter{ind}\edef\ASU{\arabic{ind}}
\affil[\arabic{ind}]{School of Earth and Space Exploration, Arizona State University, Tempe, AZ, USA}
\author[\ASU]{Lindsey S. Wiser}

\stepcounter{ind}\edef\SO{\arabic{ind}}
\affil[\arabic{ind}]{Steward Observatory, University of Arizona, Tucson, AZ, USA}
\stepcounter{ind}\edef\NHFP{\arabic{ind}}
\affil[\arabic{ind}]{NHFP Sagan Fellow}
\author[\SO,\NHFP]{Megan Mansfield} 

\stepcounter{ind}\edef\Cornell{\arabic{ind}}
\affil[\arabic{ind}]{Department of Astronomy and Carl Sagan Institute, Cornell University, NY, USA}
\author[\UofM,\Cornell,\NHFP]{Ryan J. MacDonald}

\author[\UofM]{Hayley Beltz}

\stepcounter{ind}\edef\UofC{\arabic{ind}}
\affil[\arabic{ind}]{Department of Astronomy and Astrophysics, University of Chicago, Chicago, IL, USA}
\stepcounter{ind}\edef\nsf{\arabic{ind}}
\affil[\arabic{ind}]{NSF Graduate Research Fellow}
\author[\UofC, \nsf]{Adina D. Feinstein}

\author[\UdeM]{Michael Radica}

\stepcounter{ind}\edef\UofMld{\arabic{ind}}
\affil[\arabic{ind}]{Department of Astronomy, University of Maryland, College Park, MD, USA}

\stepcounter{ind}\edef\FInst{\arabic{ind}}
\affil[\arabic{ind}]{Center for Computational Astrophysics, Flatiron Institute, New York, NY, USA}
\author[\UofMld,\FInst]{Arjun B. Savel}

\stepcounter{ind}\edef\STScI{\arabic{ind}}
\affil[\arabic{ind}]{Space Telescope Science Institute, Baltimore, MD, USA}
\author[\STScI]{Leonardo A. Dos Santos}

\author[\UofC]{Jacob L. Bean}

\stepcounter{ind}\edef\UCA{\arabic{ind}}
\affil[\arabic{ind}]{Université Côte d'Azur, Observatoire de la Côte d'Azur, CNRS, Laboratoire Lagrange, France}
\author[\UCA]{Vivien Parmentier}

\stepcounter{ind}\edef\GSFC{\arabic{ind}}
\affil[\arabic{ind}]{NASA Goddard Space Flight Center, Greenbelt, MD, USA}
\stepcounter{ind}\edef\NPP{\arabic{ind}}
\affil[\arabic{ind}]{NASA Postdoctoral Program Fellow}
\author[\GSFC,\NPP]{Ian Wong}

\author[\UofM]{Emily Rauscher}
\author[\UofMld]{Thaddeus D. Komacek}
\author[\UofMld]{Eliza M.-R. Kempton}

\stepcounter{ind}\edef\TDLI{\arabic{ind}}
\affil[\arabic{ind}]{Tsung-Dao Lee Institute, Shanghai Jiao Tong University, Shanghai, People’s Republic of China}

\stepcounter{ind}\edef\SJTU{\arabic{ind}}
\affil[\arabic{ind}]{School of Physics and Astronomy, Shanghai Jiao Tong University, Shanghai, People’s Republic of China}

\stepcounter{ind}\edef\Oxf{\arabic{ind}}
\affil[\arabic{ind}]{Atmospheric, Oceanic and Planetary Physics, Department of Physics, University of Oxford, Oxford, UK}

\author[\TDLI,\SJTU,\Oxf]{Xianyu Tan}

\author[\Oxf]{Mark Hammond}

\stepcounter{ind}\edef\UofExe{\arabic{ind}}
\affil[\arabic{ind}]{Department of Mathematics and Statistics, Faculty of Environment, Science and Economy, University of Exeter, Exeter, UK}
\author[\UofExe]{Neil T. Lewis}

\author[\ASU]{Michael R. Line}

\stepcounter{ind}\edef\Bern{\arabic{ind}}
\affil[\arabic{ind}]{Center for Space and Habitability, University of Bern, Bern, Switzerland}
\author[\Bern]{Elspeth K. H. Lee}

\stepcounter{ind}\edef\UofA{\arabic{ind}}
\affil[\arabic{ind}]{Anton Pannekoek Institute for Astronomy, University of Amsterdam, Amsterdam, The Netherlands}
\author[\UofA]{Hinna Shivkumar}

\stepcounter{ind}\edef\UofA{\arabic{ind}}
\affil[\arabic{ind}]{Department of Physics \& Astronomy, University of Kansas, Lawrence, KS, USA}
\author[\UofA]{Ian J. M. Crossfield}

\author[\UofMld]{Matthew C. Nixon}

\stepcounter{ind}\edef\EAPMIT{\arabic{ind}}
\affil[\arabic{ind}]{Department of Earth, Atmospheric and Planetary Sciences, Massachusetts Institute of Technology, Cambridge, MA, USA}

\stepcounter{ind}\edef\MIT{\arabic{ind}}
\affil[\arabic{ind}]{Kavli Institute for Astrophysics and Space Research, Massachusetts Institute of Technology, Cambridge, MA, USA}

\stepcounter{ind}\edef\Pegasi{\arabic{ind}}
\affil[\arabic{ind}]{51 Pegasi b Fellow}

\author[\EAPMIT,\MIT,\Pegasi]{Benjamin V.\ Rackham}

\stepcounter{ind}\edef\UoB{\arabic{ind}}
\affil[\arabic{ind}]{School of Physics, University of Bristol, Bristol, UK}

\author[\UoB]{Hannah R. Wakeford}

\author[\ASU,\NHFP]{Luis Welbanks}

\stepcounter{ind}\edef\UCSC{\arabic{ind}}
\affil[\arabic{ind}]{Department of Earth and Planetary Sciences, University of California Santa Cruz, Santa Cruz, California, USA}

\author[\UCSC]{Xi Zhang}

\stepcounter{ind}\edef\AUCSC{\arabic{ind}}
\affil[\arabic{ind}]{Department of Astronomy and Astrophysics, University of California Santa Cruz, Santa Cruz, California, USA}

\author[\AUCSC]{Natalie M. Batalha}

\stepcounter{ind}\edef\UoCB{\arabic{ind}}
\affil[\arabic{ind}]{Department of Astrophysical and Planetary Sciences, University of Colorado, Boulder, CO,
USA}
\author[\UoCB]{Zachory K. Berta-Thompson}

\stepcounter{ind}\edef\ESAStsci{\arabic{ind}}
\affil[\arabic{ind}]{European Space Agency, Space Telescope Science Institute, Baltimore, MD, USA}

\stepcounter{ind}\edef\UCL{\arabic{ind}}
\affil[\arabic{ind}]{Department of Physics and Astronomy, University College London, United Kingdom}

\author[\ESAStsci,\UCL]{Quentin Changeat}

\author[\UofA]{Jean-Michel D\'esert}

\author[\STScI]{N\'estor Espinoza}

\stepcounter{ind}\edef\SEPS{\arabic{ind}}
\affil[\arabic{ind}]{School of Earth and Planetary Sciences (SEPS), National Institute of Science Education and Research (NISER), HBNI, Odisha, India}

\author[\SEPS]{Jayesh M. Goyal}

\stepcounter{ind}\edef\UoCF{\arabic{ind}}
\affil[\arabic{ind}]{Planetary Sciences Group, Department of Physics and Florida Space Institute, University of Central Florida, Orlando, Florida, USA}

\author[\UoCF]{Joseph Harrington}

\stepcounter{ind}\edef\caltech{\arabic{ind}}
\affil[\arabic{ind}]{Division of Geological and Planetary Sciences, California Institute of Technology, Pasadena, CA, USA}

\author[\caltech]{Heather A. Knutson}

\stepcounter{ind}\edef\MPIA{\arabic{ind}}
\affil[\arabic{ind}]{Max Planck Institute for Astronomy, Heidelberg, Germany}

\author[\MPIA]{Laura Kreidberg}

\stepcounter{ind}\edef\CfA{\arabic{ind}}
\affil{Center for Astrophysics ${\rm \mid}$ Harvard {\rm \&} Smithsonian, Cambridge, MA, USA}

\author[\CfA]{Mercedes L\'opez-Morales}

\stepcounter{ind}\edef\DEPSJH{\arabic{ind}}
\affil[\arabic{ind}]{Department of Earth and Planetary Sciences, Johns Hopkins University, Baltimore, MD, USA}

\stepcounter{ind}\edef\DPAJH{\arabic{ind}}
\affil[\arabic{ind}]{Department of Physics \& Astronomy, Johns Hopkins University, Baltimore, MD, USA}

\author[\MIT]{Avi Shporer}

\author[\DEPSJH,\DPAJH]{David K. Sing}

\stepcounter{ind}\edef\APLJH{\arabic{ind}}
\affil[\arabic{ind}]{Johns Hopkins APL, Laurel, MD, USA}
\author[\APLJH]{Kevin B. Stevenson}

\stepcounter{ind}\edef\IIT{\arabic{ind}}
\affil[\arabic{ind}]{Indian Institute of Technology, Indore, India}
\author[\IIT]{Keshav Aggarwal}

\stepcounter{ind}\edef\CEHUW{\arabic{ind}}
\affil[\arabic{ind}]{Centre for Exoplanets and Habitability, University of Warwick, Coventry, UK}

\stepcounter{ind}\edef\DPUW{\arabic{ind}}
\affil[\arabic{ind}]{Department of Physics, University of Warwick, Coventry, UK}

\author[\CEHUW,\DPUW]{Eva-Maria Ahrer}

\author[\CarnegieEPL]{Munazza K. Alam}

\stepcounter{ind}\edef\BAER{\arabic{ind}}
\affil[\arabic{ind}]{BAER Institute, NASA Ames Research Center, Moffet Field, CA, USA}

\author[\BAER]{Taylor J. Bell}

\stepcounter{ind}\edef\NYUAD{\arabic{ind}}
\affil[\arabic{ind}]{Department of Physics, New York University Abu Dhabi, Abu Dhabi, UAE}

\stepcounter{ind}\edef\CAPNYUAD{\arabic{ind}}
\affil[\arabic{ind}]{Center for Astro, Particle and Planetary Physics (CAP3), New York University Abu Dhabi, Abu Dhabi, UAE}

\author[\NYUAD,\CAPNYUAD]{Jasmina Blecic}

\stepcounter{ind}\edef\IAUAB{\arabic{ind}}
\affil[\arabic{ind}]{Instituto de Astrofisica, Universidad Andres Bello, Santiago, Chile}

\stepcounter{ind}\edef\NPF{\arabic{ind}}
\affil[\arabic{ind}]{Nucleo Milenio de Formacion Planetaria (NPF), Chile}

\stepcounter{ind}\edef\CATA{\arabic{ind}}
\affil[\arabic{ind}]{Centro de Astrofisica y Tecnologias Afines (CATA), Casilla 36-D, Santiago, Chile}
\author[\IAUAB,\NPF,\CATA]{Claudio Caceres}

\author[\AUCSC]{Aarynn L. Carter}

\stepcounter{ind}\edef\UoL{\arabic{ind}}
\affil[\arabic{ind}]{School of Physics and Astronomy, University of Leicester, Leicester}

\author[\UoL]{Sarah L. Casewell}

\stepcounter{ind}\edef\LObs{\arabic{ind}}
\affil[\arabic{ind}]{Leiden Observatory, University of Leiden, Leiden, The Netherlands}
\author[\LObs]{Nicolas Crouzet}

\stepcounter{ind}\edef\INAF{\arabic{ind}}
\affil[\arabic{ind}]{INAF – Osservatorio Astrofisico di Torino, Pino Torinese, Italy}

\stepcounter{ind}\edef\SRIAAS{\arabic{ind}}
\affil[\arabic{ind}]{Space Research Institute, Austrian Academy of Sciences, Graz, Austria}

\author[\INAF,\SRIAAS]{Patricio E. Cubillos}

\stepcounter{ind}\edef\IoAKUL{\arabic{ind}}
\affil[\arabic{ind}]{Institute of Astronomy, Department of Physics and Astronomy, KU Leuven, Leuven, Belgium}

\author[\IoAKUL]{Leen Decin}

\author[\AUCSC]{Jonathan J. Fortney}

\stepcounter{ind}\edef\TCD{\arabic{ind}}
\affil[\arabic{ind}]{School of Physics, Trinity College Dublin, Dublin, Ireland}

\author[\TCD]{Neale P. Gibson}

\stepcounter{ind}\edef\LMU{\arabic{ind}}
\affil[\arabic{ind}]{Universitäts-Sternwarte, Ludwig-Maximilians-Universität München, München, Germany}

\stepcounter{ind}\edef\ARTORG{\arabic{ind}}
\affil[\arabic{ind}]{ARTORG Center for Biomedical Engineering Research, University of Bern, Bern, Switzerland}

\author[\LMU,\DPUW,\ARTORG]{Kevin Heng}

\author[\MPIA]{Thomas Henning}

\stepcounter{ind}\edef\PF{\arabic{ind}}
\affil[\arabic{ind}]{Institute for Planetary Research (PF), German Aerospace Centre (DLR), Berlin, Germany}
\author[\PF]{Nicolas Iro}

\author[\ESAStsci]{Sarah Kendrew}

\stepcounter{ind}\edef\UPS{\arabic{ind}}
\affil[\arabic{ind}]{Université Paris-Saclay, Université Paris Cité, CEA, CNRS, AIM, Gif-sur-Yvette, France}

\author[\UPS]{Pierre-Olivier Lagage}

\stepcounter{ind}\edef\LAB{\arabic{ind}}
\affil[\arabic{ind}]{Laboratoire d'Astrophysique de Bordeaux, Université de Bordeaux, Pessac, France}

\author[\LAB]{J\'er\'emy Leconte}

\stepcounter{ind}\edef\UdeG{\arabic{ind}}
\affil[\arabic{ind}]{Département d'Astronomie, Université de Genève, Sauverny, Switzerland}

\author[\UdeG]{Monika Lendl}

\stepcounter{ind}\edef\UVU{\arabic{ind}}
\affil[\arabic{ind}]{Department of Physics, Utah Valley University, Orem, UT, USA}

\author[\UVU]{Joshua D. Lothringer}

\stepcounter{ind}\edef\UoR{\arabic{ind}}
\affil[\arabic{ind}]{Department of Physics, University of Rome ``Tor Vergata'', Rome, Italy}

\author[\UoR,\INAF,\MPIA]{Luigi Mancini}

\author[\MPIA]{Thomas Mikal-Evans}

\stepcounter{ind}\edef\EO{\arabic{ind}}
\affil[\arabic{ind}]{Exzellenzcluster Origins, Garching, Germany}

\author[\LMU,\EO,\MPIA]{Karan Molaverdikhani}

\author[\STScI]{Nikolay K. Nikolov}

\author[\AUCSC]{Kazumasa Ohno}

\stepcounter{ind}\edef\IAC{\arabic{ind}}
\affil[\arabic{ind}]{Instituto de Astrofísica de Canarias (IAC), Tenerife, Spain}

\author[\IAC]{Enric Palle}

\author[\UdeM]{Caroline Piaulet}

\stepcounter{ind}\edef\VVO{\arabic{ind}}
\affil[\arabic{ind}]{Astronomy Department and Van Vleck Observatory, Wesleyan University, Middletown, CT, USA}

\author[\VVO]{Seth Redfield}

\author[\UdeM]{Pierre-Alexis Roy}

\stepcounter{ind}\edef\UCR{\arabic{ind}}
\affil[\arabic{ind}]{Department of Earth and Planetary Sciences, University of California, Riverside, CA, USA}

\author[\UCR]{Shang-Min Tsai}

\stepcounter{ind}\edef\UPCPE{\arabic{ind}}
\affil[\arabic{ind}]{Université de Paris Cité and Univ Paris Est Creteil, CNRS, LISA, Paris, France}

\author[\UPCPE]{Olivia Venot}

\author[\CEHUW,\DPUW]{Peter J. Wheatley}

\DeclareCaptionFormat{cont}{#1 (cont.)#2#3\par}

\begin{document}

\maketitle

\begin{center}
Status: submitted on January 18, 2023 
\end{center}

\begin{abstract}
Close-in giant exoplanets with temperatures greater than 2,000 K (``ultra-hot Jupiters'') have been the subject of extensive efforts to determine their atmospheric properties using thermal emission measurements from the Hubble and Spitzer Space Telescopes \cite{baxter20, mansfield_unique_2021, changeat22}. However, previous studies have yielded inconsistent results because the small sizes of the spectral features and the limited information content of the data resulted in high sensitivity to the varying assumptions made in the treatment of instrument systematics and the atmospheric retrieval analysis \cite{changeat22, sheppard_evidence_2017, mikal-evans17, cartier_near-infrared_2017, arcangeli_h-_2018, kreidberg_global_2018, parmentier_thermal_2018, lothringer_extremely_2018, ghandi20, mikal-evans20}. Here we present a dayside thermal emission spectrum of the ultra-hot Jupiter WASP-18b obtained with the NIRISS\cite{doyon_jwst_2012} instrument on JWST. The data span 0.85 to 2.85\,$\mu$m in wavelength at an average resolving power of 400 and exhibit minimal systematics. The spectrum shows three water emission features (at $>$6$\sigma$ confidence) and evidence for optical opacity, possibly due to H$^-$, TiO, and VO (combined significance of 3.8$\sigma$). Models that fit the data require a thermal inversion, molecular dissociation as predicted by chemical equilibrium, a solar heavy element abundance (``metallicity'', M/H = 1.03$_{-0.51}^{+1.11}$ $\times$ solar), and a carbon-to-oxygen (C/O) ratio less than unity. The data also yield a dayside brightness temperature map, which shows a peak in temperature near the sub-stellar point that decreases steeply and symmetrically with longitude toward the terminators.

\end{abstract}



We observed a secondary eclipse of WASP-18b with NIRISS/SOSS\cite{doyon_jwst_2012} as part of the JWST Transiting Exoplanet Community Early Release Science Program\cite{bean_transiting_2018}. WASP-18b is a 10.4$\pm$0.4~$M_{J}$ ultra-hot Jupiter on a 0.94 day orbit around a bright ($J$ mag = 8.4) F6V-type star\cite{hellier_orbital_2009}. Our goals were to characterize WASP-18b's atmosphere and demonstrate the capabilities of JWST observations for exoplanets orbiting bright stars. We used the SUBSTRIP96 subarray mode (96 $\times$ 2048 pixels) to avoid saturation by minimizing the individual integration times. The SUBSTRIP96 mode covers the first spectral order between 0.85 and 2.85 $\mu$m, whereas the SUBSTRIP256 subarray also provides the shorter-wavelength measurements in the second spectral order.
The time series spans 6.71 hours and consists of 2720 continuous integrations with 3 groups and 8.88 seconds per integration, delivering an integration efficiency of 67\%. We used the F277W filter in the final ten integrations to check for contamination from background stars and found none. We observed for 2.83 hours before the eclipse, and continued for 1.70 hours after the eclipse. The observations captured 107$^\circ$ of WASP-18b's orbit. Assuming it is tidally locked, the planet rotated by the same angle during the observation.

We analyzed the data using four independent pipelines: \texttt{NAMELESS}, \texttt{nirHiss}, \texttt{supreme-SPOON}, and \texttt{transitspectroscopy} (see Methods). Beginning from the raw uncalibrated data or stage 1 products, we performed custom reductions and extracted 1D spectra from each integration using either a fixed-width aperture (\texttt{NAMELESS}, \texttt{supreme-SPOON}, and \texttt{transitspectroscopy}) or an optimal extraction (\texttt{nirHiss}) technique. We put particular emphasis on the removal of 1/$f$ noise ($f$ is frequency), a signal with power spectrum inversely proportional to the frequency that is introduced through variations of the reference voltage as the detector is being read. Its removal requires careful treatment as the spectral trace covers most of the SUBSTRIP96 subarray (see Methods, Extended Data Fig. \ref{fig:reduc_steps}).
Finally, we obtained spectrophotometric light curves by summing the observed flux within 408 spectral bins, each containing five pixel columns on the detector (Extended Data Fig. \ref{fig:chrom_lcs_2d}). We also produced a white-light curve by summing the spectrophotometric light curves over all wavelengths. All light curves show a sudden decrease in flux around the 1336th integration, simultaneous to a tilt event from one of the primary mirror's segments\cite{2022commissioning,alderson_early_2023}. In the NIRISS data, this can be independently identified via a small but detectable morphological change in the spectral trace on the detector (Extended Data Fig. \ref{fig:pca_niriss}).
We also observed small variations in the spectral trace morphology throughout the time series, mainly of its position and full width at half maximum, that are correlated with the measured flux. We detrended against these morphological changes at the fitting stage.  


We analyzed the extracted white and spectrophotometric light curves by fitting the parameters for the secondary eclipse, the partial phase curve, and the systematics using ExoTEP\citep{benneke_water_2019} (see Methods). When fitting the white light curve, we allowed the semi-major axis and impact parameter to vary, constrained by Gaussian priors that were derived from an analysis of the full-orbit phase curve observed by the Transiting Exoplanet Survey Satellite (TESS; see Methods). We imposed a uniform prior on the mid-eclipse time and assumed a circular orbit.
Those parameters were subsequently fixed when fitting the spectrophotometric bins (see Methods). The maximum signal-to-noise (SNR) ratio for a single pixel spectrophotometric light curve is 617 at 1.14~$\mu$m, with the SNR curve closely following the shape of the throughput-weighted stellar spectrum and reaching a minimum of 62 at 2.83~$\mu$m. All four reductions yield consistent results (Extended Data Fig. \ref{fig:Tb_spec_reducs}), with all resulting thermal emission spectra being consistent at less than one standard deviation on average. The residuals for each spectrophotometric light curve closely follow the expected $1/\sqrt{n}$~ ($n$ is the number of events) scaling of Poisson noise when binned in time, and the white light curve bins down to 5\,ppm over one hour timescales (see Extended Data Fig. \ref{fig:binned_resids}).



The secondary eclipse spectrum was created by collating the planet-to-star flux ratio values at mid-eclipse for all the wavelength channels (see Fig. \ref{fig:fpfs_spec_plus_lcs}). We then multiplied by a PHOENIX stellar spectrum model\cite{jack_time-dependent_2009} produced using previously published parameters for \waspstar \, (i.e., $T_\mathrm{eff}$ = 6435~K, $\log g$ = 4.35, and [Fe/H] = 0.1\cite{cortes-zuleta_tramos_2020,stassun_revised_2019}) to convert the dayside secondary-eclipse spectrum into the planet's thermal emission spectrum (see Fig. \ref{fig:fpfs_spec_plus_lcs}). For clarity, we also computed the brightness temperature spectrum, commonly used in planetary science, by calculating the blackbody temperature corresponding to the observed thermal emission in each wavelength bin (see Fig. \ref{fig:TbSpectrum}). This transformation into brightness temperature facilitates identification of the various opacity sources by removing the large average slope caused by the behavior of the Planck emission across the NIRISS/SOSS wavelength range.


The observed brightness temperature spectrum shows strong deviations from a blackbody. It is dominated by the 1.4, 1.9, and 2.5$\,\mu$m water emission features and a rise in brightness temperature shortwards of 1.3$\,\mu$m. 
The rise in brightness is caused by the combined opacities of H$^-$, TiO, and VO and we infer a combined detection significance of 3.8$\sigma$ for these three species (Fig. \ref{fig:TbSpectrum}). All molecular features appear in emission, indicating a thermal inversion (i.e., temperature increases with altitude, see also Fig.\ \ref{fig:atm_ret}). The water features are consistent with a solar-composition atmosphere, as predicted by 1D radiative-convective models and 3D general circulation models (GCMs). They are strongly inconsistent with any high C/O or high metallicity scenarios \cite{sheppard_evidence_2017} (Fig.\ \ref{fig:TbSpectrum}b). This finding is further strengthened by the lack of detectable CO features at 1.6 and 2.4$\,\mu$m, which should be the dominant species in a high-metallicity carbon-rich atmosphere (Fig.\ \ref{fig:TbSpectrum}b). Using the free retrieval, we constrain the 3$\sigma$ upper limit of the CO log mixing ratio to -2.42 (see below, Table \ref{tab:retrieval_results}, Extended Data Fig. \ref{fig:mol_posteriors}).

Quantitatively, we infer an atmospheric metallicity of $0.82_{-0.37}^{+0.59}$ times solar when fitting the \texttt{NAMELESS} reduction to a grid of self-consistent 1D radiative-convective models, and, consistently, $1.19_{-0.67}^{+1.22}$ times solar when allowing for a free vertical temperature structure in the chemically-consistent retrievals. Both modeling approaches accounted for the thermal dissociation of water in the upper atmosphere and assumed chemical equilibrium. In both cases the best fits are obtained for sub-solar C/O values around 0.03--0.3, with the self-consistent models giving 3$\sigma$ upper limits of 0.2, while the free-temperature-structure retrievals allows C/O values up to 0.6 at 3$\sigma$ (Fig.\ \ref{fig:atm_ret}), consistent with the upper limit from high-resolution dayside thermal emission observations \cite{brogi2022}. We also assessed the effect of disequilibrium chemistry (see Methods) on the observed thermal emission and find the impact to be below 10\,ppm due to the short chemical timescales in this hot atmosphere, justifying the assumption of thermochemical equilibrium models. In addition, we performed a free-chemistry atmospheric retrieval\citep{Gandhi2018,gandhi_h-_2020,Piette2020b}, including the effects of thermal dissociation\citep{parmentier_thermal_2018}, and inferred a H$_2$O deep atmospheric log mixing ratio of $-3.23^{+0.45}_{-0.29}$, consistent with the models assuming chemical equilibrium (Fig.\ \ref{fig:TbSpectrum}c). We identify a strong thermal inversion with a temperature increase of 500~K in the middle atmosphere from 1~bar to 0.01~bar, which corresponds to the pressure range covered by the contribution functions (Fig.\ \ref{fig:atm_ret}a). Our best-fit radiative-convective model provides strong evidence that the temperature inversion is caused by the absorption of stellar light by TiO (see Extended Data Fig. \ref{fig:Tb_spec_samples_TiO}). At first sight this can seem at odds with high-spectral resolution observations that have detected other species able to create thermal inversions, such as atomic iron~\citep{lothringer_extremely_2018}, but have had trouble detecting TiO~\cite{Guillot2022}. This tension is easily solved when considering that both TiO and water thermally dissociate in the upper atmospheric layers of ultra-hot Jupiters. Our observations, on the other hand, are sensitive to deeper layers of the atmosphere close to the infrared photosphere, which extends from 0.01 down to 1~bar (see contribution function on Fig.~\ref{fig:atm_ret}a), in the region where molecules such as water and TiO recombine (see Extended Data Fig.~\ref{fig:mol_posteriors}). Even though our model also predicts that iron can produce a thermal inversion, its near-constant abundances means that inversions due to iron happen at pressures lower than 1 millibar and not where we detect the main thermal inversion.


The precise constraints on the atmospheric metallicity and C/O enable us to investigate possible formation scenarios of WASP-18b. Considering the core-accretion formation scenario\cite{pollack1996}, the measured atmospheric metallicity of WASP-18b, consistent with the near-solar metallicity of the host star WASP-18 ([Fe/H] = 0.1$\pm$0.1) \cite{hellier_orbital_2009,polanski2022}, indicates that accretion of protoplanetary gas, rather than rocky or icy planetesimals, dominated the planet's late-stage formation. The mass-metallicity trend derived from solar system planets\cite{lodders98,Kreidberg_2014,wakeford17,Welbanks_2019} predicts that the metallicity decreases as the mass of the planet increases, approaching the composition of the star for the most massive planets. Our finding of solar metallicity, 
three times lower than that of Jupiter, is consistent with this trend, given WASP-18b's mass of 10.4~M$_{J}$. 
The low C/O ratio furthermore disfavors forming WASP-18b beyond the CO$_2$ ice line followed by an inward migration after the disk has dispersed, as gas accretion in that region would have led to high C/O values\cite{_berg_2011}. Detailed spectroscopic observations of the 4.5$\,\mu$m CO feature, which is found within the spectral range of JWST NIRSpec G395H, could lead to a more stringent constraint on the C/O ratio, and, thus, on WASP-18b's formation and migration history.
A detailed interpretation of the atmospheric C/O ratio of WASP-18b would require knowledge of the C/O ratio of the host star. This was recently found to be significantly sub-solar (C/O=0.23$\pm$0.05)\cite{polanski2022} based on high-resolution spectroscopy. However, stellar C/O measurements are especially challenging due to stellar model inaccuracies and weak/blended absorption lines\cite{bedell_chemical_2018}, so further confirmation is warranted.

Another possible formation scenario for WASP-18b is through collapse of the disk from gravitational instability\cite{boss1997} with a disk-free migration. This process leads to an atmospheric metallicity and C/O dictated by the local disk composition, and is expected to result in planets with stellar-to-super-stellar metallicities and sub-stellar-to-stellar C/O\cite{madhusudhan2014a}, in agreement with our results.

In addition to the extracted planetary spectrum and the elemental abundances, we also recover the broadband brightness temperature distribution across WASP-18b's dayside using the eigencurves eclipse-mapping method (see Methods and refs. \cite{rauscher_more_2018,mansfield_eigenspectra_2020,challener_theresa_2022}). We begin this analysis with the systematics-corrected white-light curve. We performed two independent applications of the method, both enforcing positive flux contribution from visible locations on the planet. We find two brightness map solutions which fit the data similarly well (Fig. \ref{fig:eclipse_map}). We convert brightness maps to brightness temperature maps assuming the star emits as a blackbody at $6432 \pm 48\,$K\cite{cortes-zuleta_tramos_2020} and $R_p/R_s = 0.09783 \pm 0.00028$ (Table \ref{tab:tess_results}). The first solution (blue model) shows a brightness temperature plateau stretching from approximately -40$^\circ$ to +40$^\circ$ of longitude relative to the substellar point, with a virtually constant latitudinally averaged brightness temperature of $2781^{+25}_{-13}\,$K. The second solution (red model) shows a more concentrated hot spot at the substellar point with a maximum brightness temperature of $2925^{+16}_{-18}\,$K and a consistent decrease in temperature both eastward and westward of the substellar point. Both solutions consistently reveal a steep temperature drop with longitude toward the terminators, with the inferred brightness temperature falling to $1686^{+70}_{-27}\,$K at the western terminator and $1869^{+50}_{-14}\,$K at the eastern terminator (blue model), and neither shows any significant shift of the brightest region away from the substellar point. This is consistent with what was measured from the HST phase curve of WASP-18b\cite{arcangeli_climate_2019}. 
The high temperatures covering most of the dayside in both solutions, along with the steep decrease in temperature near the limbs, are consistent with the atmospheric retrieval results (Fig.\ \ref{fig:atm_ret}d). 

Beyond the terminators and leading to the nightside, we infer a continued drop in the thermal emission.
Our JWST observations have the sensitivity to probe part of the night side because the planet rotates by 107$^\circ$ during the time series, providing a view of up to 62.5$^\circ$ of the night side east of the eastern terminator at the beginning and ending 44.5$^\circ$ west away from the western terminator. 
The lack of a hot-spot offset and the large center-to-limb brightness temperature contrast suggest heat transport by winds moving radially away from the sub-stellar point and toward the nightside, rather than redistributing heat to the nightside through the formation of an equatorial jet\cite{komacek_DN1,komacek_atmospheric_2017}. Lorentz forces are expected to play an important role in atmospheric dynamics of ultra-hot Jupiters, due to their high dayside temperatures\cite{perna_magnetic_2010,batygin_magnetically_2013,komacek_atmospheric_2017}. Thermally ionized alkali metals coupled to an internal dynamo-driven planetary magnetic field interact with the neutral species and are expected to prevent the formation of an eastward equatorial jet\cite{Menou:2012fu}. By approximating the effects of the Lorentz force as a locally calculated magnetic drag force in GCMs\cite{beltz_exploring_2021} (see Methods), we find that the observed white-light curve is best explained by an internal planetary field strength larger than 5~G, as this field strength is sufficient to prevent a discernible longitudinal shift of the hot spot from the substellar point (Fig.\ \ref{fig:eclipse_map}). This is further confirmed by a separate GCM considering spatially uniform drag timescales, for which we find that the case with the highest drag strength ($\tau_{\rm drag}$ = 10$^3$~s) produces white-light curves that best fit the observations (Fig.\ \ref{fig:eclipse_map}, Extended Data Fig. \ref{fig:gcm-lightcurves}). 
Furthermore, self-consistent magnetohydrodynamic (MHD) models of ultra-hot Jupiters considering the response of the magnetic field to the circulation have shown the possibility of time variability in the longitudinal hot spot offset, oscillating between the western and eastern hemispheres over timescales of 10--100 days\cite{Rogers:2017,hindle_shallow-water_2019,hindle_observational_2021}, but additional observations are needed to test this possibility.

The large wavelength coverage and high spectral and photometric precision of JWST's NIRISS/SOSS mode present many opportunities for the study and detailed characterization of atmospheric processes through thermal emission spectroscopy. Furthermore, planets with high signal to noise eclipses such as WASP-18b allow for the three-dimensional mapping of their atmospheres to retrieve the temperature structure across the dayside as well as variations in properties such as molecular abundances\cite{mansfield_eigenspectra_2020,challener_theresa_2022}. JWST will enable these measurements for most bright transiting exoplanets, giving rise to the possibility of studying the dynamics and chemistry of a wide range of exoplanets directly from secondary eclipse observations.

\begin{methods}

\subsection{NIRISS/SOSS Reduction and Spectrophotometric Extraction.}\label{sec:extraction}
We perform four separate reductions of the NIRISS/SOSS eclipse observations of WASP-18b using the \texttt{NAMELESS}, \texttt{nirHiss}\footnote{\url{https://github.com/afeinstein20/nirhiss}}, \texttt{supreme-SPOON}\footnote{\url{https://github.com/radicamc/supreme-spoon}}, and \texttt{transitspectroscopy}\footnote{\url{https://github.com/nespinoza/transitspectroscopy}} pipelines for inter-comparison of individual reduction steps\cite{feinstein_early_2023}. All pipelines are built around the official STScI \texttt{jwst} reduction pipeline\footnote{\url{https://jwst-pipeline.readthedocs.io/}} with the addition of custom correction steps for systematics such as 1/$f$ noise, zodiacal background, and cosmic rays. Reductions are performed from either the raw uncalibrated data (\texttt{NAMELESS}, \texttt{nirHiss}, and \texttt{supreme-SPOON}) or stage 1 (\texttt{transitspectroscopy}) products up to the extraction of the spectrophotometric light curves.

\subsection{\texttt{NAMELESS} Reduction}\label{sec:nameless}
We use the \texttt{NAMELESS} pipeline\cite{feinstein_early_2023} to reduce the WASP-18\,b observations from the uncalibrated data products through spectral extraction. We used the \texttt{jwst} pipeline version 1.6.0, CRDS (Calibration Reference Data System) version 11.16.5, and CRDS context \texttt{jwst\_0977.pmap} for the reduction. First, we go through all steps of the \texttt{jwst} pipeline Stage 1, with the exception of the \texttt{dark\_current} step. We skip the dark subtraction step to avoid introducing additional noise due to the lack of a high-fidelity reference file. After the ramp-fitting step, we go through the \texttt{assign\_wcs}, \texttt{srctype}, and \texttt{flat\_field} steps of the \texttt{jwst} pipeline Stage 2; we skip the \texttt{background} step and apply our own custom routine for handling the background. We skip the \texttt{pathloss} and \texttt{photom} steps, as an absolute flux calibration is not needed. We perform outlier detection by computing the product of the second derivatives in the column and row directions for all frames\cite{feinstein_early_2023}. We divide the frames into windows of $4{\times}4$ pixels, where we then compute the local median and standard deviation of the second derivative. We flag all pixels that are $\geq 4 \sigma$ away from the window median. Furthermore, we flag pixels that show null or negative counts. All flagged pixels are set equal to the local median of their window. We correct for background systematics using the following routine. First, we identified section (x$\in$[5,400], y$\in$[0,20]) as the region of the SUBSTRIP96 sub-array where the contribution from the spectral orders to the counts is minimal. Next, we compute the scaling factor between the median frame and the  model background provided on the STScI JDox User Documentation\footnote{\url{https://jwst-docs.stsci.edu/}} within the aforementioned region. We consider the 16$\textsuperscript{th}$ percentile of the distribution as the scaling value and subtract the scaled background from all integrations. We pay close attention to the 1/$f$ correction for these observations, as the magnitude of the spectral trace variation is highly wavelength-dependent in the secondary eclipse. Therefore, we consider all columns independently when scaling the trace to compute the 1/$f$ noise. Furthermore, we treat this correction in two parts as we observe a tilt event\cite{2022commissioning,alderson_early_2023} around the 1336th integration (Extended Data Fig. \ref{fig:pca_niriss}), possibly due to a sudden movement in one of the primary mirror's segments, resulting in a change in the morphology of the trace that manifests as a sudden decrease in flux. First, we compute the median columns $\bar{c}$ before and after the tilt event; we use integrations 300--900 and 1350--1900. Then, we define a given column $j$ and row $k$ at an integration $i$ as the sum of the scaled median column $m_{i,j}\bar{c}_{j,k}$ and the 1/$f$ noise $n_{i,j} = c_{i,j,k} - m_{i,j}\bar{c}_{j,k}$. Using the errors $\epsilon_{i,j,k}$ returned by the \texttt{jwst} pipeline, we minimize the $\chi^2$ to find the value of the 1/$f$ noise that best fits the observed column, such that 

\begin{equation}\label{eq:1overfcorr}
m_{i,j}  = \frac{\sum_k\frac{\bar{c}_{j,k}}{\epsilon_{i,j,k}^2}\left(\sum_k\frac{1}{\epsilon_{i,j,k}^2}\right)^{-1} \sum_k\frac{c_{i,j,k}}{\epsilon_{i,j,k}^2} - \sum_k\frac{c_{i,j,k} \bar{c}_{j,k}}{\epsilon_{i,j,k}^2}}{\left(\sum_k\frac{\bar{c}_{j,k}}{\epsilon_{i,j,k}^2}\right)^2\left(\sum_k\frac{1}{\epsilon_{i,j,k}^2}\right)^{-1}  - \sum_k\left[\frac{\bar{c}_{j,k}}{\epsilon_{i,j,k}}\right]^2}
\end{equation}

and

\begin{equation}\label{eq:1overfcorr}
n_{i,j}  = \sum_k\frac{c_{i,j,k}}{\epsilon_{i,j,k}^2} \left(\sum_k\frac{1}{\epsilon_{i,j,k}^2}\right)^{-1} - m_{i,j} \sum_k\frac{\bar{c}_{j,k}}{\epsilon_{i,j,k}^2}\left(\sum_k\frac{1}{\epsilon_{i,j,k}^2}\right)^{-1}.
\end{equation}

\noindent
We then subtract this value from all columns and integrations. We set the error $\epsilon_{i,j,k}$ to $\infty$ for all pixels that have non-zero data quality flags returned by the \texttt{jwst} pipeline such that they are not considered in the fit of the 1/$f$ noise. We also set the errors to $\infty$ in the region $x \in [76,96]$, $y \in[530,1350]$ of the detector, where a portion of the second spectral order is visible. This is appropriate treatment as the 1/$f$ noise scales independently across each order due to the difference in wavelength coverage. After correction of the 1/$f$ noise, we trace the location of Order 1 on the detector by computing the maximum of the trace convolved with a Gaussian filter for all columns. We further smooth the positions of the trace centroids using a spline function. Finally, we perform a box spectral extraction of the first order using the \texttt{transitspectroscopy.spectroscopy.getSimpleSpectrum} routine with an aperture diameter of 30 pixels.

\subsection{\texttt{nirHiss} Reduction}\label{sec:eureka!} 
We use the \texttt{nirHiss} Python open-source data reduction pipeline as described in Feinstein et al. (2023)\cite{feinstein_early_2023}. To summarize, this pipeline uses \eureka\ \citep{bell_eureka_2022} to go from the Stage 0 JWST outputs to Stage 2 calibration, which applies detector-level corrections, produce count rate images, and calibrates individual exposures. From the Stage 2 ``calints" FITS files, we use \texttt{nirHiss} to correct for background sources, 1/$f$ noise, cosmic ray removal, trace identification, and spectral extraction. We follow two steps for trace identification. First, we use the (x,y) position of the trace from the ``jwst\_niriss\_spectrace\_0022.fits" reference file, with y-values offset by ${\sim}25$ pixels, to identify Order 2. We mask this region, such that it does not contaminate the trace identification or background routine later on. Next, we use the \texttt{nirHiss.tracing.mask\_method\_edges} function. This technique identifies the edges of Order 1 using a canny edge detection method from \texttt{scikit-image}, an open-source image processing package \citep{scikit-image}. This method uses the derivative of a Gaussian function in order to identify regions with the maximum gradient. From this step, the potential edges are narrowed down to 1-pixel curves along the maxima. This results in an image where the outline of Order 1 is presented. We identify the median location along the column from the top and bottom edges of Order 1, and smooth the trace by fitting a 4\textsuperscript{th}-order polynomial. We use the trace to mask the location of Order 1 when stepping through the \texttt{nirHiss} background routines. For background treatment, we follow a similar method presented in Feinstein et al. (2023)\cite{feinstein_early_2023}, namely we identify a region without significant contamination from the spectral trace, and scale this region to the same region on the model background on the STScI JDox User Documentation. We use the region $x \in [4, 250]$ and $y \in [0, 30]$ and find an average scaling factor of 0.6007. We apply this scaling factor to the model background and subtract it from all integrations. Next, we remove $1/f$ noise in a similar manner to \texttt{transitspectroscopy} and scale this $1/f$ noise treatment to the out-of-eclipse integrations (0 -- 1250 and 1900 -- 2500). We remove cosmic rays using the L.A. Cosmic technique \citep{vandokkum01, ccdproc}. Finally, we extract the spectra using the optimal extraction routine, which is a robust means to simultaneously remove additional bad pixels/cosmic ray events while placing non-uniform weighting on each pixel in order to negate distortion produced by the spatial profile \cite{horne86}. We use a normalized median image to best capture the unique NIRISS/SOSS spatial profile.

\subsection{\texttt{supreme-SPOON} Reduction}\label{sec:sSpoon_ex}
We follow a similar approach with \texttt{supreme-SPOON} as presented in Feinstein et al. (2023)\cite{feinstein_early_2023}. We start from the raw uncalibrated data files, which we downloaded from the MAST archive, and process them through the \texttt{supreme-SPOON} Stage 1, which performs the detector level calibrations including superbias subtraction, saturation flagging, jump detection, and ramp fitting. As with the previous pipelines, we do not perform any dark current subtraction. \texttt{supreme-SPOON} additionally treats the 1/$f$ noise at the group level. This is done by subtracting a median stack of all in-eclipse integrations, scaled to the flux level of each individual integration via the white light curve, to create a difference image revealing the characteristic 1/$f$ striping. A column-wise median of the n$\rm ^{th}$ difference image is then subtracted from the corresponding integration. \texttt{supreme-SPOON} also removes the zodiacal background signal directly before the 1/$f$ correction step via scaling the SUBSTRIP96 SOSS background model provided by the STScI to the flux level of each integration, as is described in Feinstein et al. (2023)\cite{feinstein_early_2023}. We then pass the Stage 1 processed files through \texttt{supreme-SPOON} Stage 2, which performs additional calibrations such as flat fielding and hot pixel interpolation. We extract the stellar spectra at the pixel level using a simple box aperture extraction with a width of 30 pixels centered on the order 1 trace, as the dilution resulting from the order overlap with the second order has been shown to be negligible \cite{darveau-bernier_atoca_2022, radica_applesoss_2022}. The y-pixel positions of the trace are determined via the edgetrigger algorithm \cite{radica_applesoss_2022}. We find that the extracted trace positions match with those measured during commissioning and included in the \texttt{spectrace} reference file, and we therefore use the default JWST wavelength solution.

\subsection{\texttt{transitspectroscopy} Reduction}\label{sec:STScI_ex}
We follow a similar approach adopted by the {\tt transitspectroscopy} pipeline discussed in Feinstein et al. (2023)\cite{feinstein_early_2023}. This reduction starts from the {\tt *\_rateints.fits} files that were processed by the {\tt jwst} pipeline from STScI. We scaled the zodiacal background model provided on STScI JDox User Documentation to the observed two-dimensional spectra in the box delimited by pixels $x \in [10, 250]$ and $y \in [10, 30]$. The scaled background is then subtracted from each integration. In summary, the procedure to remove the 1/$f$ noise is as follows: We take the median of all integrations and subtract it from each integration, which leaves predominantly the 1/$f$ noise. We then take the column-by-column median of this residual noise, considering only the pixels that are 20 pixels away from the center of the trace, and assume it is representative of the structure of the 1/$f$ noise of the images. These values are then subtracted from each column. For the spectral extraction, we used the {\tt transitspectroscopy.spectroscopy.getSimpleSpectrum} routine with an aperture width of 30 pixels. We removed the outliers of the extracted spectra caused by cosmic rays or deviating pixel by taking the combined median of all spectra and flagging outlier points that deviate by more than 5$\sigma$ from this median spectrum. The flagged wavelength bins are then corrected by taking the mean of the neighboring bins.

\subsection{Spectrophotometric Light Curve Creation.}\label{sec:lc_fit}
From the aforementioned stellar spectral extraction pipelines, we create and fit models to the spectrophotometric light curves $F(t,\lambda)$. The light curves are composed of three distinct signals: the planetary flux throughout partial phase curve and eclipse $F_p$, the stellar flux $F_*$, and the systematics $S$, which we model as a function of time $t$ and wavelength $\lambda$ via equation~\ref{eq:flux}. 

\begin{equation}\label{eq:flux}
F(t,\lambda) = S(t,\lambda)\left[\frac{F_p(t,\lambda,\theta)}{F_*(t,\lambda)}+ 1\right]
\end{equation}

\noindent
As the main scientific quantity of interest is the planetary signal $F_p(t,\lambda,\theta)$, where $\theta$ represents the planetary orbital parameters, we aim to properly characterize and correct for the stellar flux and systematics. Light curve fitting is performed in two separate steps: (I) we fit the white light curve (II) we run individual fitting for each spectrophotometric bin. The values of the orbital parameters retrieved from the white light curve are fixed in the spectrophotometric light curves. The following sections describe our treatment of the planetary flux $F_p(t,\lambda,\theta)$, the stellar variability $F_*(t,\lambda)$, and the systematics $S(t,\lambda)$.

\subsection{Light Curve Component I: Planetary Flux}\label{sec:pla_model}

Despite the fact that the main target of these observations was the secondary eclipse of WASP-18\,b, we also capture a portion of its phase curve during the before- and after-eclipse baseline. Over the course of the observations, the planet rotates 107$^\circ$, revealing significant information on the spatial distribution of its atmosphere. To model the planetary flux in time, we consider a second-order harmonic function

\begin{equation}\label{eq:pla_flux}
F_p(t,\lambda,\theta) = f(t,\theta) \sum_{n=1}^2\left[ F_n + F_n \cos\left(\frac{2\pi n}{P}[t-t_n(\lambda)]\right)\right], 
\end{equation}

\noindent 
where $0\leq f(t,\theta)\leq 1$ is the time-dependent, visible fraction of the planetary projected disk as a function of the orbital parameters $\theta$ and $P$ is the orbital period. The harmonics consist of a term describing the semi-amplitude $F_n$ of the planetary flux variation, as well as the time $t_n$ where the harmonic reaches its maximum. The visible fraction is computed using the normalized secondary eclipse light curve modeled with the \texttt{batman} python package\cite{kreidberg_batman_2015}. The second-order harmonic function provides sufficient precision for the noise floor of the JWST\cite{cowan_inverting_2008}. We fit for the orbital parameters that dictate the shape and duration of the eclipse: the time of superior conjunction $T_{sec}$ ($\mathcal{U}[2459802.28,2459802.48]$), the impact parameter $b$ ($\mathcal{N}[0.360,0.026^2]$), as well as the semi-major axis to stellar radius ratio $a/R_*$ ($\mathcal{N}[3.496,0.029^2]$). We assume the orbit to be circular, which is justified by TESS analysis as it finds a strong Bayesian Information Criterion\cite{raftery_bayesian_1995} (BIC) and Aikake Information Criterion\cite{1974aikake} (AIC) preference for the non-eccentric orbit. Given the close proximity of WASP-18b to its host star, strong tidal interactions lead to the ellipsoidal deformation of the planet and its host. Past studies have shown that this deformation for WASP-18b is of $\sim$2.5$\times$10$^{-3}$ R$_{p}$, leading to a variation of the flux of order unity ppm, and is thus negligible\cite{leconte_distorted_2011,arcangeli_climate_2019}. Finally, near the lower end of the first order (0.85$\mu$m), there is also a contribution to the observed flux from the stellar light reflected by WASP-18b. However, the upper limits of the geometric albedo ($A_g<$0.048\cite{shporer_tess_2019} and $A_g$ = 0.025$\pm$0.027\cite{blazek_constraints_2022}) obtained from TESS correspond to a reflected light contribution of $<$35 ppm near 0.8$\mu$m. We therefore do not consider a specific term for the reflected light and instead assume that this will be fit by the second-order harmonic function. 


\subsection{Light Curve Component II: Stellar Variability}\label{sec:stellar_act}
We consider three phenomena that can lead to temporal changes in the observed stellar flux: stellar activity, $A$, ellipsoidal variations, $E$, and Doppler boosting, $D$. Stellar activity, generally caused by the presence and movement of star spots on the stellar hemisphere visible to the observer, leads to variations in the observed stellar spectrum on a timescale that is of the order of the stellar rotation period. Past observations of WASP-18 have constrained this period to be $P_{rot} \sim 5.5$ days\cite{han_exoplanet_2014}. Despite this relatively short period with respect to stars of similar physical properties (e.g., the effective temperature, $T_\textrm{eff}$, and luminosity, $L_\odot$)\cite{fetherolf22}, the star shows abnormally low activity in the UV and X-ray domains, possibly due to tidal interactions with WASP-18b disrupting its dynamo\cite{miller_detectability_2012,pillitteri_no_2014,shkolnik_signatures_2018}. As we expect the rotational modulation to be on a relatively long timescale compared to our observations and its amplitude to be quite low, we do not directly fit for this term and instead assume it to be handled by the systematics model. As for the ellipsoidal variation and Doppler boosting, they are both caused by the influence of WASP-18b on its host star. Although the ellipsoidal deformation of WASP-18b leads to a negligible impact on the phase curve, the same is not true for its host. The stellar ellipsoidal effect, with maxima fixed at quadrature when the projected area is at its highest, is found to be of semi-amplitude 172.2 ppm from the TESS analysis. Following previous analyses of HST observations\cite{arcangeli_climate_2019}, we consider ellipsoidal variation to be achromatic and fix its amplitude to the TESS value for the full NIRISS/SOSS wavelength range. The Doppler boosting effect is a result of the Doppler shift of the stellar spectral energy distribution as its radial velocity varies throughout its orbit. We fix this amplitude to 21.8 ppm, with the maximum at phase 0.25, as done in Arcangeli et al. (2019)\cite{arcangeli_climate_2019}. The observed stellar flux is therefore described as the sum of the ellipsoidal variation and Doppler boosting to the mean stellar flux in time $\bar{F_*}(\lambda)$.

\begin{equation}\label{eq:star_flux}
F_*(t,\lambda) = \bar{F}_*(\lambda) - D \sin\left(\frac{2\pi}{P}[t - T_{sec}]\right) - E\cos\left(\frac{4\pi}{P}[t-T_{sec}]\right) 
\end{equation}

\subsection{Light Curve Component III: Systematics Model}\label{sec:sys_mod}
The white and spectrophotometric light curves show two distinct important systematics: a sudden drop in flux caused by a tilt event shortly after the beginning of full-eclipse as well as high-frequency variations in the flux throughout the observations caused by small changes in the morphology of the trace. We track the trend of these systematics throughout the observations by performing incremental principal components analysis (IPCA) with the open-source \texttt{scikit-learn}\cite{scikit-learn} package on the processed detector images (Extended Data Fig. \ref{fig:pca_niriss}). The first principal component is the tilt event, which we use to determine the exact integration where it occurs. We handle the tilt event in the white and spectrophotometric light curves by fitting for an offset in flux for the data after the 1336th integration. We also observe two principal components analogous to the y-position and full width at half maximum (FWHM) of the trace in time. We find that these two components are correlated to short-frequency variations in the light curves and therefore detrend linearly against them at the fitting stage. We find that, despite having a lower variance than the y-position, the variation of the FWHM has a larger effect on the light curves when using box extraction. Finally, we fit for a linear trend in time to account for any further potential stellar activity and instrumental effect.

We note that considering a second order polynomial trend or higher in time for the systematics results in significant correlation with the partial phase curve information. However, a linear trend systematics model has been found sufficient to fit past NIRISS/SOSS observations (Feinstein et al. (2023)\cite{feinstein_early_2023}, Radica et al. submitted). Furthermore, we find that the curvature around secondary eclipse increases monotonically with wavelength, which is expected from the planetary signal and inconsistent with stellar activity as well as instrumental effects.

\subsection{Light Curve Fitting}
Light curve fitting is performed on the extracted spectrophotometric observations using the ExoTEP framework\cite{benneke_water_2019}. With the orbital parameter values constrained from the white light curve ($T_{sec}$ = 2459802.381867$_{-0.000089}^{+0.000092}$, $a/R_* = 3.483_{-0.020}^{+0.021}$, $b = 0.340_{-0.018}^{+0.016}$), we solely fit for the planetary flux and systematics for each spectrophotometric light curve. We chose a resolution of 5 pixels/bin for our spectrum, corresponding to 408 spectrophotometric bins, to mitigate potential correlation between wavelengths in the atmospheric retrievals as pixels in the spectral direction are not independent. All fits for the 408 bins are then done independently. Retrievals are performed using the Affine Invariant Markov chain Monte Carlo Ensemble sampler \texttt{emcee}\cite{foreman-mackey_emcee_2013}, using 20,000 steps and 4 walkers per free parameter. The first 12,000 steps, 60\% of the total amount, are discarded as burn-in to ensure that the samples are taken after the walkers have converged. The samples from the white and spectrophotometric light curves are used to produce two science products: the detrended white light curve and dayside thermal emission spectrum. The detrended white light curve is obtained by dividing out the best-fit systematics model and subtracting the stellar variability from the light curve to isolate the planetary signal. For the dayside thermal emission, the median values and uncertainties of $F_p(T_{sec},\lambda)$ are computed from the samples of the parameters of equation \ref{eq:pla_flux}.

\subsection{TESS Phase Curve Analysis}\label{sec:TESS}

During the TESS Primary Mission, the WASP-18 system was observed in Sectors 2 and 3 (2018 August 22 to October 18). The full-orbit phase curve was analyzed in several earlier publications\cite{shporer_tess_2019,wong_tess_2020}, which reported a robust detection of the planet's secondary eclipse and high signal-to-noise measurements of the planet's phase curve variation and signals corresponding to the host star's ellipsoidal distortion and Doppler boosting. During the ongoing TESS Extended Mission, the spacecraft reobserved WASP-18 in Sectors 29 and 30 (2020 August 26 to October 21). We carried out a follow-up phase curve analysis of all four Sectors' worth of TESS data, following the same methods used previously.

The data from the TESS observations were processed by the Science Processing Operations Center (SPOC) pipeline, which yielded near-continuous light curves at a 2-minute cadence. In addition to the raw extracted flux measurements contained in the simple aperture photometry (SAP) light curves, the SPOC pipeline also produced the pre-search conditioning (PDC) light curves, which have been corrected for common-mode instrumental systematics trends that are shared among all sources on the corresponding detector. We used these PDC light curves for our phase curve analysis. After dividing the light curves into individual segments that are separated by the spacecraft's scheduled momentum dumps, we fit each segment to a combined phase curve and systematics model. The astrophysical phase curve model consists of two components describing the planetary and stellar fluxes:
\begin{equation}\label{planet}
F_{p}(t) = \bar{f_{p}} - F_{\rm atm} \cos(\phi)
\end{equation}
\begin{equation}\label{star}
F_{*}(t) = 1 + D\sin(\phi) - E\cos(2\phi)
\end{equation}
The Doppler boosting $D$ and ellipsoidal distortion $E$ semi-amplitudes are defined as before. Here, the planet's orbital phase is defined relative to the mid-transit time $T_0$: $\phi = 2\pi(t-T_0)$. The planet's phase curve contribution has a single mode with a semi-amplitude of $F_{\rm atm}$ and oscillates around the average relative planetary flux $\bar{f_{p}}$. The transit and secondary eclipse light curves $\phi_t$ and $\phi_e$ were modeled using \texttt{batman} with quadratic limb-darkening. In this parameterization, the secondary eclipse depth (i.e., dayside flux) and nightside flux are $\bar{f_{p}} + F_{\rm atm}$ and $\bar{f_{p}} - F_{\rm atm}$, respectively. For the systematics model, we used polynomials in time and chose the optimal polynomial order for each segment individually that minimized the BIC.

We used ExoTEP to calculate the posterior distributions of the free parameters through a joint MCMC fit of all light curve segments. In order to reduce the number of free parameters in the joint fit, we first carried out fits to smaller groups of light curve segments corresponding to each TESS Sector and then divided the light curve segments by the best-fit systematics model. In the final joint fit of the systematics-corrected TESS light curve, no additional systematics parameters were included. We accounted for time-correlated noise (i.e., red noise) by fixing the uncertainty of all datapoints within each Sector to the standard deviation of the residuals, multiplied by the fractional enhancement of the average binned residual scatter from the expected Poisson noise scaling across bin sizes ranging from 30 minutes to 8 hours\cite{wong_tess_2020}. In addition to the phase curve parameters described above ($\bar{f_{p}}$, $F_{\rm atm}$, $D$, $E$), we allowed the mid-transit time $T_0$, orbital period $P$, relative planetary radius $R_{p}/R_{*}$, scaled orbital semi-major axis $a/R_*$, impact parameter $b$, and quadratic limb-darkening coefficients $u_1$ and $u_2$ to vary freely.

The results of our TESS phase curve fit are listed in Table~\ref{tab:tess_results}. The updated values for the orbital ephemeris and transit shape parameters are statistically consistent with the published results from the previous TESS phase curve analyses\cite{shporer_tess_2019,wong_tess_2020}, while being significantly more precise. We used the median and $1\sigma$ uncertainties of $a/R_*$ and $b$ as Gaussian priors in the NIRISS/SOSS white light curve fit. We also used the median values of $P$, $R_p/R_*$, $D$, and $E$ as fixed parameters for the NIRISS/SOSS white and spectrophotometric light curves fit. We obtained a secondary eclipse depth of $357 \pm 14$ ppm and a nightside flux that is consistent with zero. All three phase curve amplitudes were measured at high signal-to-noise: $F_{\rm atm} = 177.5 \pm 5.7$ ppm, $D = 21.8 \pm 5.2$ ppm, and $E = 172.2 \pm 5.6$ ppm.

\subsection{Secondary Eclipse Mapping.}\label{sec:sec_ecl_map}

To perform eclipse mapping, we use both ThERESA\cite{challener_theresa_2022} and the methods of Mansfield et al. (2020)\cite{mansfield_eigenspectra_2020}, which are separate implementations of the same process introduced by Rauscher et al. (2018)\cite{rauscher_more_2018} when applied to white light curves, to cross-check our results. First, we generate a basis set of light curves from spherical harmonic maps\cite{louden_spiderman_2018,luger_starry_2019} with degree $l \leq l_{\rm max}$, then transform these light curves to a new, orthogonal basis set of ``eigencurves'' using principal component analysis (PCA). Each eigencurve corresponds to an ``eigenmap,'' the planetary flux map which, when integrated over the visible hemisphere at each exposure time, generates the corresponding eigencurve.

We then fit the white light curve with a linear combination of a uniform-map light curve, the $N$ most informative (largest eigenvalue) eigencurves, and a constant offset term to adjust for the fact that the observed planetary flux during eclipse (when the planet is entirely blocked by the star) must be equal to 0 and to allow for adjustments to light-curve normalization. Since the eigenmaps represent differences from a uniform map, it is possible to recover a fit which contains regions of negative planet emission. 
This is physically impossible, so we impose a positivity constraint on the total flux map. While the eigenmaps are mathematically defined across the entire planetary sphere, our observations only constrain the portion of the planet that is visible during the observation, so we only enforce the flux positivity condition in the visible region of the planet. 
We test all combinations of $l_{\rm max} \leq 6$ and $N \leq 8$ using a least-squares minimization and select the optimal values by minimizing the BIC.

We find that the fit with the lowest BIC to the broadband light curve has $l_{\rm max}=5$ and $N=5$. However, the fit with $l_{\rm max}=2$ and $N=5$ was only slightly less preferred, so here we explore the inferred brightness distribution from both solutions. Figure \ref{fig:eigenmaps} shows the resulting light curve for the $l_{\rm max}=5$, $N=5$ solution after sequential subtraction of each eigencurve. The preference for a fit with five eigencurves is driven by the residuals in ingress and egress, which can be seen by eye and are not sufficiently corrected for with a uniform map. Including the uniform-map light curve and the constant term, the fit thus contained a total of seven free parameters. We used a Markov Chain Monte Carlo (MCMC) procedure to estimate parameter uncertainties. For the analysis following Mansfield et al. (2020)\cite{mansfield_eigenspectra_2020}, we test for convergence of the MCMC by ensuring that the chain length is 50 times the autocorrelation timescale, while for the analysis using ThERESA\cite{challener_theresa_2022} we use the Gelman-Rubin convergence test\cite{Gelman1992}. 

The resulting weights of each eigencurve are then applied to the corresponding eigenmaps to generate a flux map of the planet. 
We convert the star-normalized flux map to brightness temperature by assuming the planet and star are blackbodies emitting at the NIRISS/SOSS throughput-weighted mean wavelength of 1.6 microns\cite{rauscher_more_2018}.
We estimate temperature map uncertainties by computing a subsample of maps from the MCMC posterior distribution and calculating 68.3\%, 95.5\%, and 99.7\% quantiles at each location.

Figure~\ref{fig:eclipse_map} shows the resulting broadband brightness temperature map for the $l_{\rm max}=5$, $N=5$ case and longitudinal brightness temperature profiles for both the $l_{\rm max}=5$, $N=5$ and $l_{\rm max}=2$, $N=5$ cases, calculated by averaging meridian flux at each longitude weighted by $\cos{\rm (latitude)}^2$. Additionally, we compare the equatorial slices to predictions from several GCMs (see General Circulation Models section). We note that not all structures on a planetary map will leave an observable signature in a secondary eclipse light curve. When comparing GCMs to secondary eclipse maps, it is important to only compare the components of GCM maps which can be physically accessed with eclipse mapping. Therefore, we use the methods of Luger et al. (2021)\cite{Luger2021} to separate each GCM map into the ``null space'', or components that are inaccessible to eclipse mapping observations, and the ``preimage'', or components that are accessible through mapping. Figure~\ref{fig:eclipse_map} compares the longitudinal temperature trends from only the preimage of each GCM to the observed map. We find that both map solutions agree on a steep gradient in temperature near the limbs, which is well matched by GCM predictions. Additionally, both maps show a temperature distribution roughly symmetrical in longitude about the substellar point. However, the two maps disagree on the exact shape of the brightness distribution. The $l_{\rm max}=5$, $N=5$ map shows an extended hot plateau region of roughly constant temperature from -40$^\circ$ to +40$^\circ$ in longitude, while the $l_{\rm max}=2$, $N=5$ map shows a more concentrated hot spot with a steady decrease in temperature away from the substellar point. As these maps both fit the data with similar BIC, the current data do not give us the necessary precision to determine which solution represents the true temperature distribution of WASP-18b. Future observations at higher precision may distinguish between these two modes of solutions. 

To test our ability to constrain latitudinal structures, we performed two eclipse-mapping fits: the eigenmapping fit presented above and a fit where the initial basis set of maps is a longitudinal Fourier series that is constant with latitude (see Extended Data Fig.~\ref{fig:latitude}). Both methods retrieve similar longitudinal temperature structures, with steep gradients near the limb and a extended hot plateau. However, the constant-with-latitude map is also able to fit the data well, indicating a lack of constraints on latitudinal features within the uncertainties on the data. This is not unexpected as the relatively low impact parameter ($b$ = 0.36) of WASP-18b results in a lower amount of latitudinal information contained in the secondary eclipse signal. Further observations of WASP-18b, or of planets with higher impact parameter, may enable us to pull latitudinal signals out of the noise.

Our eclipse mapping assumes the orbital parameters of the system are precisely known relative to data uncertainties, a safe assumption with Spitzer data\cite{rauscher_more_2018}. With JWST, data quality may be high enough that uncertainties on orbital parameters impart significant uncertainty on the mapping results. To test this, we ran analyses with impact parameter, orbital semi-major axis, and eclipse time fixed to values $\pm 1\sigma$. In some cases, this led to a "hotspot" model like the red one in Figure \ref{fig:eclipse_map} being preferred over a "plateau" model like the blue one, which is unsurprising given their similar statistical preference. However, all resulting maps were well within the uncertainties of one of the two models presented. We note that, while the eccentricity is kept fixed to zero throughout this analysis, considering an eccentric orbit would allow to first order for variations in mid-eclipse time and eclipse duration\cite{winn2011}. Past radial velocity measurements have found a small but non-zero eccentricity for WASP-18b of $e = 0.0076\pm0.0010$\cite{Bonomo_2017}, corresponding to an offset of the time of mid-eclipse of 9 seconds, as well as a difference of 120 seconds between the transit and eclipse durations. These differences in eclipse timing and duration are of the same magnitude as those induced while varying $T_{sec}$, $a/R$, and $b$ (8 seconds for the mid-eclipse time and 90 seconds for the eclipse duration). Therefore, performing the mapping considering a circular orbit while varying $T_{sec}$, $a/R$, and $b$ is analogous to effects that could be expected from an eccentric orbit.



\subsection{Atmospheric Retrieval.}\label{sec:atm_retrieval}

We perform 1D atmospheric retrievals on the \texttt{NAMELESS} reduction at a resolution of 5 pixels per bin (408 bins) using four techniques with varying levels of physical assumptions: a self-consistent radiative-convective-thermochemical equilibrium grid retrieval (ScCHIMERA), a chemical equilibrium with free temperature-pressure profile retrieval (SCARLET), a free chemistry retrieval with thermal dissociation (HyDRA), and a free chemistry retrieval with abundances assumed constant with altitude (POSEIDON). All retrieval methods considered the same PHOENIX stellar spectrum\cite{jack_time-dependent_2009}, produced using previously published parameters for WASP-18 (i.e., $T_{eff}$ = 6435~K, $\log g$ = 4.35, and [Fe/H] = 0.1\cite{cortes-zuleta_tramos_2020,stassun_revised_2019}), to convert from model planet flux spectra to $F_p/F_s$ values. We chose to use a model stellar spectrum instead of the extracted spectrum to avoid the possible introduction of systematic errors in the through the process of absolute flux calibration.

\subsection{1D Radiative-Convective-Thermochemical Equilibrium (1D-RCTE) Grid Retrieval}\label{sec:grid_ret}
We use a 1D-RCTE grid retrieval-based method, ScCHIMERA\cite{Piskorz2018,arcangeli_h-_2018}, with the opacity sources described in Mansfield et al. (2021)\cite{mansfield_unique_2021} and Glidic et al. (2022)\cite{Glidic2022}.
These 1D-RCTE models assume cloud-free one-dimensional radiative-convective-thermochemical equilibrium using the methods described in Toon et al. (1989)\cite{Toon1989} to solve for the net flux divergence across each layer of the atmosphere, and the Newton-Raphson iteration scheme of McKay et al. (1989)\cite{McKay1989} to march towards an equilibrium vertical temperature structure. The NASA Chemical Equilibrium with Applications-2 (CEA2) routine\cite{Gordon1994} is used to compute the thermochemical equilibrium gas and condensate mole fractions for hundreds of relevant species. Opacities are computed with the correlated-K random-overlap-resort-rebin method\cite{Amundsen2017}.
Input elemental abundances from Lodders \& Palme (2009)\cite{LoddersPalme2009} are scaled to a given metallicity ([M/H]) and carbon-to-oxygen ratio (C/O). 

Using WASP-18b's planetary and stellar parameters, we produced a grid of 2730 1D-RCTE models and resulting top-of-atmosphere thermal emission spectra spanning the atmospheric C/O (0.01 -- 2.0) , [M/H] (-2.0 -- 2.0, where 0 is solar, 1 is 10$\times$, etc), and heat redistribution ($f$, 1.0 -- 2.8, where 1=full, 2=dayside, and 2.67 is the max value, as defined in Arcangeli et al. (2019)\cite{arcangeli_climate_2019}). We additionally include a scale factor, $A_{HS}$ (allowed to vary from 0.5 -- 2.0), that multiplies the planetary flux by a constant to account for a hot spot area fraction emitting most of the observed flux\cite{Taylor2020}. The PyMultiNest\cite{Buchner2014} routine is used to sample the 1D-RCTE spectra via interpolation (and subsequent binning to the data wavelength bins) to obtain posterior probability constraints on the above parameters. We have made public our grid models, including temperature-pressure profiles, molecular abundances, and emission spectra, as well as additional figures showing the posteriors of retrieved parameters and the impact of each parameter on the spectrum. This can be found on Zenodo under the DOI: 10.5281/zenodo.7332105.

\subsection{Chemical Equilibrium \& Free Temperature Retrieval}\label{sec:chem_equi}
We use the SCARLET atmospheric retrieval framework\cite{benneke_atmospheric_2012,benneke_how_2013,benneke_strict_2015,benneke_water_2019} to perform a chemical equilibrium retrieval with a free temperature-pressure profile on our retrieved dayside thermal emission spectrum. The SCARLET forward model computes the emergent disk-integrated thermal emission for a given set of molecular abundances, temperature structure, and cloud properties. The forward model is then coupled to the Affine Invariant Markov chain Monte Carlo Ensemble sampler emcee\cite{foreman-mackey_emcee_2013} to constrain the atmospheric properties. Due to the high dayside temperature and large pressures probed through thermal emission spectroscopy, we assume the atmosphere is in thermochemical equilibrium. For the equilibrium chemistry, we consider the following species: H$_2$, H, H$^-$\cite{Bell1987,John1988}, He, Na, K\cite{vald1995,Burrows_2003}, Fe, H$_2$O\cite{Polyansky2018}, OH\cite{Rothman2010}, CO\cite{Rothman2010}, CO$_2$\cite{Rothman2010}, CH$_4$\cite{Yurchenko_2014}, NH$_3$\cite{Coles_2019}, HCN\cite{Barber_2013}, TiO\cite{McKemmish2019}, VO \cite{McKemmish2016}, and FeH\cite{Wende2010}.
The abundances of these species are interpolated in temperature and pressure using a grid of chemical equilibrium abundances from FastChem2\cite{fastchem2}, which includes the effects of thermal dissociation for all the species included in the model. These abundances also vary with the atmospheric metallicity, [M/H] ($\mathcal{U}[-3,3]$), and carbon-to-oxygen ratio, C/O ($\mathcal{U}[0,1]$), which are considered as free parameters in the retrieval. As for the temperature structure, we use a free parametrization\cite{pelletier_where_2021} which here fits for $N = 10$ temperature points ($\mathcal{U}[100,4500]$~K) with fixed spacing in log-pressure (P = 10$^2$ -- 10$^{-6}$ bar). Although this parametrization is free, it is regularized by a prior punishing for the second derivative of the profile using a physical hyper-parameter, $\sigma_s$, with units of kelvin per pressure decade squared (K dex$^{-2}$). This prior is implemented to prevent over-fitting and nonphysical temperature oscillations at short pressure scale lengths. For this work, we use a hyper-parameter value of $\sigma_s =$ 1000 K dex$^{-2}$, corresponding to a low punishment against second derivatives as we want the retrieval to explore freely the temperature-pressure profile parameter space. We note that further lowering this punishment does not affect the retrieved TP profile. Finally, we fit for an area fraction $A_{HS}$ ($\mathcal{U}[0,1]$) that is multiplied directly with the thermal emission spectrum, for a total of 14 free parameters. This factor is used to compensate for the presence of a hot spot which, while taking up only a portion of the planetary disk, contributes almost completely to the observed emission\cite{Taylor2020}. For the retrieval, we use four walkers per free parameter and consider the standard chi-square likelihood for the spectrum fit. We run the retrieval for 25,000 steps and discard the first 15,000 steps, 60\% of the total amount, to ensure that the samples are taken after the walkers have converged. Spectra are initially computed using opacity sampling at a resolving power of R = 15,625, which is sufficient to simulate JWST observations\cite{rocchetto_exploring_2016}, convolved to the instrument resolution and subsequently binned to the retrieved wavelength bins.

\subsection{Free Chemistry \& Free Temperature-Pressure Profile Retrieval}\label{sec:freechem_ret}

We use two independent atmospheric retrieval codes to perform free-chemistry retrievals on WASP-18b's dayside thermal emission spectrum: \textsc{H}{\small y}\textsc{DRA} \citep{Gandhi2018, gandhi_h-_2020, Piette2020b, Piette2022}, including the effects of thermal dissociation, and \textsc{POSEIDON} \citep{MacDonald2017,MacDonald2023}, which here assumes constant-with-depth abundances for all chemical species.

\textsc{H}{\small y}\textsc{DRA} consists of a parametric forward atmospheric model coupled to a \textsc{Python} implementation of the \textsc{multinest}\citep{Feroz2009} Nested Sampling Bayesian parameter estimation algorithm \citep{Skilling2006}, \textsc{PyMultiNest} \citep{Buchner2014}. The inputs to the parametric model are the deep-atmosphere mixing ratios for each of the chemical species included, the temperature-pressure profile parameters (6 free parameters), and a dilution parameter (area fraction) to account for 3D effects on the dayside \citep{Taylor2020}. Given the high dayside temperatures of WASP-18b, we consider high-temperature opacity sources and the effects of thermal dissociation, as described in Gandhi et al. (2020)\cite{gandhi_h-_2020}. We include opacity due to the molecular, atomic and ionic species with spectral features in the 0.8--2.8~$\mu$m range which are expected in a high-temperature, H$_2$-rich atmosphere: collision-induced absorption (CIA) due to H$_2$-H$_2$ and H$_2$-He \citep{Richard2012}, H$_2$O \citep{Rothman2010}, CO \citep{Rothman2010}, CO$_2$ \citep{Rothman2010}, HCN \citep{Harris2006}, OH \citep{Rothman2010}, TiO \citep{McKemmish2019}, VO \citep{McKemmish2016}, FeH \citep{Dulick2003}, Na \citep{Burrows2003}, K \citep{Burrows2003} and H$^-$\citep{Bell1987,John1988}. The line-by-line absorption cross sections for these species are calculated following the methods in Gandhi \& Madhusudhan (2017)\cite{Gandhi2017}, using data from the references listed. The H$^-$ free-free and bound-free opacity is calculated using the methods of Bell \& Berrington (1987)\cite{Bell1987} and John (1988)\cite{John1988}, respectively. \textsc{H}{\small y}\textsc{DRA} includes the effects of thermal dissociation of H$_2$O, TiO, VO and H$^-$ as a function of pressure and temperature, following the method of Parmentier et al. (2018)\citep{parmentier_thermal_2018}. In particular, the abundance profiles of these species are calculated following Equations 1 and 2 of ref.\,\citep{parmentier_thermal_2018}, where the deep abundance of each species ($A_0$) is a parameter in the retrieval, and the $\alpha$, $\beta$ and $\gamma$ parameters are those given in Table 1 of that same work. 
The abundance profiles of the remaining chemical species are assumed to be constant with depth.

\textsc{H}{\small y}\textsc{DRA} uses the parametric temperature-pressure profile of Madhusudhan \& Seager (2009)\citep{Madhusudhan2009}, which has been used extensively in exoplanet atmosphere retrievals, including ultra-hot Jupiters such as WASP-18b \citep{gandhi_h-_2020}. The temperature parameterization is able to capture thermally inverted, non-inverted, and isothermal profiles, spanning the range of possible thermal structures for ultra-hot Jupiters. The \textsc{H}{\small y}\textsc{DRA} retrievals also include an area fraction parameter, $A_{\rm HS}$, which multiplies the emergent emission spectrum by a constant factor to account for the dominant contribution of the hot spot \citep{Taylor2020}. Given the input chemical abundances, temperature-pressure profile parameters and area fraction, the model thermal emission spectrum is calculated at a resolving power of $R \sim$15,000, convolved to the instrument resolution, binned to the data resolution, and compared to the data to calculate the likelihood of the model instance. Detection significances are calculated for specific chemical species by comparing the Bayesian Evidences of retrievals which include/exclude the species in question \citep{trotta2008,benneke_how_2013}. These detection significances factor in the ability of the retrieval to fit the observations with a different temperature profile and/or other chemical species, when the species in question is not included. Since thermal emission spectra are very sensitive to the atmospheric temperature profile, changes in the temperature structure can, in some cases, somewhat compensate for the absence of a particular chemical species, contributing to a lower detection significance.

We additionally use \textsc{H}{\small y}\textsc{DRA} to test the sensitivity of the free chemistry retrievals to the limits of the log-normal prior distributions assumed for the chemical abundances. For the \textsc{H}{\small y}\textsc{DRA} retrieval including thermal dissociation effects, we test two scenarios: wide, uninformative priors for all 11 species included (log mixing ratio ranging from -12 to -1), and slightly more restricted priors for the refractory species included (log mixing ratio ranging from -12 to -4 for TiO, VO and FeH, -12 to -2 for Na and K, and -12 to -1 for the remaining species). The more restricted prior limits for the refractory species are motivated by the relatively lower abundances expected for these species compared to the volatile species, across a range of metallicities and C/O ratios \citep{Burrows1999,Madhusudhan2011,Piette2020a}. 

We find that the atmospheric properties retrieved with \textsc{H}{\small y}\textsc{DRA} are consistent within 1$\sigma$ for the two choices of prior limits. With the wide priors, the \textsc{H}{\small y}\textsc{DRA} retrieval infers a H$_2$O log mixing ratio of -3.09$^{+1.28}_{-0.32}$, with double-peaked posterior probability distributions for the H$_2$O and TiO abundances. While the dominant posterior peaks correspond to approximately solar H$_2$O and TiO abundances, the second, lower-likelihood peaks correspond to $\sim 30 \times$ and $\sim 10^4 \times$ super-solar H$_2$O and TiO abundances, respectively. Such an extreme TiO abundance warrants skepticism, and may, for example, be a result of the well-known degeneracy between chemical abundances and the atmospheric temperature gradient (see also Evans et al. (2017)\citep{Evans2017}). The retrieved H$_2$O abundance in this case is consistent with the chemical equilibrium retrievals and self-consistent 1D radiative-convective models described above, though with a larger uncertainty due to the double-peaked posterior distribution. When the restricted priors are used, the low-likelihood, high-abundance peaks in the H$_2$O and TiO posterior distributions are no longer present, and the retrieved H$_2$O abundance is -3.23$^{+0.45}_{-0.29}$, in excellent agreement with the chemical equilibrium retrievals and self-consistent 1D radiative-convective models. We note that the retrieved temperature-pressure profiles and abundances of CO, CO$_2$, HCN, OH, H$^-$ and FeH are unaffected by the choice of prior limits discussed above. The abundances of Na and K are unconstrained in both cases. While the two choices of prior limits give consistent results, the expectation of chemical equilibrium in the atmosphere of WASP-18b, in addition to the unlikelihood of a $\sim 10^4 \times$ super-solar TiO abundance, motivate the use of the somewhat restricted priors on the refractory chemical abundances.

We note that for either choice of prior, the retrieved deep-atmosphere abundance of VO is significantly higher than that inferred by the chemical equilibrium retrieval (Extended Data Fig. \ref{fig:mol_posteriors}). This is due to a difference in the thermal dissociation prescriptions; in the \textsc{H}{\small y}\textsc{DRA} retrieval, thermal dissociation results in a significantly depleted VO abundance at the photosphere, while in the chemical equilibrium retrieval thermal dissociation begins at lower pressures. Furthermore, the posterior distribution for the VO abundance peaks at highly super-solar values ($\sim 10^4 \times$ solar). Such a high abundance is physically unlikely, and may indicate the presence of further sources of optical opacity not included in the retrieval.

We also conduct free-chemistry retrievals, without the inclusion of thermal dissociation, using \textsc{POSEIDON}. \textsc{POSEIDON} is an atmospheric retrieval code originally designed for the interpretation of exoplanet transmission spectra \citep{MacDonald2017}. We have recently extended \textsc{POSEIDON} to include secondary eclipse emission spectra modelling and retrieval capabilities. For an ultra-hot Jupiter such as WASP-18b, where the dayside can be assumed clear, \textsc{POSEIDON}'s emission forward model calculates the emergent flux via a standard single stream prescription without scattering
\begin{equation}\label{eq:POSEIDON_rad_transfer_1}
I_{\rm{layer \, top}} (\mu, \lambda) = I_{\rm{layer \, bot}} (\mu, \lambda) e^{-d\tau_{\rm{layer}} (\lambda) / \mu} + (1 - e^{-d\tau_{\rm{layer}} (\lambda) / \mu}) \, B (T_{\rm{layer}}, \lambda)
\end{equation}
where $I_{\rm{layer \, bot}}$ and $I_{\rm{layer \, top}}$ are, respectively, the upwards specific intensity incident on the lower layer boundary and the intensity leaving the upper layer boundary,  $d\tau_{\rm{layer}}$ is the differential vertical optical depth across the layer, $\mu = \cos\theta$ specifies the ray direction, and $B (T_{\rm{layer}}, \lambda)$ is the black body spectral radiance at the layer temperature. Using the boundary condition $I_{\rm{deep}} (\mu, \lambda) = B (T_{\rm{layer}}, \lambda)$, \textsc{POSEIDON} propagates Equation \ref{eq:POSEIDON_rad_transfer_1} upwards to determine the emergent intensity at the top of the atmosphere. The emergent planetary flux is determined via 
\begin{equation}\label{eq:POSEIDON_rad_transfer_2}
F_{p, \, \rm{emergent}} (\lambda) = 2 \pi \sum_{\mu} \mu \, I_{\rm{emergent}} (\mu, \lambda) \, W (\mu)
\end{equation}
where $W$ are the Gaussian quadrature weights corresponding to each $\mu$ (here taken as second order quadrature over the interval $\mu \in [0, 1]$). Finally, the planet-star flux ratio seen at Earth is given by
\begin{equation}\label{eq:POSEIDON_rad_transfer_3}
\left(\frac{F_p}{F_*}\right)_{\rm{obs}} (\lambda) = \left(\frac{R_{p, \, \rm{phot}}  (\lambda)}{R_*}\right)^2 \frac{F_{p, \, \rm{emergent}} (\lambda)}{F_* (\lambda)}
\end{equation}
where $R_{p, \, \rm{phot}}$ is the effective photosphere radius\cite{Fortney2019,Taylor2022} at wavelength $\lambda$ (evaluated at $\tau (\lambda) = 2/3$). Since the calculation of $R_{p, \, \rm{phot}}$ requires a reference radius boundary condition to solve the equation of hydrostatic equilibrium, we prescribe $r (P = \rm{10 \, mbar})$ as a free parameter.

For the WASP-18b \textsc{POSEIDON} retrieval analysis, we calculated emission spectra via opacity sampling and explored the parameter space using \textsc{MultiNest} via its Python wrapper \textsc{PyMultiNest} \citep{Feroz2009,Buchner2014}. \textsc{POSEIDON} solves the radiative transfer on an intermediate resolution wavelength grid (here, $R =$ 20,000 from 0.8 to 3.0\,$\mu$m), onto which high-spectral-resolution ($R \sim 10^6$), pre-computed cross sections\citep{MacDonald2022} are down-sampled. For WASP-18b, we consider the following opacity sources: H$_2$-H$_2$ \citep{Karman2019} and H$_2$-He\citep{Karman2019} CIA, H$_2$O\citep{Polyansky2018}, CO\citep{Li2015}, CO$_2$\citep{Tashkun2011}, HCN\citep{Barber2014}, H$^{-}$\citep{John1988}, OH\citep{Brooke2016}, FeH\citep{Wende2010}, TiO\citep{McKemmish2019}, VO\citep{McKemmish2016}, Na\citep{Ryabchikova2015}, and K\citep{Ryabchikova2015}. We prescribed uniform-in-altitude mixing ratios, defined by a single free parameter for each of the chemical species included. The \textsc{PyMultiNest} retrievals with \textsc{POSEIDON} use 2,000 live points, and the six-parameter temperature-pressure profile\cite{Madhusudhan2009} outlined above in the description of \textsc{H}{\small y}\textsc{DRA}. \textsc{POSEIDON} accounts for the dominant contribution of the hot spot by prescribing the 10\,millibar radius as a free parameter, which is subsequently converted into an equivalent $A_{\rm HS}$ posterior 
by comparison to the white light planet radius. 

 We note that both \textsc{H}{\small y}\textsc{DRA} and \textsc{POSEIDON} yield consistent retrieval results when thermal dissociation is not considered in the \textsc{H}{\small y}\textsc{DRA} retrievals. However, the inclusion of thermal dissociation results in significantly different retrieved H$_2$O abundances and temperature-pressure profiles, which are in agreement with the chemical equilibrium retrievals and self-consistent 1D radiative-convective models described above.

\subsection{Disequilibrium Chemistry Model.}\label{sec:diseq_chem}
To further justify the use of chemical equilibrium models in our analysis of WASP-18b, we produce a grid of disequilibrium chemistry forward models to assess whether disequilibrium effects might strongly shape our observations. We begin by calculating the atmospheric temperature--pressure structure of WASP-18b under radiative-convective equilibrium with the \texttt{HELIOS} \cite{malik2017helios,malik2019helios} radiative transfer code. Next, we calculate altitude-dependent mixing ratios of chemical species under this temperature--pressure structure with the \texttt{VULCAN}\cite{tsai2017vulcan} 1-dimensional chemical kinetics code, using an N-C-H-O reaction network that includes ionization and recombination of Fe, Mg, Ca, Na, K, H, and He. We use the current version of \texttt{VULCAN} (\texttt{VULCAN2}\cite{Tsai_2021}), which includes optional photochemistry and parameterizes the transport flux of chemical species with eddy diffusion, molecular diffusion, thermal diffusion, and vertical advection. We update this code to include the effect of photoionization. Finally, we generate emission spectra with the \texttt{PLATON} radiative transfer code \citep{Zhang_2019} at the resolution and wavelength range of NIRISS/SOSS.\footnote{Our \texttt{PLATON} emission spectrum calculations use the code branch that allows varying chemical mixing ratios as a function of altitude: \url{https://github.com/ideasrule/platon/tree/custom_abundances}} We modify \texttt{PLATON} to calculate bound-free and free-free $\rm H^-$ opacity as a function of altitude; this addition is necessary to assess whether disequilibrium abundance $\rm H^-$ opacity could mute spectral features more strongly than predicted by equilibrium chemistry. We furthermore modify \texttt{PLATON} to accept higher-temperature (T $>$ 3000~K) opacity files that we calculate with the \texttt{HELIOS-K} code \citep{grimm2015helios, grimm2021helios}.

For our set of models, we vary the eddy diffusion coefficient, \kzz, from $10^7$~cm$^{-2}$\,s$^{-1}$ to $10^{13}$~cm$^{-2}$\,s$^{-1}$, holding it constant at all altitudes for a given simulation. We perform this sweep over many orders of magnitude of \kzz\, to understand the maximum effect that disequilibrium chemistry could have on the observed emission spectrum. While \kzz\, is a limited descriptor of vertical mixing and is expected to vary as a function of altitude (e.g., Parmentier et al. (2013)\cite{parmentier20133d}), we assume that our forward models bracket the expected vertical mixing behavior of this planet. This statement is further motivated by our GCM models, if we approximate $K_{\rm zz} = vH$ for vertical wind velocity $v$ and atmospheric scale height $H$ \cite{smith1998estimation, lewis2010atmospheric}. The minimum dayside-average \kzz\, for our kinematic MHD GCM (see General Circulation Models section) is $\sim 10^8$~cm$^{-2}$\,s$^{-1}$, and the maximum dayside-average \kzz\, is $\sim 10^9$~cm$^{-2}$\,s$^{-1}$, well within our \texttt{VULCAN} grid range. Our model grid also toggles the inclusion of molecular diffusion and photochemistry. As input to \texttt{VULCAN} when photochemistry is included, we use a stellar spectrum that is appropriate for WASP-18 from Rugheimer et al. (2013)\cite{rugheimer2013spectral}. The spectrum is constructed by joining synthetic spectra from Kurucz (1979)\cite{kurucz1979model} and UV flux measurements of Piscium HD 222368 by the International Ultraviolet Explorer\footnote{\url{https://archive.stsci.edu/iue/}} at 300 nm.

We find that our inclusion of disequilibrium chemistry effects---photochemistry, molecular diffusion, thermal diffusion, and eddy diffusion---produces spectra that are not strongly discrepant from spectra computed assuming chemical equilibrium. Indeed, all discrepancies between spectra produced under chemical equilibrium and spectra produced under chemical disequilibrium spectra are less than 10\,ppm. This agreement is expected, as chemistry at the pressure levels probed by low-resolution emission spectra are not predicted to be strongly modified by photochemistry or mixing (e.g., Tsai et al. (2021)\cite{Tsai_2021}). Furthermore, the high temperature of WASP-18b implies that the chemical timescales in the atmosphere are quite short, allowing chemical reactions to occur more quickly than the relevant disequilibrium timescales.


In sum, our grid of chemical disequilibrium forward models indicates that disequilibrium chemistry effects considered here do not strongly affect the emission spectrum of WASP-18b in the NIRISS/SOSS waveband.

\subsection{General Circulation Models.}\label{sec:gcms}
A suite of General Circulation Models (GCMs) are compared to the retrieved dayside spectrum and dayside temperature map.

We use the SPARC/MITgcm~\citep{Showman2009} to model the 3D atmospheric structure of WASP-18b. The model solves the primitive equations in spherical geometry using the MITgcm~\citep{Adcroft2004} and the radiative transfer equations using a current 1D radiative transfer model~\citep{Marley1999}. We use the correlated-k framework to generate opacities based on the line-by-line opacities\citep{Freedman2014}. Our model assumes a solar composition for the elemental abundances~\citep{Lodders2002} and chemical equilibrium gas-phase composition~\citep{Visscher2006}. Our model naturally takes into account the effect of thermal dissociation~\citep{parmentier_thermal_2018}. 
We used a time step of 25 s and ran the simulations for about 300 Earth days, averaging all quantities over the last 100 days.

We include additional sources of drag through a Rayleigh drag parametrization with a single constant timescale per model that determines the efficiency with which the flow is damped. The drag is constant over the whole planetary atmosphere. We vary this timescale between models from $\tau_{\rm drag} = 10^3$ to $10^6$~s (efficient drag) and a no drag model with $\tau_{\rm drag}=\infty$. Our range of drag strengths cover the transition from a drag-free, wind-circulation case to a drag-dominated circulation. The specific WASP-18b simulations that we use are described in more detail in Arcangeli et al. (2019)\citep{arcangeli_climate_2019}.

The second model we use is the kinematic magnetohydrodynamics (MHD) GCM (described in detail in ref.\,\cite{beltz_exploring_2021})  with an updated picket fence radiative transfer scheme \cite{Lee2021picketfence,Malsky2022}. Due to the high gravity of this planet, we chose to model the planet from 100 bar to $10^{-4}$ bar over 65 layers at a horizontal resolution of T31 (corresponding to roughly 3 degree resolution at the equator). We use the kinematic MHD prescription described in Perna et al. (2010)\cite{perna_magnetic_2010}, which has been used in models of hot Jupiters HD~209458b and HD~189733b \cite{RauscherMenou2013} as well as the ultra-hot Jupiter WASP-76b \cite{beltz_exploring_2021,beltz_magnetic_2022}. This drag prescription assumes a global dipole magnetic field, generated by an interior dynamo. Because of this geometry, our drag timescale is applied as a Rayleigh drag term solely to the east-west momentum equation (influencing flow perpendicular to magnetic field lines) and is calculated as
\begin{equation} \label{tdrag}
    \tau_{mag}(B,\rho,T, \phi) = \frac{4 \pi \rho \ \eta (\rho, T)}{B^{2} |sin(\phi) | },
\end{equation}

\bigskip 
\noindent 
where $B$ is the chosen global magnetic field strength, $\phi$ is the latitude, $\rho$ is the density, and $\eta$ is magnetic resistivity. This timescale is calculated locally and often, allowing the timescale to vary by over 10 orders of magnitude throughout the model and respond to changes in atmospheric temperatures. A minimum timescale cutoff  ($\sim 10^{3}$s) is applied in locations where $\tau_{mag}$ would be less than 1/20 of the the planet's orbit, for numerical stability. We chose to model a range of magnetic field strengths (0 G, 5 G, 10 G, and 20 G) as its true value is not known. 

\end{methods}


\noindent\textbf{Data availability}\\
The data used in this work are publicly available in the Mikulski Archive for Space Telescopes ({\small \url{https://archive.stsci.edu/}}). The data which was used to create all of the figures in this manuscript are freely available on Zenodo (Zenodo Link). 

\noindent{\textbf{Code availability}}\\
The open-source pipelines that were used throughout this work are listed below:

\noindent 
NIRISS/SOSS data reduction: \\
\indent nirHiss ({\small \url{https://github.com/afeinstein20/nirhiss}});\\ 
\indent supreme-SPOON ({\small \url{https://github.com/radicamc/supreme-spoon}});\\
\indent transitspectroscopy ({\small \url{https://github.com/nespinoza/transitspectroscopy}})

\noindent 
Light curve fitting: \\
\indent batman ({\small \url{https://github.com/lkreidberg/batman}}); \\ 
\indent emcee ({\small \url{https://emcee.readthedocs.io/en/stable/}})  

\noindent 
Atmospheric retrievals: \\
\indent CHIMERA ({\small\url{https://github.com/mrline/CHIMERA}});\\
\indent POSEIDON ({\small\url{https://github.com/MartianColonist/POSEIDON}});\\
\indent MultiNest ({\small\url{https://github.com/JohannesBuchner/MultiNest}});\\
\indent PyMultiNest ({\small\url{https://github.com/JohannesBuchner/PyMultiNest}}) 

\noindent 
Eclipse mapping: \\ 
\indent ThERESA ({\small \url{https://github.com/rychallener/ThERESA}}); \\
\indent Eigenspectra  ({\small \url{https://github.com/multidworlds/eigenspectra})

\noindent 
Atmospheric modeling: \\
\indent HELIOS ({\small\url{https://github.com/exoclime/HELIOS}});\\
\indent HELIOS-K ({\small\url{https://github.com/exoclime/HELIOS-K}});\\
\indent PLATON ({\small\url{https://github.com/ideasrule/platon}});\\ 
\indent VULCAN ({\small\url{https://github.com/exoclime/VULCAN}})

\begin{addendum}
 \item This work is based on observations made with the NASA/ESA/CSA JWST. The data were obtained from the Mikulski Archive for Space Telescopes at the Space Telescope Science Institute, which is operated by the Association of Universities for Research in Astronomy, Inc., under NASA contract NAS 5-03127. These observations are associated with program JWST-ERS-01366. Support for this program was provided by NASA through a grant from the Space Telescope Science Institute. The results reported herein benefited during the design phase from collaborations and/or information exchange within NASA’s Nexus for Exoplanet System Science (NExSS) research coordination network sponsored by NASA’s Science Mission Directorate.
 
 \subsection{Author contributions} All
 authors played a significant role in one or more of the following: development of the original proposal, management of the project, definition of the target list and observation plan, analysis of the data, theoretical modeling, and preparation of this manuscript. Some specific contributions are listed as follows. NMB, JLB, and KBS provided overall program leadership and management. LPC and BB led the efforts for this manuscript. DS, EK, HRW, IC, JLB, KBS, LK, MLM, MRL, NMB, VP, and ZBT made significant contributions to the design of the program. KBS generated the observing plan with input from the team. NE provided instrument expertise. BB, EK, HRW, IC, JLB, LK, MLM, MRL, NMB, and ZBT led or co-led working groups and/or contributed to significant strategic planning efforts like the design and implementation of the pre-launch Data Challenges. AC, DS, ES, NE, NG, and VP generated simulated data for pre-launch testing of methods. BB, ER, JLB, LPC, TDK, and VP contributed significantly to the writing of this manuscript, along with contributions in the Methods from AAAP, ABS, ADF, HB, IW, LDS, LSW, MM, MR, RC, RJM, and XT. ADF, LDS, LPC, and MR contributed to the development of data analysis pipelines and/or provided the data analysis products used in this analysis i.e., reduced the data, modeled the light curves, and/or produced the planetary spectrum. IW performed the updated TESS analysis. AAAP, LPC, LSW, and RJM performed atmospheric retrievals on the planetary spectrum. MH, MM, NTL and RC performed the secondary eclipe mapping analysis. LPC, LSW, and RC generated figures for this manuscript. BVR, HRW, IC, LW, MCN, and XZ provided significant feedback to the manuscript coordinating comments from all other authors.
 
 \item[Competing Interests] The authors declare that they have no competing financial interests.
 \item[Correspondence] Correspondence and requests for materials should be addressed to Louis-Philippe Coulombe.~(email: louis-philippe.coulombe@umontreal.ca).
\end{addendum}

\newpage


\vspace*{-25mm}
\begin{figure}[H]
    \centering
    \includegraphics[width=1.\columnwidth]{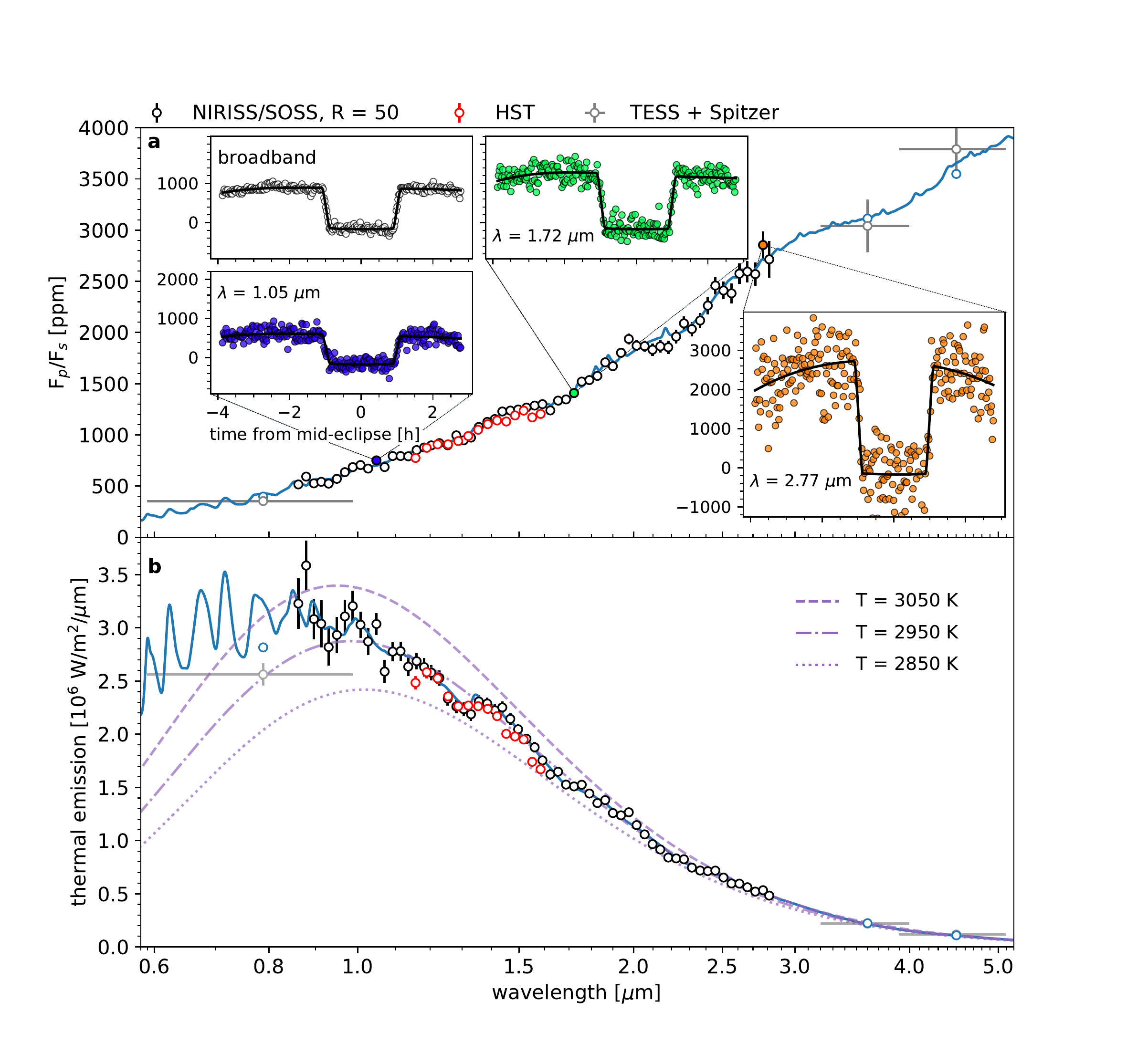}
    \vspace*{-10mm}
    \caption{\sep \textbf{Dayside thermal emission spectrum of WASP-18b. a,} Observed dayside planet-to-star flux ratio spectrum (black points), binned at a fixed resolving power of R = 50 for visual clarity. Past HST\cite{arcangeli_h-_2018} (red points), TESS (see Methods), and Spitzer\cite{2013maxted} (gray points) are shown for comparison. We show the best-fit model (blue line) from the SCARLET chemical equilibrium retrieval
    , extrapolated to the TESS and Spitzer wavelengths considering the same atmospheric parameters. We find that the retrieved spectrum is in good agreement with the past HST observations. The throughput-integrated model is shown for the TESS and Spitzer points (blue points). The white (broadband) light curve (white points) and three example spectrophotometric light curves (blue, green, and orange points at 1.05, 1.72, and 2.77~$\mu$m respectively), along with their best fitting models (black line), are shown to scale. The phase variation of the measured planetary flux around the secondary eclipse is clearly visible. \textbf{b,} Planetary thermal emission spectrum of WASP-18b, as computed from the $F_p/F_s$ spectrum and the PHOENIX stellar spectrum. The shortest wavelengths of the NIRISS/SOSS first order reach the maximum of the planetary spectral energy distribution, thereby enclosing 65\% of the total thermal energy emitted by the planet. Blackbody spectra for temperatures T = 2850 (dotted), 2950 (dash-dotted), and 3050~K (dashed) are shown in purple, with the best-fitting blackbody spectrum to the NIRISS data being T = 2950$\pm$3~K.}
    \label{fig:fpfs_spec_plus_lcs}
\end{figure}

\vspace*{-20mm}
\begin{figure}[H]
    \centering
    \includegraphics[width=1.0\columnwidth]{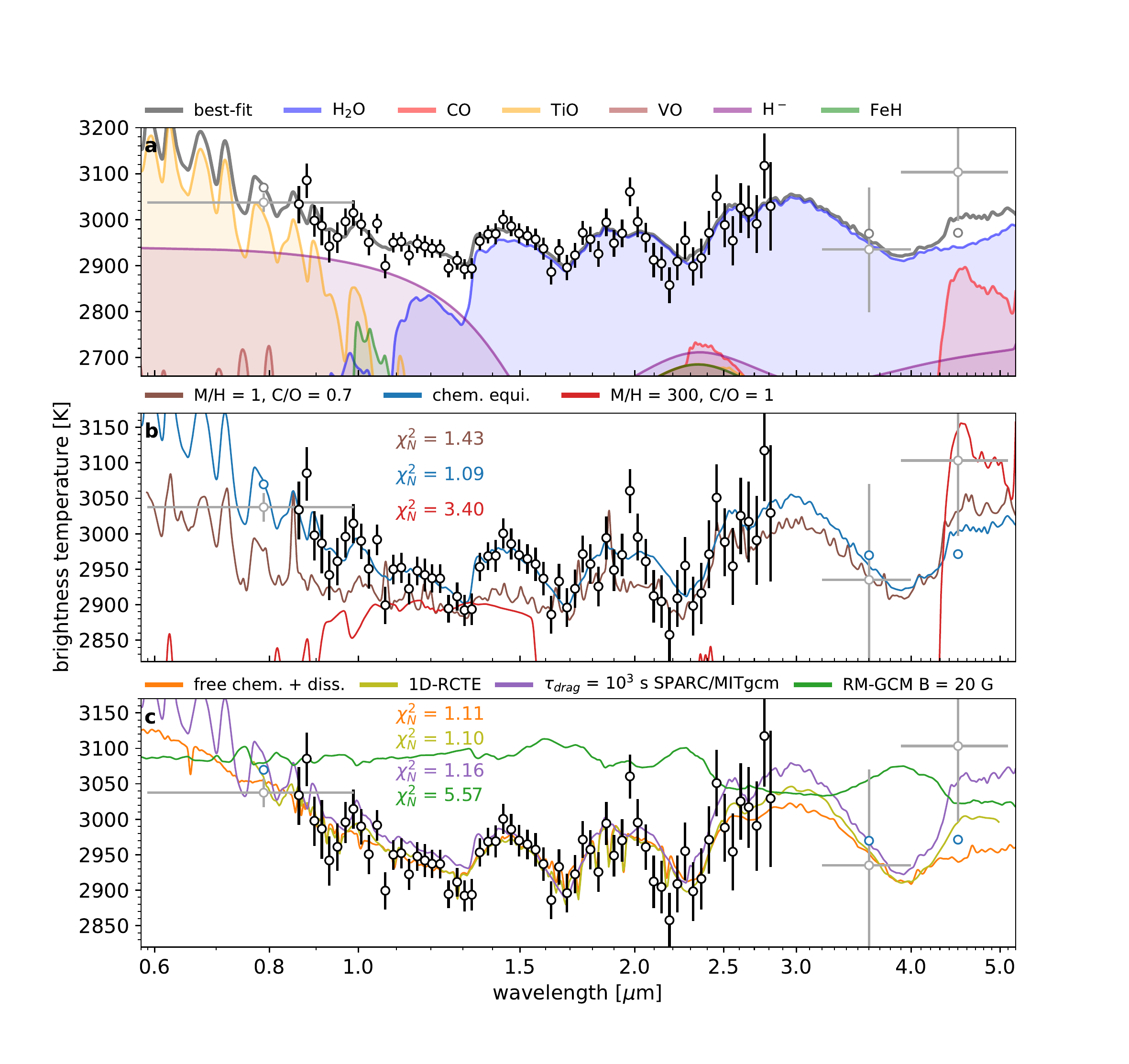}
    \vspace*{-10mm}
    \caption{\sep \textbf{Brightness temperature spectrum of WASP-18b.} \textbf{a,} Brightness temperature of WASP-18b as a function of wavelength, with models extrapolated to the TESS and Spitzer points considering the same atmospheric parameters. The H$_2$O emission features at 1.4, 1.9, and 2.5 $\mu$m are clearly visible. The rise in brightness temperature observed in the water features is indicative of a thermal inversion. We also observe a downward slope in the spectrum from 0.8--1.3$\,\mu$m as the opacities of H$^-$, TiO, and VO decrease. We find that the precision of the observations at 2.4$\,\mu$m is not sufficient to detect the small expected contribution from CO. \textbf{b,} Comparison of the high M/H and C/O case\cite{sheppard_evidence_2017} (red) as well as the solar metallicity case with H$^-$ opacity and H$_2$O dissociation\cite{arcangeli_h-_2018} (brown, best-fit to the HST data shown in Fig. \ref{fig:fpfs_spec_plus_lcs}) that could both explain the past HST observations. We also show the SCARLET best-fit model to the NIRISS observations (blue). \textbf{c,} Median fits of the free chemistry retrieval (orange) and of the self-consistent chemical equilibrium grid retrieval (green). We also show the dayside spectra obtained by post-processing the SPARC/MITgcm (purple) and RM-GCM (green) for a drag timescale of $\tau_{\rm drag}$ = 10$^3$~s and a magnetic field strength of B~=~20~G respectively. We find that the SPARC/MITgcm  better reproduces the observed features as the RM-GCM is more isothermal.}
    \label{fig:TbSpectrum}
\end{figure}

\begin{figure}[H]
    \centering
    \includegraphics[width=1.0\columnwidth]{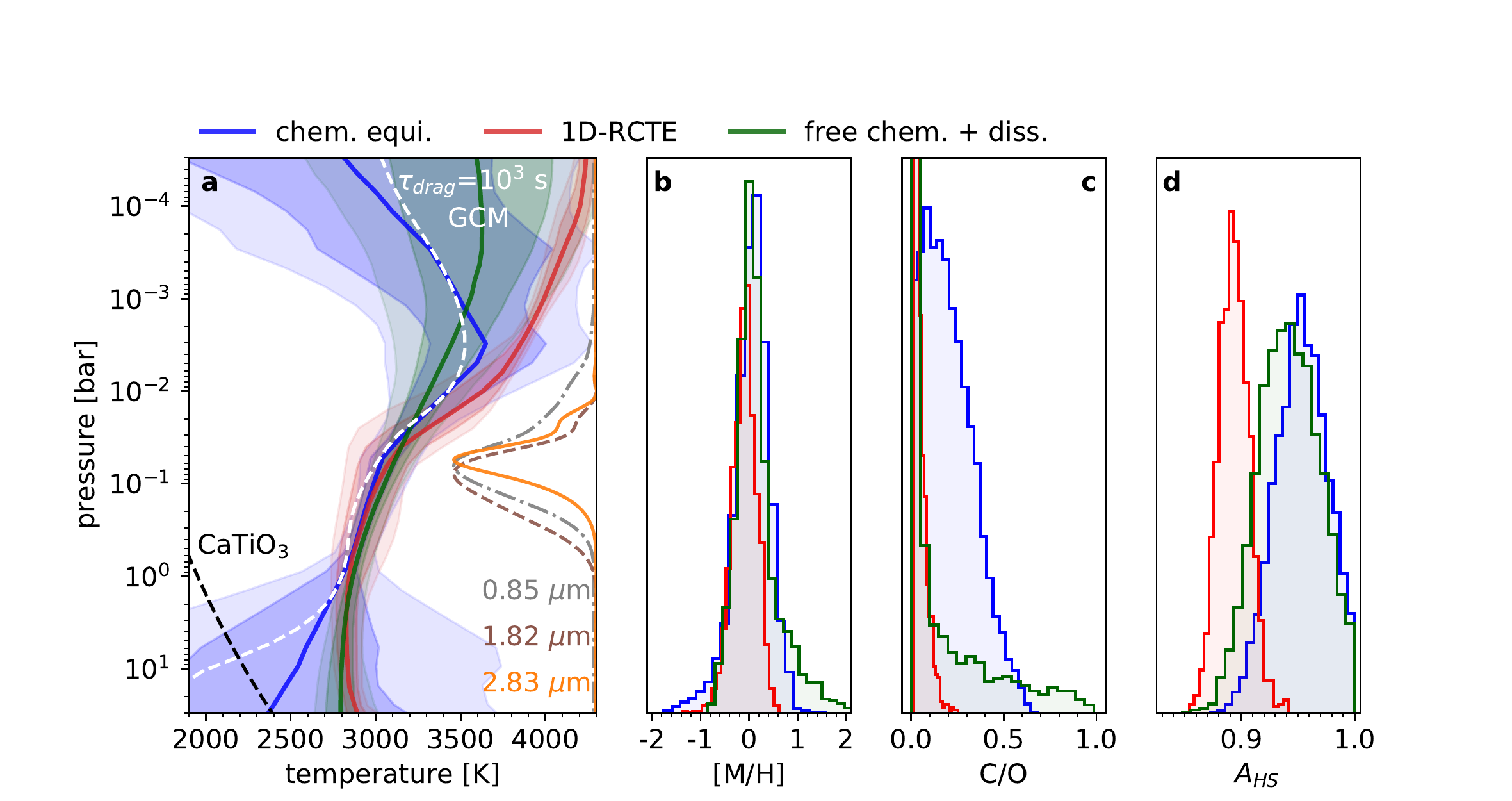}
     \caption{\sep \textbf{Atmospheric constraints from the chemical equilibrium and free chemistry retrievals. a,} Retrieved temperature-pressure profiles with 1 and 2 $\sigma$ contours for the chemical equilibrium with free temperature-pressure profile (blue), radiative-convective thermochemical equilibrium (1D-RCTE, red), and free chemistry with thermal dissociation (green) retrievals. The retrieved temperature-pressure profiles are consistent between the retrievals and show an inversion in the pressure range that is constrained from the observations, as shown by the contribution functions at 0.85 (dot-dashed gray line), 1.82 (dashed brown line), and 2.83~$\mu$m (orange line). The temperature-pressure profile of WASP-18b is above the CaTiO$_3$ condensation curve\cite{wakeford_high-temperature_2016} (black dashed line) at almost all pressures, which motivates the presence of a temperature inversion caused by TiO as Ti is available in gas form. The dayside average temperature-pressure profile of the $\tau_{drag}$ = 10$^3$~s SPARC/MITgcm (white dashed line) is computed from the viewing angle average of $T(P)^4$ and shown for comparison. We also show the posterior probability distributions of the atmospheric metallicity [M/H] (\textbf{b}), C/O ratio (\textbf{c}), and area fraction $A_{HS}$ (\textbf{d}). The area fraction $A_{HS}$ is a scaling factor applied to the thermal emission spectrum to compensate for the possible presence of a concentrated hot spot contributing to most of the observed emission \cite{Taylor2020}. All methods retrieve metallicities consistent with solar at 1$\sigma$. The retrieved C/O 3$\sigma$ upper limits are of 0.6 and 0.2 for the chemical equilibrium with free temperature-pressure profile and the 1D-RCTE retrievals respectively. Finally, we find the area fraction $A_{HS}$ is consistent with 1 when allowing the temperature-pressure profile to vary freely, indicating the lack of a concentrated hot spot on the dayside contributing to the majority of the observed emission.}
    \label{fig:atm_ret}
\end{figure}

\begin{figure}
    \centering
    \includegraphics{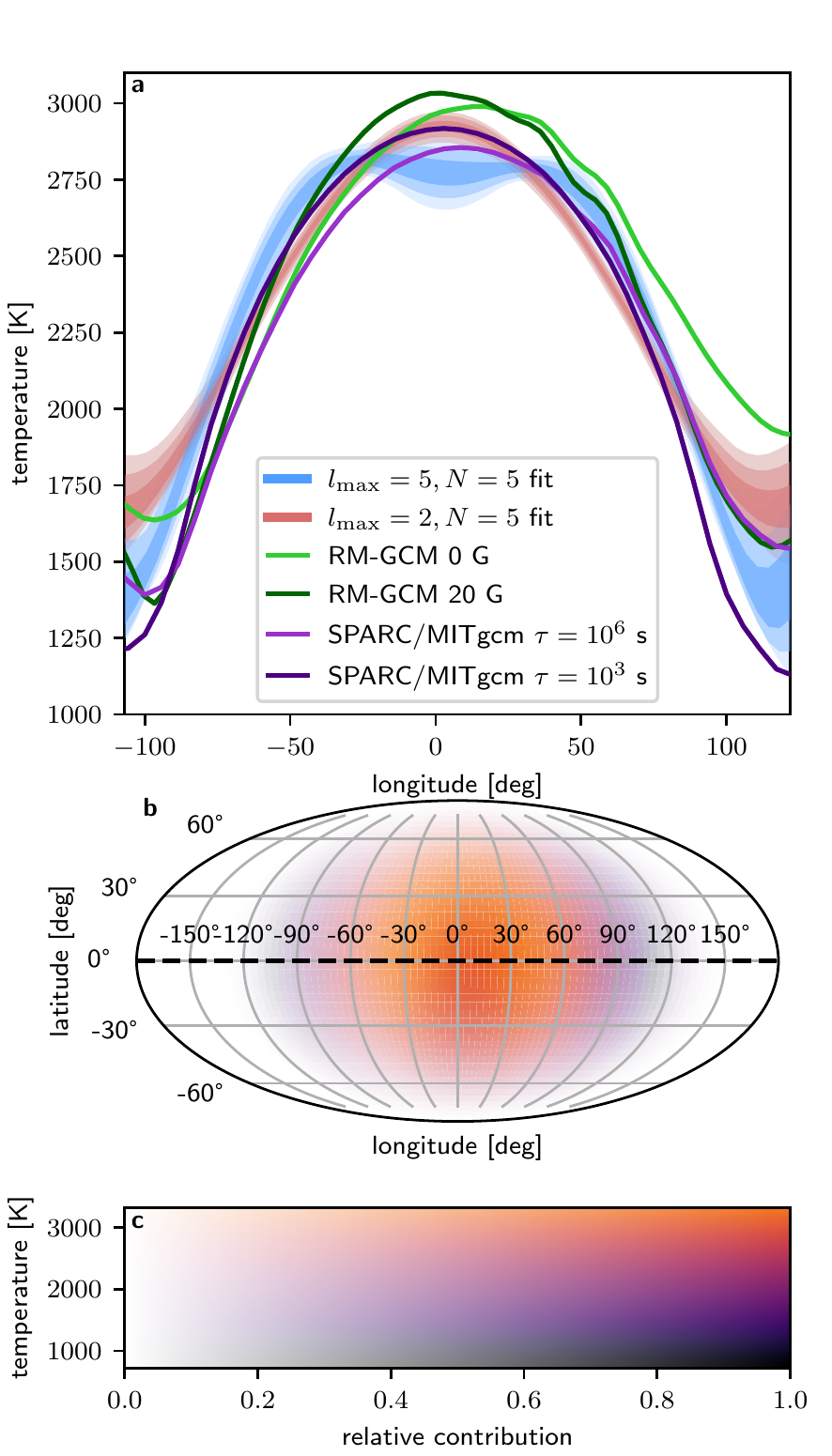}
    \caption{\sep \textbf{Retrieved temperature map of WASP-18b.} \textbf{a,} Latitudinally-averaged brightness temperatures (see Methods) of the planet along the equator. The blue and red shaded areas show solutions for $l_{\rm max}=5$, $N=5$ and $l_{\rm max}=2$, $N=5$ (see Methods), respectively. Statistically, the blue model is marginally preferred. Dark, medium, and light shading denote the 1, 2, and 3 $\sigma$ confidence regions respectively, showing the range of model possibilities. Overplotted are several predictions from general circulation models with magnetic-field\cite{beltz_magnetic_2022} (green) or uniform\cite{Showman2009} (purple) drag timescales (see Methods). The plot only shows longitudes emitting at least 10\% of the substellar flux. \textbf{b,} The temperature map of WASP-18b for the $l_{\rm max}=5$, $N=5$ solution. Along the equator at -90$^\circ$, 0$^\circ$, and 90$^\circ$ longitude the temperatures are 1744 K, 3121 K, and 2009 K, respectively.  \textbf{c,} The colorbar for the map shown in \textbf{b}. Color represents the brightness temperature and saturation represents the relative contribution to the light curve based on its visibility.}
    \label{fig:eclipse_map}
\end{figure}

\clearpage
\setcounter{page}{1}
\setcounter{figure}{0}
\renewcommand{\figurename}{Extended Data Fig.}
\renewcommand{\tablename}{Extended Data Table}

\begin{center}
\textbf{\huge{}Extended Data}{\Huge\par}
\par\end{center}

\begin{figure}[H]
    \centering
    \includegraphics[width=1.0\columnwidth]{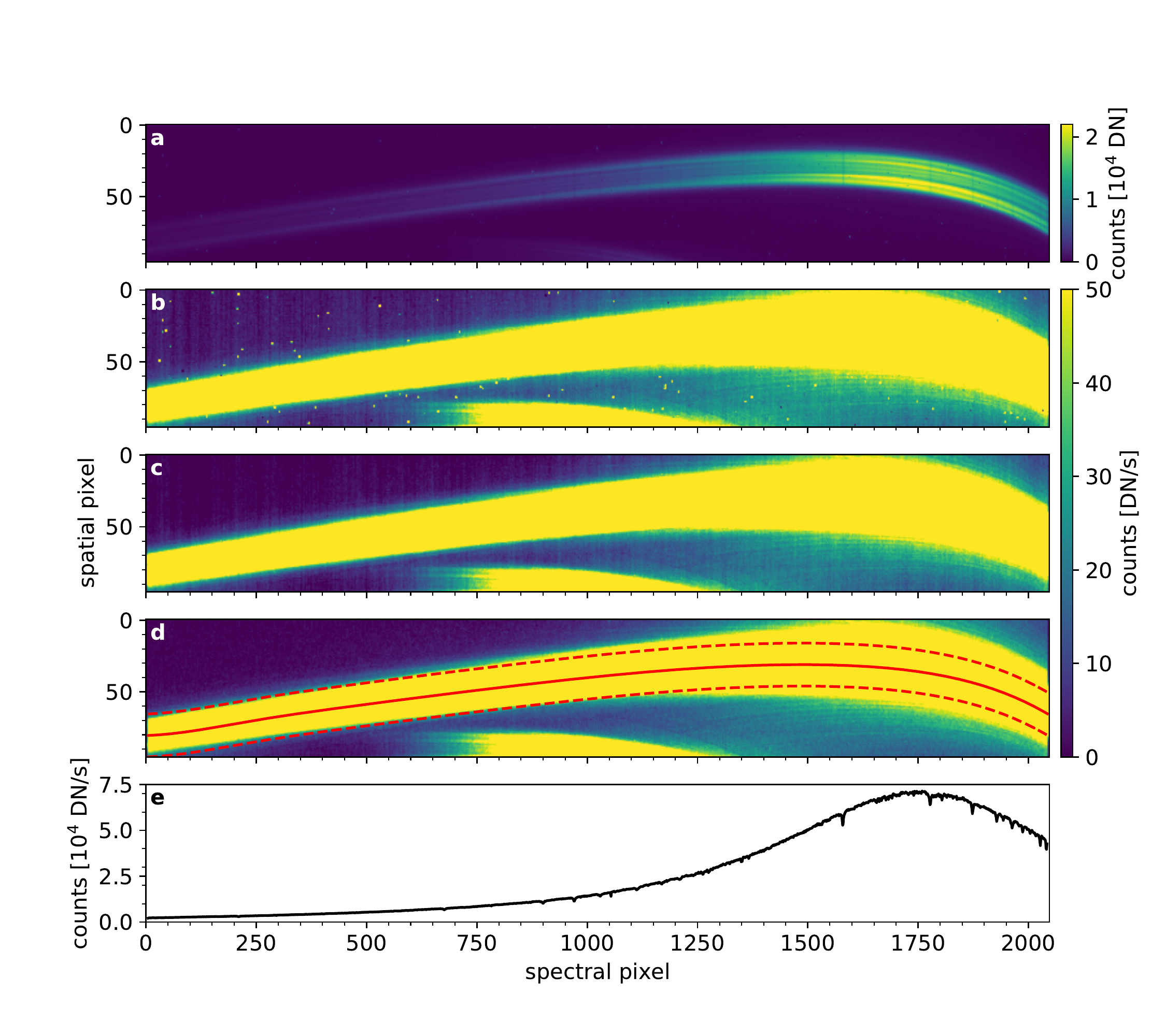}
    \caption{\sep \textbf{Reduction from uncalibrated data to spectral extraction}. \textbf{a,} Last group of the 100$^{\text{th}}$ integration after super-bias subtraction and linearity correction steps, which are the main reduction steps applied for stage 1 of the \texttt{jwst} pipeline before ramp-fitting. \textbf{b,} Frame of the 100$^{th}$ integration after ramp-fitting, where the data are converted from counts to rates. \textbf{c,} Frame after outlier/bad pixel correction and background subtraction. The spectral trace extends in the spatial direction over most of the SUBSTRIP96 subarray. \textbf{d,} Frame after 1/$f$ correction, the striped structure visible in \textbf{c} is removed. The center of the trace and size of the extraction box are shown by the red full and dashed lines respectively. \textbf{e,} Total counts per column obtained from the spectral extraction step.}
    \label{fig:reduc_steps}
\end{figure}


\begin{figure}[H]
    \centering
    \includegraphics[width=1.0\columnwidth]{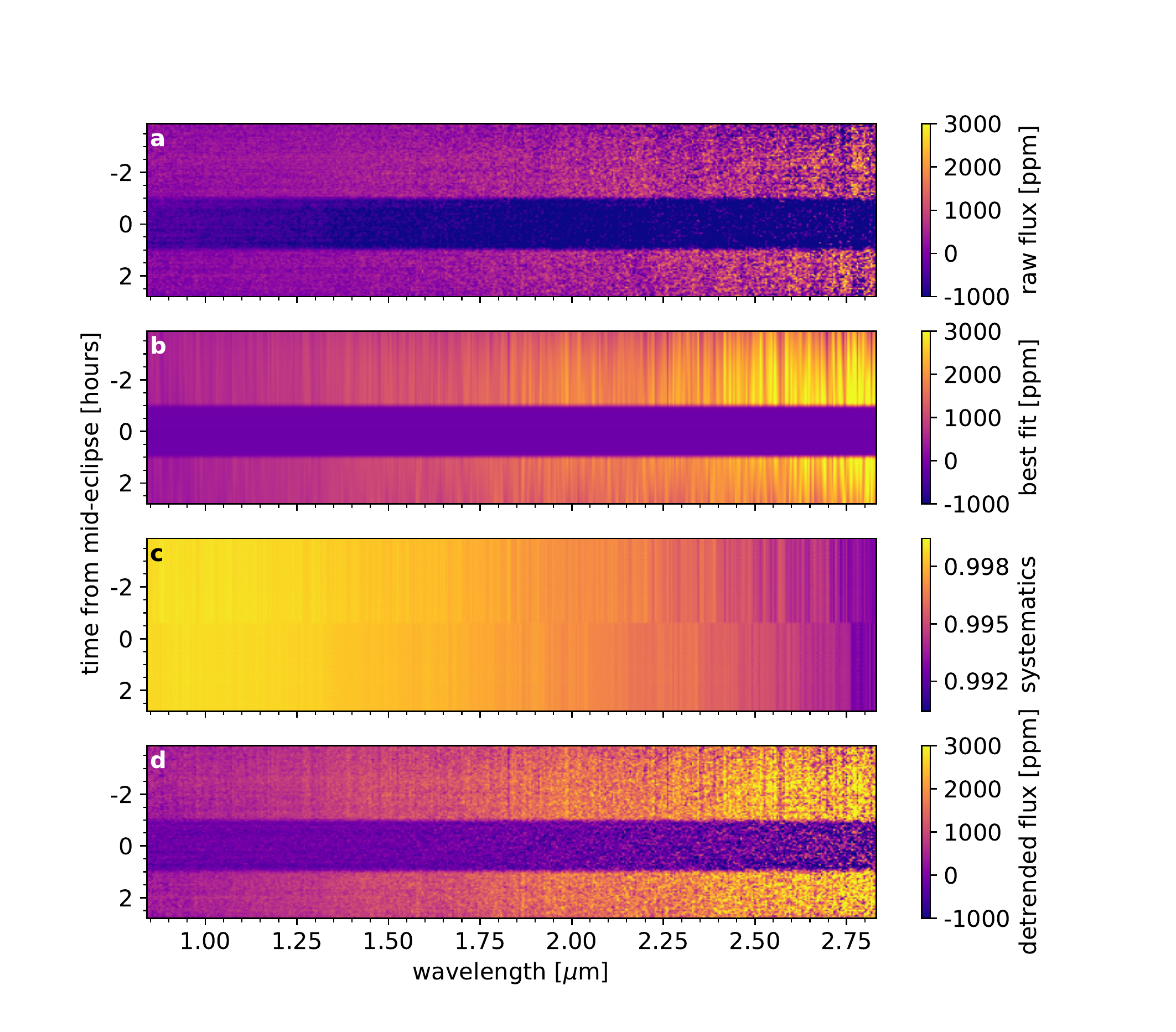}
    \caption{\sep \textbf{Spectrophotometric secondary eclipse light curves of WASP 18b}. \textbf{a,} Raw light curves for all 408 spectrophotometric bins. \textbf{b,} Best-fit planetary flux retrieved from the light curve fits. \textbf{c,} Systematics subtracted from \textbf{a}, consisting of a linear trend and the detrending against the tilt event and the trace morphology changes. The jump around $\sim$0.7 hours before mid-eclipse comes from the fit of the flux offset caused by the tilt event. \textbf{d,} Raw light curves after subtraction of the best-fit systematics model. Some of the detrended light curves show sudden flux variations between wavelength bins outside of eclipse caused by correlations between the astrophysical and systematics models. Those correlations are however considered when computing the spectrum as the $F_p/F_s$ values are marginalized over the range of systematics model that fit the light curves.}
    \label{fig:chrom_lcs_2d}
\end{figure}

\begin{figure}[H]
    \centering
    \includegraphics[width=1.0\columnwidth]{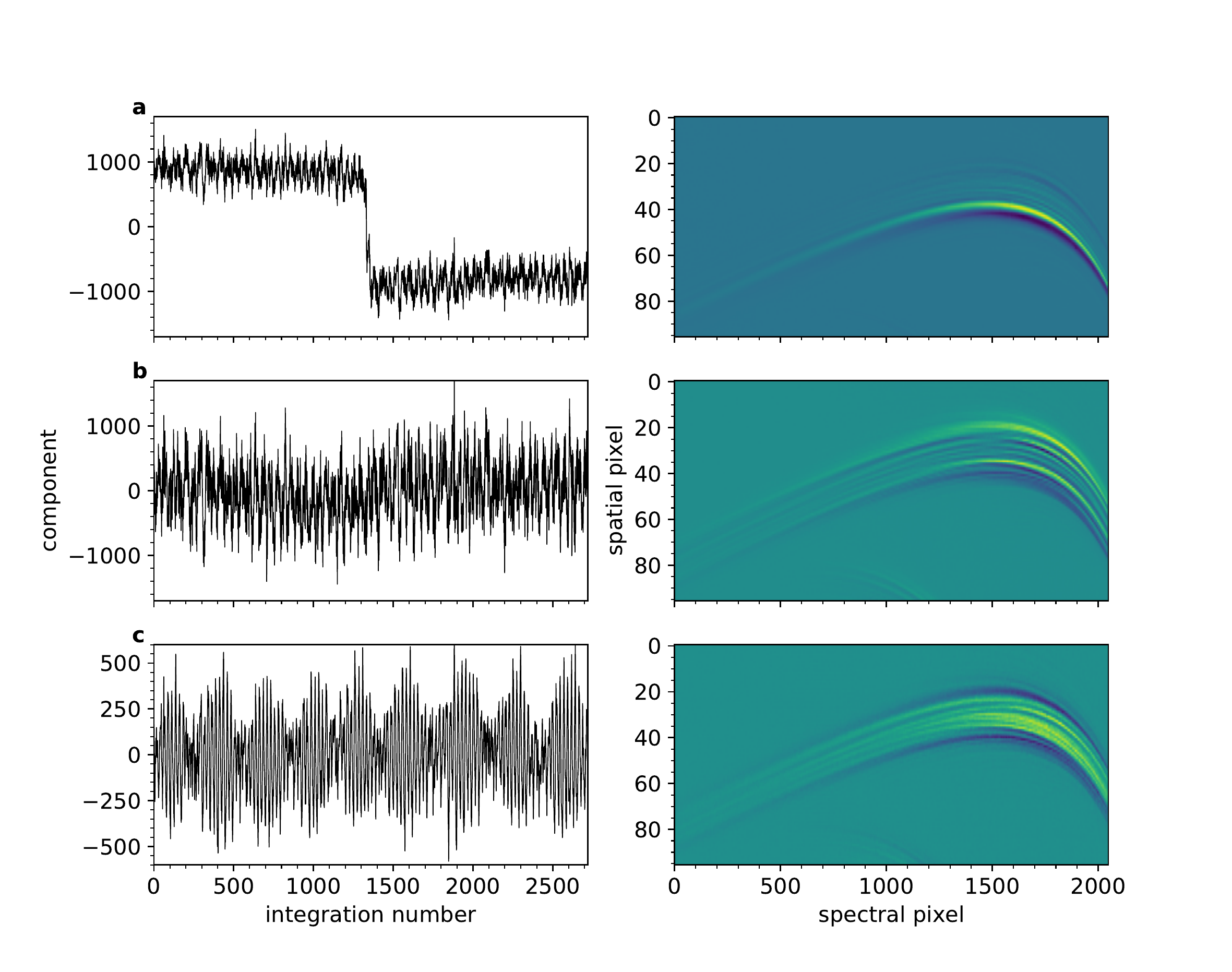}
    \caption{\sep \textbf{Morphological changes of the spectral trace on the NIRISS detector as identified through PCA on the time series of the detector images.} \textbf{a,} First principal component with its eigenvalues (left) and its corresponding eigenimage (right). The tilt event occurring near the 1336th integration can clearly be identified as the largest source of variance to the detector images. It results in a subtle change to the trace profile in the cross-dispersion direction, predominately visible near its lower edge of the trace. \textbf{b,} Second principal component with its eigenvalues (left) and its corresponding eigenimage (right). The second principal component represents subtle changes in the y-position of the trace throughout the time series, with the two edges of the trace trading flux. \textbf{c,} Third principal component with its eigenvalues (left) and its corresponding eigenimage (right). The third component represents changes in the FWHM of the trace and shows a clear beat pattern in time. The eigenimage for this component shows a trade of flux between the center and the edges of the trace throughout the time series.}
    \label{fig:pca_niriss}
\end{figure}

\begin{figure}[H]
    \centering
    \includegraphics[width=1.\columnwidth]{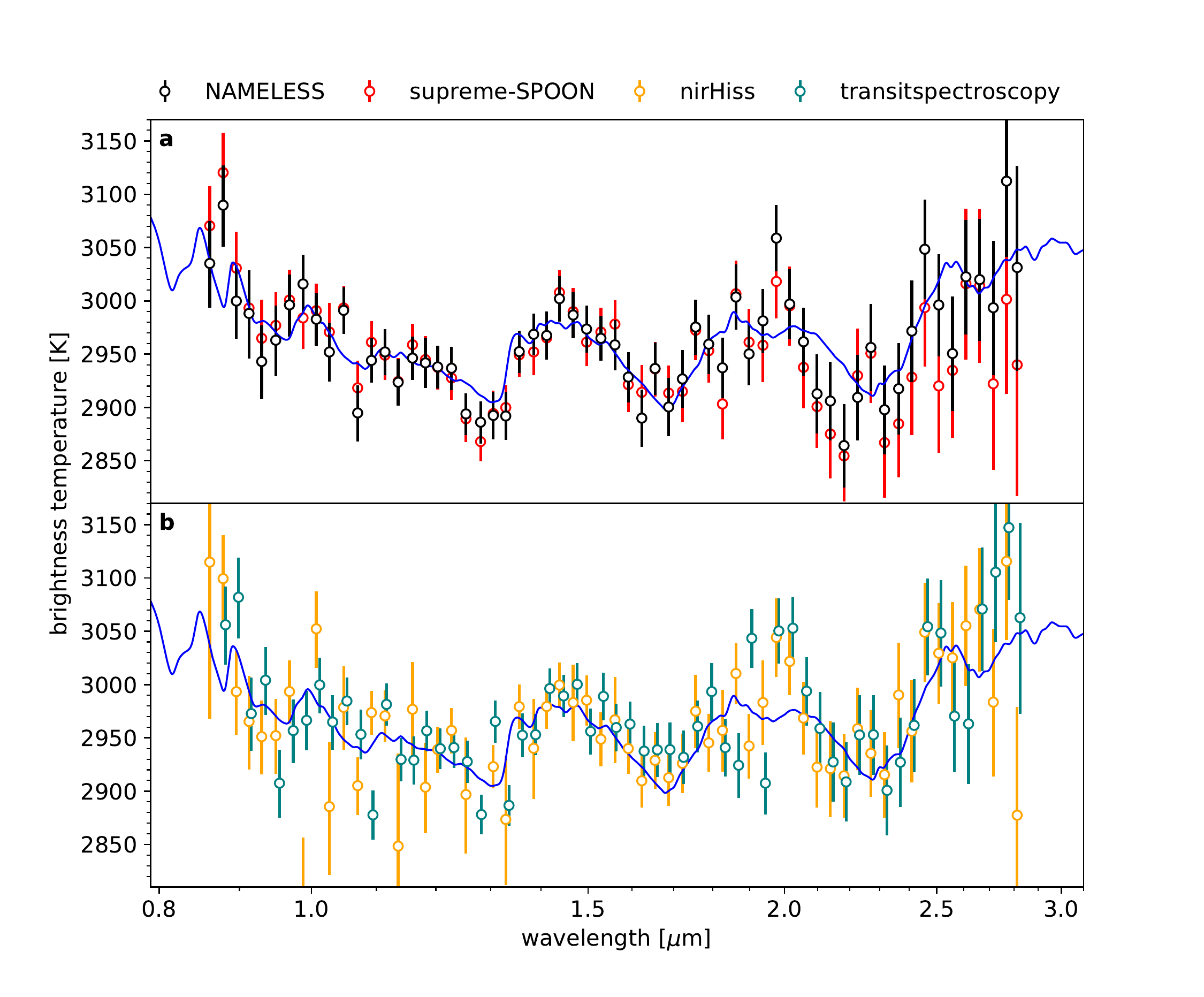}
    \caption{\sep \textbf{Spectra from the four individual reductions}. Comparison of the brightness temperature spectra obtained by fitting with ExoTEP the four separate reductions and binned at a resolving power of R = 50. We overplot the best-fit SCARLET model (blue line) to the reductions for further comparison. All reductions are consistent within less than one standard deviation on average when compared at full resolution (408 bins).}
    \label{fig:Tb_spec_reducs}
\end{figure}

\begin{figure}[H]
    \centering
    \includegraphics[width=.8\columnwidth]{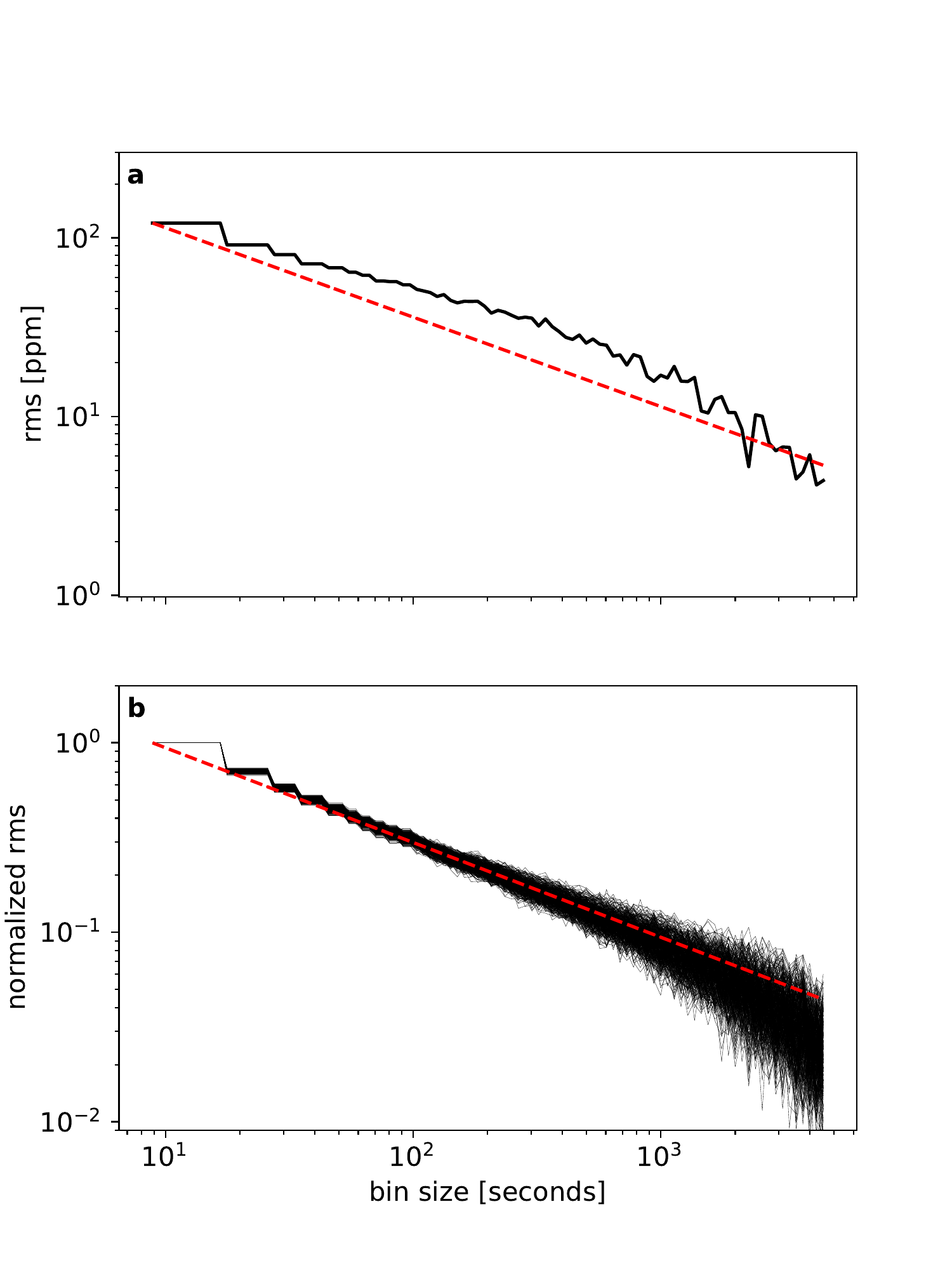}
    \caption{\sep \textbf{Light curve residuals binned in time}. \textbf{a,} Absolute root mean square (RMS) of the residuals as a function of bin size (black line) for the white light curve. The RMS values are plotted against the Poisson noise limit (red dashed line), which goes down as the square root of the number of integrations contained in a single bin. The residuals bin down to $\sim$5 ppm for bins of 1 hour. The broadband residuals do not follow perfectly the Poisson noise, which is indicative of remaining time correlations. \textbf{b,} Normalized RMS of the 408 spectrophotometric light curves considered in the analysis. We observe that the residuals follow the Poisson noise limit from bin sizes of a single integration up to bins of $\sim$75 minutes, indicating that there are no time correlations in the residuals.}
    \label{fig:binned_resids}
\end{figure}

\begin{figure}[H]
    \centering
    \includegraphics[width=1.\columnwidth]{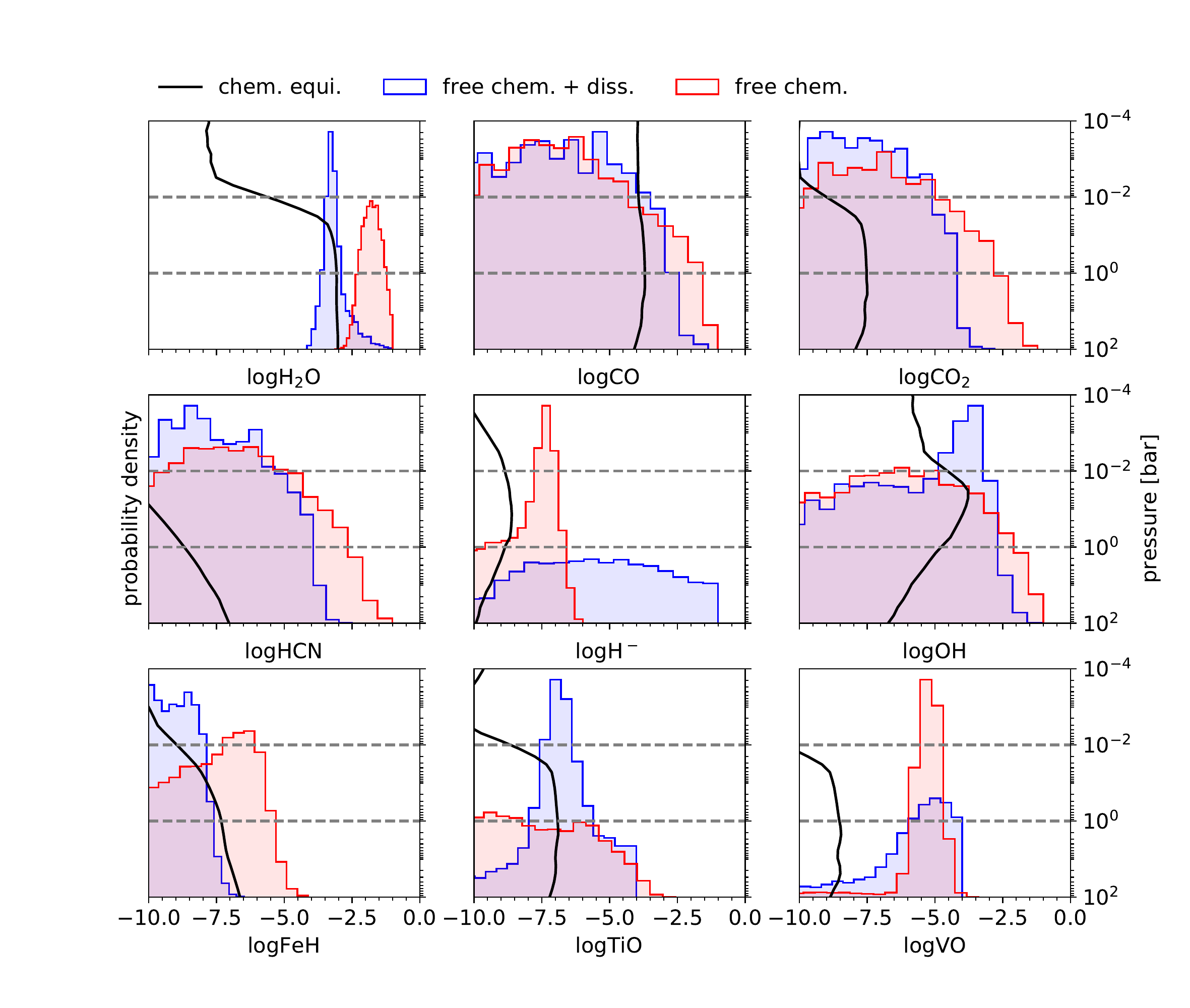}
    \caption{\sep \textbf{Abundance constraints from the free chemistry retrievals}. Probability posteriors of the deep abundance of various species considered for the free chemistry and temperature with (blue, HyDRA) and without (red, POSEIDON) thermal dissociation. We also show the median retrieved VMR profiles from the chemical equilibrium with free temperature-pressure profile retrieval (black line, SCARLET). The pressure range probed by the observations ($\sim$0.01--1 bars, see \ref{fig:atm_ret}) is indicated by the dashed gray lines. The only species independently detected is H$_2$O, which is found to be consistent with the retrieved chemical equilibrium abundance when considering the effect of thermal dissociation. All other species considered are found to be unconstrained although consistent with chemical equilibrium predictions. The photosphere as predicted by our radiative-convective model is around 50 millibar, but the retrievals infer the deep molecular abundances.}
    \label{fig:mol_posteriors}
\end{figure}

\begin{figure}[H]
    \centering
    \includegraphics[width=1.\columnwidth]{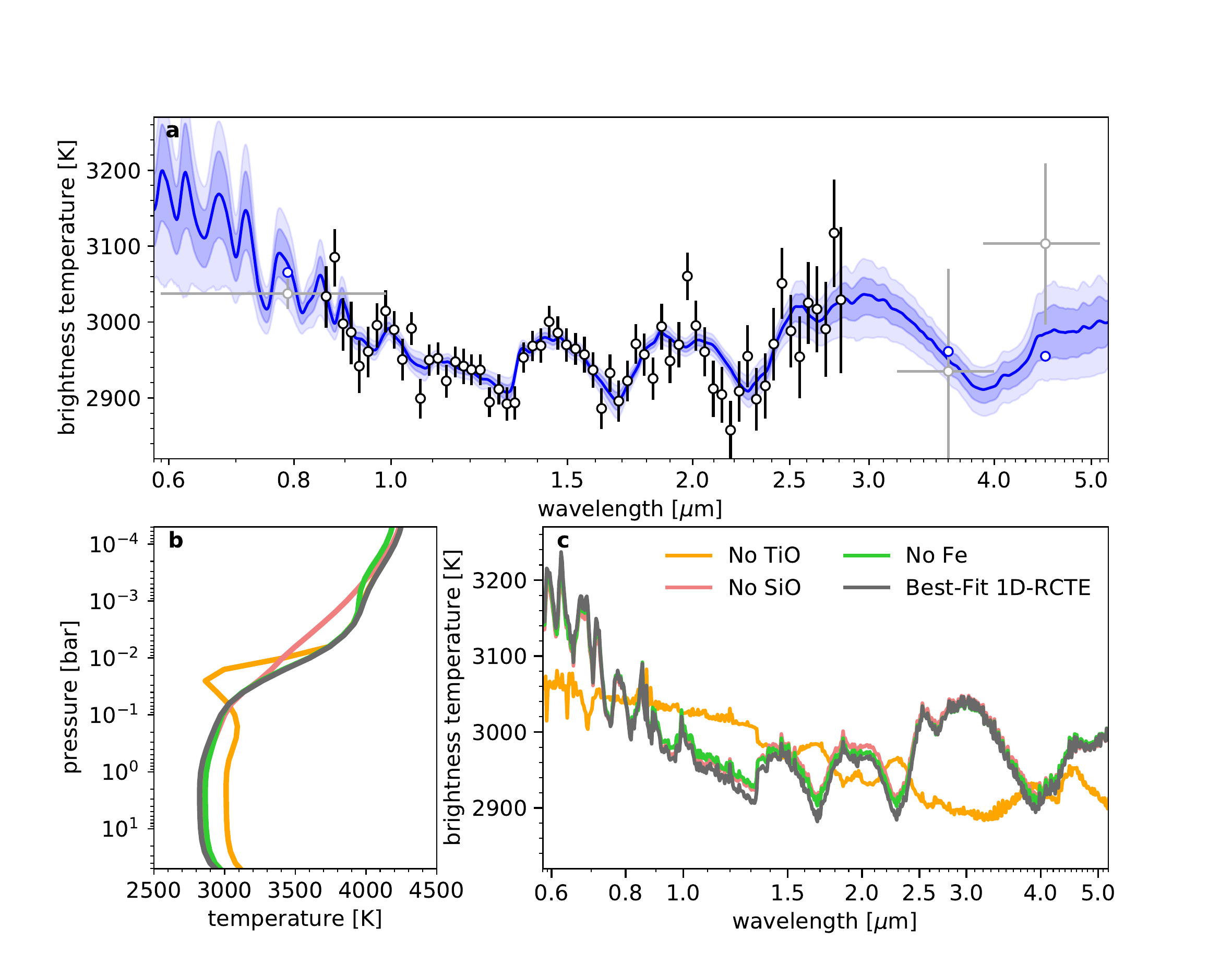}
    \caption{\sep \textbf{WASP-18b's brightness temperature spectrum fit and source of the thermal inversion.} \textbf{a,} The dark blue line indicates the chemical equilibrium median fit to the NIRISS data (black points), with shaded blue regions showing the 1$\sigma$ and 2$\sigma$ credible intervals in the retrieved spectrum (medium and light blue, respectively). The spectra are extrapolated to the TESS (visible wavelengths) and the \textit{Spitzer}/IRAC measurements (3.6 and 4.5~$\mu$m) observations (gray points) considering the same atmospheric parameters. \textbf{b,} Best-fit radiative-convective model temperature-pressure profile together with radiative-convective solutions where specific species known to create a thermal inversion are removed from the atmosphere. Absorption by atomic iron contributes to the thermal inversion at pressures lower than 1.0 millibar, whereas TiO is responsible for the thermal inversion seen between 0.1 and 0.01 bar. SiO contributes at pressures lower than 0.1 bar. \textbf{c,} Best-fit radiative-convective brightness temperature spectrum (excluding area fraction) and resulting spectra when removing specific species. As shown by the change from emission to absorption features in the spectra when TiO is removed, the TiO-induced thermal inversion is the one probed by our observations.}
    \label{fig:Tb_spec_samples_TiO}
\end{figure}

\begin{figure}[H]
    \centering
    \includegraphics[width=1.0\columnwidth]{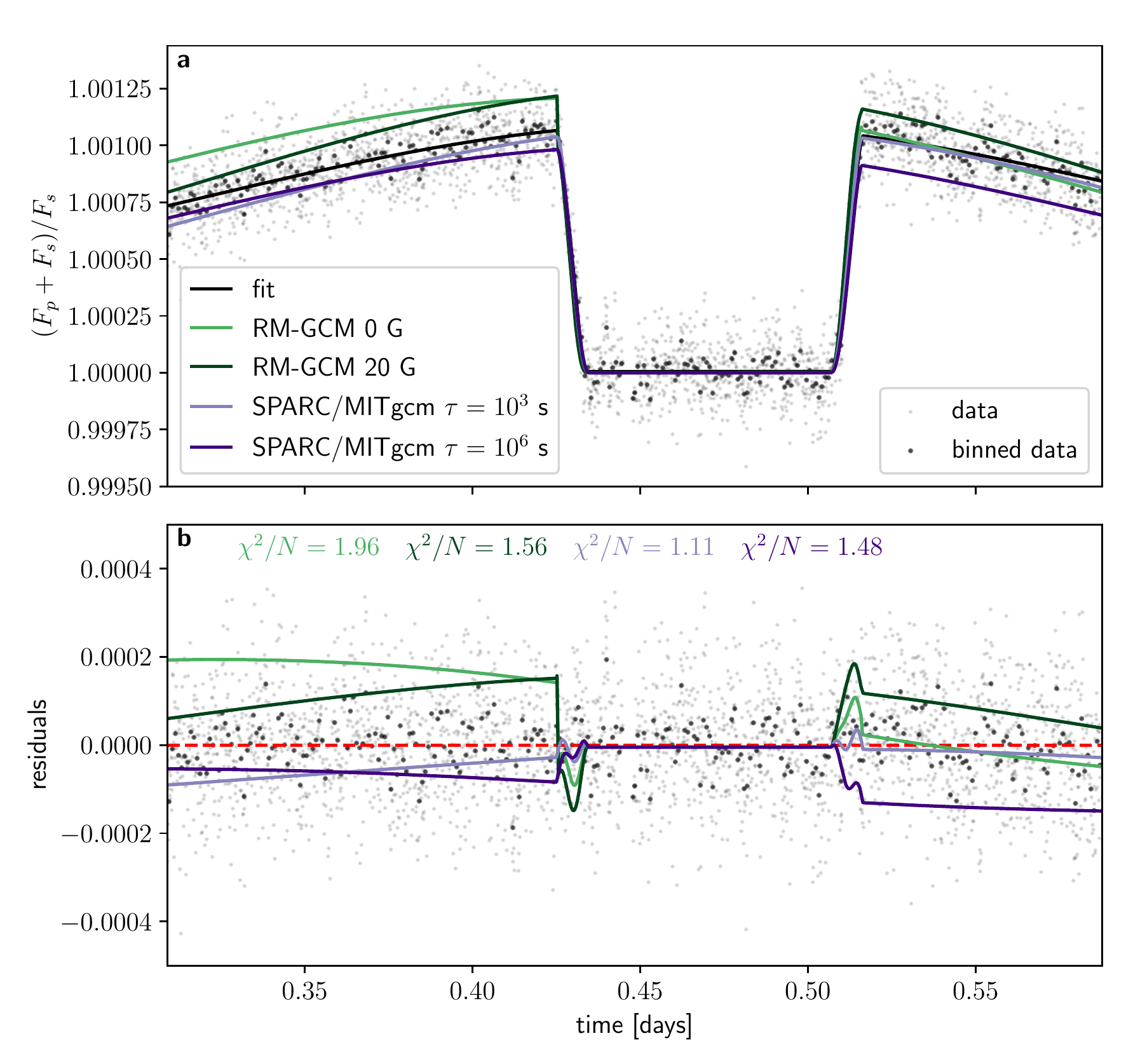}
    \caption{\sep \textbf{Light-curve predictions from GCMs}. \textbf{a,} GCM light curves compared against the data and the best-fitting eclipse-mapping model. \textbf{b,} Data and the GCM light curves with the eclipse-mapping model subtracted. The GCMs with strong atmospheric drag (RM-GCM 20 G and SPARC/MITgcm $\tau = 10^3$ s) match the data better than their counterparts that have little to no drag.}
    \label{fig:gcm-lightcurves}
\end{figure}

\begin{figure}[H]
    \centering
    \includegraphics[width=1.0\columnwidth]{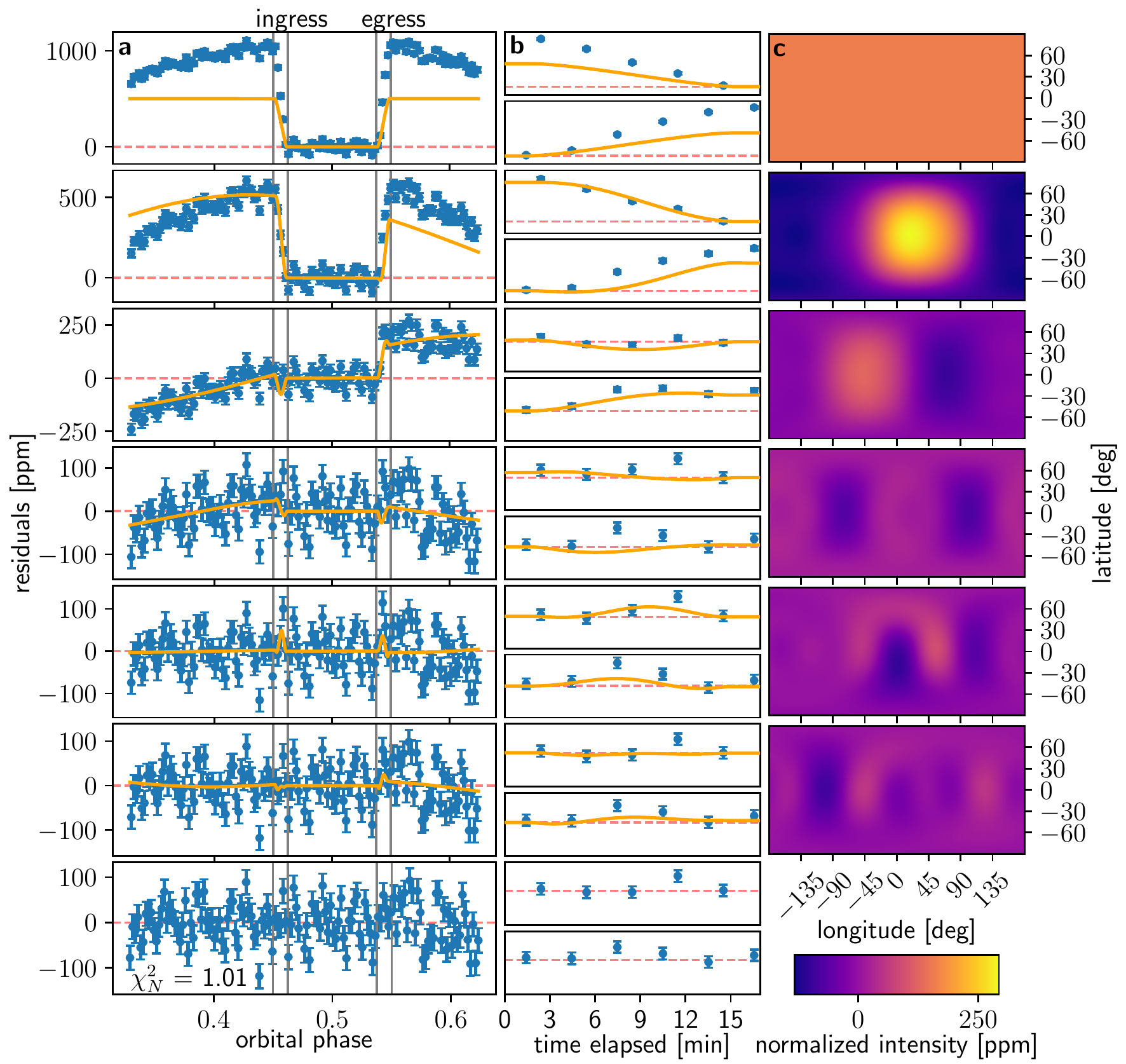}
    \caption{\sep \textbf{Components of the eclipse mapping fit}. \textbf{a,} This column shows the light-curve components of the eclipse-mapping fit for $l_{\rm max}=5$, $N=5$, overplotted on the data, which have been binned by a factor of 20 for clarity. From top to bottom, each light-curve component is subtracted off from the data to illustrate the features which are fit by each component, such that the top row is the full white-light light curve and the bottom row is the model residuals. Note that all components are fit simultaneously. \textbf{b,} The same as column \textbf{a}, zoomed in to the ingress and egress of the eclipse to highlight the fine features fit by each component. \textbf{c,} The eigenmaps associated with the corresponding components in column \textbf{a} and \textbf{b} that, when integrated, generate those light curves. Each map has been scaled by its best-fitting weight, such that a sum of this column would produce the best-fitting map.}
    \label{fig:eigenmaps}
\end{figure}

\begin{figure}[H]
    \centering
    \includegraphics[width=1.0\columnwidth]{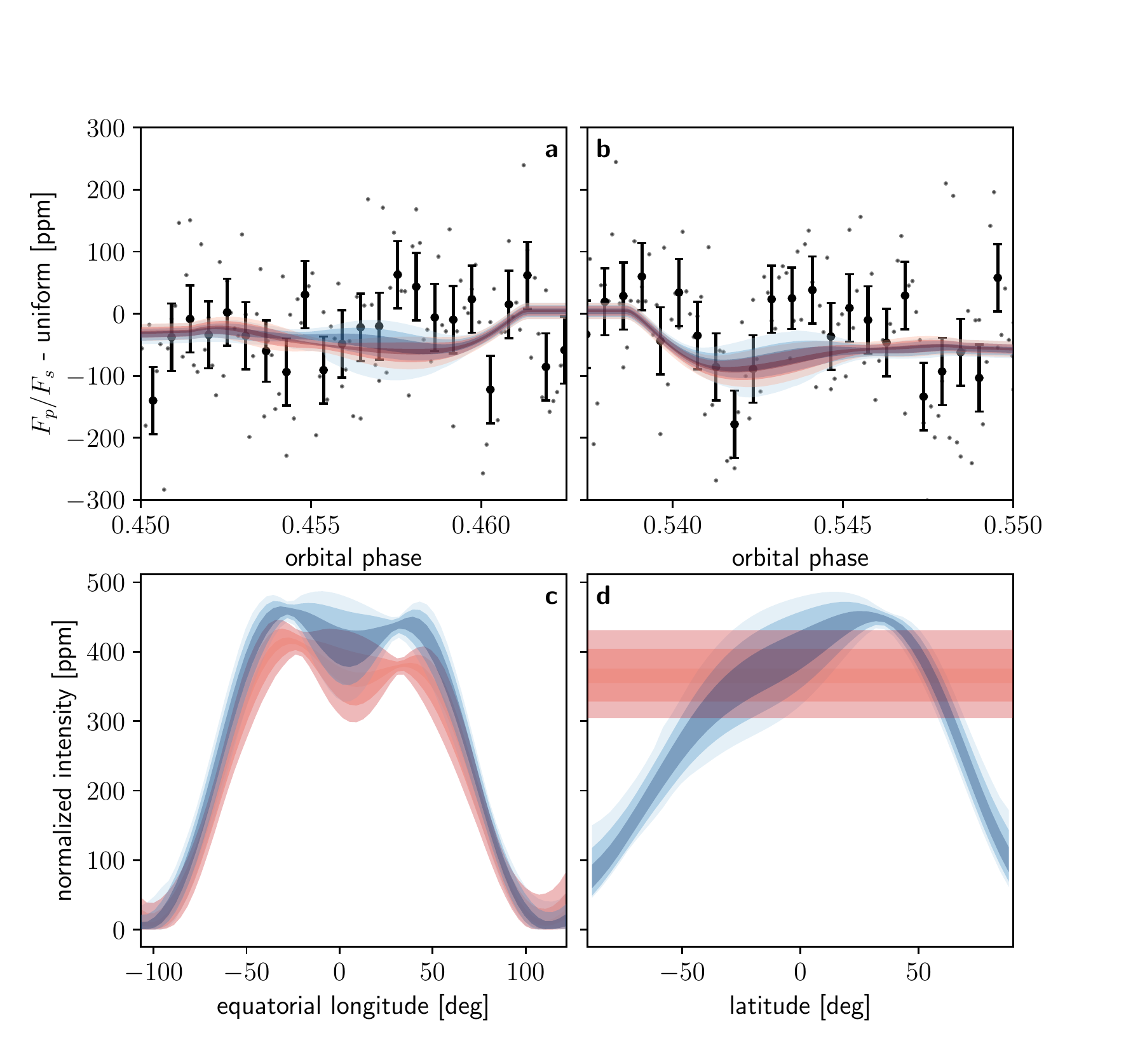}
    \caption{\sep \textbf{Latitudinal structure in the eclipse map}. \textbf{a,} Ingress of the eclipse, with two models overplotted and a 1096 ppm (white-light planet flux at mid-eclipse) uniform planet model subtracted to highlight deviations. The data (small dots) have been binned by a factor of five (dots with error bars) for clarity. The blue model is the eclipse map for $l_{\rm max}=5$, $N=5$ presented in the text. The red model uses a constant-with-latitude Fourier series as the basis set, rather than spherical harmonics, to investigate constraints on latitudinal aspects of the map. Shaded regions denote 1, 2, and 3 $\sigma$ quantiles. \textbf{b,} Same as \textbf{a} but for the eclipse egress. \textbf{c,} Planetary flux along the equator for the same two models. Note that, regardless of basis functions, we retrieve the same longitudinal structure, giving us confidence in the longitudinal brightness distribution. \textbf{d,} Same as \textbf{c} but along the substellar meridian. Both models fit the data well but find different latitudinal structure, indicating that we are unable to constrain latitudinal variation.}
    \label{fig:latitude}
\end{figure}

\begin{table}
    \footnotesize
    \centering
    \setlength{\tabcolsep}{6pt}
    \renewcommand*{\arraystretch}{1.4}
    \begin{tabular}{c c c c c c}
        \hline \hline
        Retrieval &  M/H [times solar] & C/O & A$_{\text{HS}}$ & $\log$H$_2$O & $\log$CO \\
        \hline
		SCARLET & $1.19_{-0.67}^{+1.22}$ & $< 0.60$ & $0.953^{+0.023}_{-0.024}$ & -- & -- \\
        \hline
		ScCHIMERA  & $0.82_{-0.37}^{+0.59}$ &  $<0.20$ & $0.893_{-0.013}^{+0.014}$  & -- & -- \\
		\hline
		HyDRA  & $1.19_{-0.57}^{+2.25}$ & $<0.91$ & $0.942_{-0.27}^{+0.29}$ & $-3.22_{-0.28}^{+0.46}$ & $<-2.15$ \\
		\hline
		POSEIDON  & $32.7_{-19.7}^{+49.7}$ & $<0.76$ & $0.962^{+0.018}_{-0.020} $ & $-1.78^{-0.40}_{-0.40}$ & $<-1.34$\\
		\hline
		\hline
    \end{tabular}
     \caption{\sep \textbf{Atmospheric retrieval constraints.} Constraints on the metallicity M/H, carbon-to-oxygen ratio C/O, area fraction A$_{HS}$, water log mixing ratio $\log$H$_2$O, and carbon monoxide log mixing ratio $\log$CO. For the free chemistry retrievals, water is the only species that is individually detected with statistical significance, and is therefore used as a proxy for metallicity (considering $\log$H$_2$O = -3.3 as the solar abundance). TiO, VO and H$^-$ are jointly detected at 3.8$\sigma$ significance, but are not individually detected with statistical significance. We also give 3$\sigma$ upper limits on the C/O ratio from the free chemistry retrievals by considering H$_2$O and CO as the main carbon and oxygen bearing species (C/O = CO/[H$_2$O + CO]).}
    \label{tab:retrieval_results}
\end{table}

\begin{table}
\footnotesize
\centering
\setlength{\tabcolsep}{6pt}
\renewcommand{\arraystretch}{1.1}
\begin{tabular}{c | c}
\hline \hline
Parameter & Value \\
\hline
$R_{p}/R_{*}$ &  $0.09783 \pm 0.00028$  \\
\hline
$T_{0}$  & $2458747.985032 \pm 0.000027$  \\
\hline
$P$ (days) & $0.941452382 \pm 0.000000069$ \\
\hline
$b$ & $0.360 \pm 0.026$ \\
\hline
$a/R_{*}$ & $3.496 \pm 0.029$ \\
\hline
$u_1$ & $0.290 \pm 0.032$ \\
\hline
$u_2$ & $0.169 \pm 0.061$\\
\hline
$\bar{f_p}$ (ppm)    & $180 \pm 13$  \\
\hline
$F_{\mathrm{atm}}$ (ppm)   & $177.5 \pm 5.7$  \\
\hline
$D$ (ppm)  & $21.8 \pm 5.2$ \\
\hline
$E$ (ppm)  & $172.2 \pm 5.6$  \\
\hline
Eclipse depth (ppm) &  $357 \pm 14$  \\
\hline
Nightside flux (ppm)   &  $2 \pm 14$ \\
\hline
$i$ (deg) & $84.09 \pm 0.47$  \\
\hline \hline
\end{tabular}
\caption{\sep \textbf{TESS phase curve fit results.} The median and $1\sigma$ uncertainties of the astrophysical parameters from a joint fit to all four Sectors of TESS photometry for WASP-18. In our phase curve parameterization, the secondary eclipse depth and nightside flux are derived parameters calculated from the average relative planetary flux $\bar{f}_{p}$ and the planet's phase curve semi-amplitude $F_{\rm atm}$. Likewise, the orbital inclination $i$ is derived from the scaled semi-major axis $a/R_{*}$ and the impact parameter $b$.}
    \label{tab:tess_results}
\end{table}

\clearpage





\bibliography{main}

\begin{thebibliography}{100}
\expandafter\ifx\csname url\endcsname\relax
  \def\url#1{\texttt{#1}}\fi
\expandafter\ifx\csname urlprefix\endcsname\relax\def\urlprefix{URL }\fi
\providecommand{\bibinfo}[2]{#2}
\providecommand{\eprint}[2][]{\url{#2}}

\bibitem{baxter20}
\bibinfo{author}{{Baxter}, C.} \emph{et~al.}
\newblock \bibinfo{title}{{A transition between the hot and the ultra-hot
  Jupiter atmospheres}}.
\newblock \emph{\bibinfo{journal}{\aap}} \textbf{\bibinfo{volume}{639}},
  \bibinfo{pages}{A36} (\bibinfo{year}{2020}).
\newblock \eprint{2007.15287}.

\bibitem{mansfield_unique_2021}
\bibinfo{author}{Mansfield, M.} \emph{et~al.}
\newblock \bibinfo{title}{A unique hot {Jupiter} spectral sequence with
  evidence for compositional diversity}.
\newblock \emph{\bibinfo{journal}{Nature Astronomy}}
  \textbf{\bibinfo{volume}{5}}, \bibinfo{pages}{1224--1232}
  (\bibinfo{year}{2021}).
\newblock
  \urlprefix\url{https://ui.adsabs.harvard.edu/abs/2021NatAs...5.1224M}.
\newblock \bibinfo{note}{ADS Bibcode: 2021NatAs...5.1224M}.

\bibitem{changeat22}
\bibinfo{author}{{Changeat}, Q.} \emph{et~al.}
\newblock \bibinfo{title}{{Five Key Exoplanet Questions Answered via the
  Analysis of 25 Hot-Jupiter Atmospheres in Eclipse}}.
\newblock \emph{\bibinfo{journal}{\apjs}} \textbf{\bibinfo{volume}{260}},
  \bibinfo{pages}{3} (\bibinfo{year}{2022}).
\newblock \eprint{2204.11729}.

\bibitem{sheppard_evidence_2017}
\bibinfo{author}{{Sheppard}, K.~B.} \emph{et~al.}
\newblock \bibinfo{title}{{Evidence for a Dayside Thermal Inversion and High
  Metallicity for the Hot Jupiter WASP-18b}}.
\newblock \emph{\bibinfo{journal}{\apjl}} \textbf{\bibinfo{volume}{850}},
  \bibinfo{pages}{L32} (\bibinfo{year}{2017}).
\newblock \eprint{1711.10491}.

\bibitem{mikal-evans17}
\bibinfo{author}{{Evans}, T.~M.} \emph{et~al.}
\newblock \bibinfo{title}{{An ultrahot gas-giant exoplanet with a
  stratosphere}}.
\newblock \emph{\bibinfo{journal}{\nat}} \textbf{\bibinfo{volume}{548}},
  \bibinfo{pages}{58--61} (\bibinfo{year}{2017}).
\newblock \eprint{1708.01076}.

\bibitem{cartier_near-infrared_2017}
\bibinfo{author}{Cartier, K. M.~S.} \emph{et~al.}
\newblock \bibinfo{title}{Near-infrared {Emission} {Spectrum} of {WASP}-103b
  {Using} {Hubble} {Space} {Telescope}/{Wide} {Field} {Camera} 3}.
\newblock \emph{\bibinfo{journal}{The Astronomical Journal}}
  \textbf{\bibinfo{volume}{153}}, \bibinfo{pages}{34} (\bibinfo{year}{2017}).
\newblock
  \urlprefix\url{https://ui.adsabs.harvard.edu/abs/2017AJ....153...34C}.
\newblock \bibinfo{note}{ADS Bibcode: 2017AJ....153...34C}.

\bibitem{arcangeli_h-_2018}
\bibinfo{author}{Arcangeli, J.} \emph{et~al.}
\newblock \bibinfo{title}{H- {Opacity} and {Water} {Dissociation} in the
  {Dayside} {Atmosphere} of the {Very} {Hot} {Gas} {Giant} {WASP}-18b}.
\newblock \emph{\bibinfo{journal}{The Astrophysical Journal}}
  \textbf{\bibinfo{volume}{855}}, \bibinfo{pages}{L30} (\bibinfo{year}{2018}).
\newblock
  \urlprefix\url{https://ui.adsabs.harvard.edu/abs/2018ApJ...855L..30A}.
\newblock \bibinfo{note}{ADS Bibcode: 2018ApJ...855L..30A}.

\bibitem{kreidberg_global_2018}
\bibinfo{author}{Kreidberg, L.} \emph{et~al.}
\newblock \bibinfo{title}{Global {Climate} and {Atmospheric} {Composition} of
  the {Ultra}-hot {Jupiter} {WASP}-103b from {HST} and {Spitzer} {Phase}
  {Curve} {Observations}}.
\newblock \emph{\bibinfo{journal}{The Astronomical Journal}}
  \textbf{\bibinfo{volume}{156}}, \bibinfo{pages}{17} (\bibinfo{year}{2018}).
\newblock
  \urlprefix\url{https://ui.adsabs.harvard.edu/abs/2018AJ....156...17K}.
\newblock \bibinfo{note}{ADS Bibcode: 2018AJ....156...17K}.

\bibitem{parmentier_thermal_2018}
\bibinfo{author}{Parmentier, V.} \emph{et~al.}
\newblock \bibinfo{title}{From thermal dissociation to condensation in the
  atmospheres of ultra hot {Jupiters}: {WASP}-121b in context}.
\newblock \emph{\bibinfo{journal}{Astronomy \& Astrophysics}}
  \textbf{\bibinfo{volume}{617}}, \bibinfo{pages}{A110} (\bibinfo{year}{2018}).
\newblock \urlprefix\url{https://www.aanda.org/10.1051/0004-6361/201833059}.

\bibitem{lothringer_extremely_2018}
\bibinfo{author}{Lothringer, J.~D.}, \bibinfo{author}{Barman, T.} \&
  \bibinfo{author}{Koskinen, T.}
\newblock \bibinfo{title}{Extremely {Irradiated} {Hot} {Jupiters}: {Non}-oxide
  {Inversions}, {H}- {Opacity}, and {Thermal} {Dissociation} of {Molecules}}.
\newblock \emph{\bibinfo{journal}{The Astrophysical Journal}}
  \textbf{\bibinfo{volume}{866}}, \bibinfo{pages}{27} (\bibinfo{year}{2018}).
\newblock
  \urlprefix\url{https://ui.adsabs.harvard.edu/abs/2018ApJ...866...27L}.
\newblock \bibinfo{note}{ADS Bibcode: 2018ApJ...866...27L}.

\bibitem{ghandi20}
\bibinfo{author}{{Gandhi}, S.}, \bibinfo{author}{{Madhusudhan}, N.} \&
  \bibinfo{author}{{Mandell}, A.}
\newblock \bibinfo{title}{{H- and Dissociation in Ultra-hot Jupiters: A
  Retrieval Case Study of WASP-18b}}.
\newblock \emph{\bibinfo{journal}{\aj}} \textbf{\bibinfo{volume}{159}},
  \bibinfo{pages}{232} (\bibinfo{year}{2020}).
\newblock \eprint{2004.07252}.

\bibitem{mikal-evans20}
\bibinfo{author}{{Mikal-Evans}, T.} \emph{et~al.}
\newblock \bibinfo{title}{{Confirmation of water emission in the dayside
  spectrum of the ultrahot Jupiter WASP-121b}}.
\newblock \emph{\bibinfo{journal}{\mnras}} \textbf{\bibinfo{volume}{496}},
  \bibinfo{pages}{1638--1644} (\bibinfo{year}{2020}).
\newblock \eprint{2005.09631}.

\bibitem{doyon_jwst_2012}
\bibinfo{author}{Doyon, R.} \emph{et~al.}
\newblock \bibinfo{title}{The {JWST} {Fine} {Guidance} {Sensor} ({FGS}) and
  {Near}-{Infrared} {Imager} and {Slitless} {Spectrograph} ({NIRISS})}.
\newblock In \emph{\bibinfo{booktitle}{Space {Telescopes} and {Instrumentation}
  2012: {Optical}, {Infrared}, and {Millimeter} {Wave}}}, vol.
  \bibinfo{volume}{8442}, \bibinfo{pages}{1005--1017}
  (\bibinfo{publisher}{SPIE}, \bibinfo{year}{2012}).
\newblock
  \urlprefix\url{https://www.spiedigitallibrary.org/conference-proceedings-of-spie/8442/84422R/The-JWST-Fine-Guidance-Sensor-FGS-and-Near-Infrared-Imager/10.1117/12.926578.full}.

\bibitem{bean_transiting_2018}
\bibinfo{author}{Bean, J.~L.} \emph{et~al.}
\newblock \bibinfo{title}{The {Transiting} {Exoplanet} {Community} {Early}
  {Release} {Science} {Program} for {JWST}}.
\newblock \emph{\bibinfo{journal}{Publications of the Astronomical Society of
  the Pacific}} \textbf{\bibinfo{volume}{130}}, \bibinfo{pages}{114402}
  (\bibinfo{year}{2018}).
\newblock \urlprefix\url{https://doi.org/10.1088/1538-3873/aadbf3}.
\newblock \bibinfo{note}{Publisher: IOP Publishing}.

\bibitem{hellier_orbital_2009}
\bibinfo{author}{Hellier, C.} \emph{et~al.}
\newblock \bibinfo{title}{An orbital period of 0.94days for the hot-{Jupiter}
  planet {WASP}-18b}.
\newblock \emph{\bibinfo{journal}{Nature}} \textbf{\bibinfo{volume}{460}},
  \bibinfo{pages}{1098--1100} (\bibinfo{year}{2009}).
\newblock
  \urlprefix\url{https://ui.adsabs.harvard.edu/abs/2009Natur.460.1098H}.
\newblock \bibinfo{note}{ADS Bibcode: 2009Natur.460.1098H}.

\bibitem{2022commissioning}
\bibinfo{author}{{Rigby}, J.} \emph{et~al.}
\newblock \bibinfo{title}{{Characterization of JWST science performance from
  commissioning}}.
\newblock \emph{\bibinfo{journal}{arXiv e-prints}}
  \bibinfo{pages}{arXiv:2207.05632} (\bibinfo{year}{2022}).
\newblock \eprint{2207.05632}.

\bibitem{alderson_early_2023}
\bibinfo{author}{Alderson, L.} \emph{et~al.}
\newblock \bibinfo{title}{Early {Release} {Science} of the {Exoplanet}
  {WASP}-39b with {JWST} {NIRSpec} {G395H}}.
\newblock \emph{\bibinfo{journal}{Nature}}  (\bibinfo{year}{2023}).
\newblock \urlprefix\url{https://doi.org/10.1038/s41586-022-05591-3}.

\bibitem{benneke_water_2019}
\bibinfo{author}{Benneke, B.} \emph{et~al.}
\newblock \bibinfo{title}{Water {Vapor} and {Clouds} on the {Habitable}-zone
  {Sub}-{Neptune} {Exoplanet} {K2}-18b}.
\newblock \emph{\bibinfo{journal}{The Astrophysical Journal Letters}}
  \textbf{\bibinfo{volume}{887}}, \bibinfo{pages}{L14} (\bibinfo{year}{2019}).
\newblock \urlprefix\url{https://doi.org/10.3847/2041-8213/ab59dc}.
\newblock \bibinfo{note}{Publisher: American Astronomical Society}.

\bibitem{jack_time-dependent_2009}
\bibinfo{author}{Jack, D.}, \bibinfo{author}{Hauschildt, P.~H.} \&
  \bibinfo{author}{Baron, E.}
\newblock \bibinfo{title}{Time-dependent radiative transfer with {PHOENIX}}.
\newblock \emph{\bibinfo{journal}{Astronomy \& Astrophysics}}
  \textbf{\bibinfo{volume}{502}}, \bibinfo{pages}{1043--1049}
  (\bibinfo{year}{2009}).
\newblock
  \urlprefix\url{https://www.aanda.org/articles/aa/abs/2009/30/aa10982-08/aa10982-08.html}.
\newblock \bibinfo{note}{Number: 3 Publisher: EDP Sciences}.

\bibitem{cortes-zuleta_tramos_2020}
\bibinfo{author}{Cortés-Zuleta, P.} \emph{et~al.}
\newblock \bibinfo{title}{{TraMoS} - {V}. {Updated} ephemeris and multi-epoch
  monitoring of the hot {Jupiters} {WASP}-{18Ab}, {WASP}-19b, and
  {WASP}-{77Ab}}.
\newblock \emph{\bibinfo{journal}{Astronomy \& Astrophysics}}
  \textbf{\bibinfo{volume}{636}}, \bibinfo{pages}{A98} (\bibinfo{year}{2020}).
\newblock
  \urlprefix\url{https://www.aanda.org/articles/aa/abs/2020/04/aa36279-19/aa36279-19.html}.
\newblock \bibinfo{note}{Publisher: EDP Sciences}.

\bibitem{stassun_revised_2019}
\bibinfo{author}{Stassun, K.~G.} \emph{et~al.}
\newblock \bibinfo{title}{The {Revised} {TESS} {Input} {Catalog} and
  {Candidate} {Target} {List}}.
\newblock \emph{\bibinfo{journal}{The Astronomical Journal}}
  \textbf{\bibinfo{volume}{158}}, \bibinfo{pages}{138} (\bibinfo{year}{2019}).
\newblock
  \urlprefix\url{https://ui.adsabs.harvard.edu/abs/2019AJ....158..138S}.
\newblock \bibinfo{note}{ADS Bibcode: 2019AJ....158..138S}.

\bibitem{brogi2022}
\bibinfo{author}{Brogi, M.} \emph{et~al.}
\newblock \bibinfo{title}{The roasting marshmallows program with igrins on
  gemini south i: Composition and climate of the ultra hot jupiter wasp-18 b}
  (\bibinfo{year}{2022}).
\newblock \urlprefix\url{https://arxiv.org/abs/2209.15548}.

\bibitem{Gandhi2018}
\bibinfo{author}{{Gandhi}, S.} \& \bibinfo{author}{{Madhusudhan}, N.}
\newblock \bibinfo{title}{{Retrieval of exoplanet emission spectra with
  HyDRA}}.
\newblock \emph{\bibinfo{journal}{\mnras}} \textbf{\bibinfo{volume}{474}},
  \bibinfo{pages}{271--288} (\bibinfo{year}{2018}).
\newblock \eprint{1710.06433}.

\bibitem{gandhi_h-_2020}
\bibinfo{author}{Gandhi, S.}, \bibinfo{author}{Madhusudhan, N.} \&
  \bibinfo{author}{Mandell, A.}
\newblock \bibinfo{title}{H- and {Dissociation} in {Ultra}-hot {Jupiters}: {A}
  {Retrieval} {Case} {Study} of {WASP}-18b}.
\newblock \emph{\bibinfo{journal}{The Astronomical Journal}}
  \textbf{\bibinfo{volume}{159}}, \bibinfo{pages}{232} (\bibinfo{year}{2020}).
\newblock
  \urlprefix\url{https://ui.adsabs.harvard.edu/abs/2020AJ....159..232G}.
\newblock \bibinfo{note}{ADS Bibcode: 2020AJ....159..232G}.

\bibitem{Piette2020b}
\bibinfo{author}{{Piette}, A. A.~A.} \& \bibinfo{author}{{Madhusudhan}, N.}
\newblock \bibinfo{title}{{Considerations for atmospheric retrieval of
  high-precision brown dwarf spectra}}.
\newblock \emph{\bibinfo{journal}{\mnras}} \textbf{\bibinfo{volume}{497}},
  \bibinfo{pages}{5136--5154} (\bibinfo{year}{2020}).
\newblock \eprint{2007.15004}.

\bibitem{Guillot2022}
\bibinfo{author}{Guillot, T.} \emph{et~al.}
\newblock \bibinfo{title}{Giant {Planets} from the {Inside}-{Out}}.
\newblock \bibinfo{type}{Tech. Rep.} (\bibinfo{year}{2022}).
\newblock
  \urlprefix\url{https://ui.adsabs.harvard.edu/abs/2022arXiv220504100G}.
\newblock \bibinfo{note}{Publication Title: arXiv e-prints ADS Bibcode:
  2022arXiv220504100G Type: article}.

\bibitem{pollack1996}
\bibinfo{author}{{Pollack}, J.~B.} \emph{et~al.}
\newblock \bibinfo{title}{{Formation of the Giant Planets by Concurrent
  Accretion of Solids and Gas}}.
\newblock \emph{\bibinfo{journal}{\icarus}} \textbf{\bibinfo{volume}{124}},
  \bibinfo{pages}{62--85} (\bibinfo{year}{1996}).

\bibitem{polanski2022}
\bibinfo{author}{Polanski, A.~S.}, \bibinfo{author}{Crossfield, I. J.~M.},
  \bibinfo{author}{Howard, A.~W.}, \bibinfo{author}{Isaacson, H.} \&
  \bibinfo{author}{Rice, M.}
\newblock \bibinfo{title}{Chemical abundances for 25 jwst exoplanet host stars
  with keckspec} (\bibinfo{year}{2022}).
\newblock \urlprefix\url{https://arxiv.org/abs/2207.13662}.

\bibitem{lodders98}
\bibinfo{author}{{Lodders}, K.} \& \bibinfo{author}{{Fegley}, B.}
\newblock \emph{\bibinfo{title}{{The planetary scientist's companion /
  Katharina Lodders, Bruce Fegley.}}} (\bibinfo{year}{1998}).

\bibitem{Kreidberg_2014}
\bibinfo{author}{Kreidberg, L.} \emph{et~al.}
\newblock \bibinfo{title}{A {PRECISE} {WATER} {ABUNDANCE} {MEASUREMENT} {FOR}
  {THE} {HOT} {JUPITER} {WASP}-43b}.
\newblock \emph{\bibinfo{journal}{The Astrophysical Journal}}
  \textbf{\bibinfo{volume}{793}}, \bibinfo{pages}{L27} (\bibinfo{year}{2014}).
\newblock \urlprefix\url{https://doi.org/10.1088%2F2041-8205%2F793%2F2%2Fl27}.

\bibitem{wakeford17}
\bibinfo{author}{{Wakeford}, H.~R.} \emph{et~al.}
\newblock \bibinfo{title}{{HAT-P-26b: A Neptune-mass exoplanet with a
  well-constrained heavy element abundance}}.
\newblock \emph{\bibinfo{journal}{Science}} \textbf{\bibinfo{volume}{356}},
  \bibinfo{pages}{628--631} (\bibinfo{year}{2017}).
\newblock \eprint{1705.04354}.

\bibitem{Welbanks_2019}
\bibinfo{author}{Welbanks, L.} \emph{et~al.}
\newblock \bibinfo{title}{Mass{\textendash}metallicity trends in transiting
  exoplanets from atmospheric abundances of h$_2$o, na, and k}.
\newblock \emph{\bibinfo{journal}{The Astrophysical Journal Letters}}
  \textbf{\bibinfo{volume}{887}}, \bibinfo{pages}{L20} (\bibinfo{year}{2019}).
\newblock \urlprefix\url{https://doi.org/10.3847%2F2041-8213%2Fab5a89}.

\bibitem{_berg_2011}
\bibinfo{author}{Öberg, K.~I.}, \bibinfo{author}{Murray-Clay, R.} \&
  \bibinfo{author}{Bergin, E.~A.}
\newblock \bibinfo{title}{{THE} {EFFECTS} {OF} {SNOWLINES} {ON} c/o {IN}
  {PLANETARY} {ATMOSPHERES}}.
\newblock \emph{\bibinfo{journal}{The Astrophysical Journal}}
  \textbf{\bibinfo{volume}{743}}, \bibinfo{pages}{L16} (\bibinfo{year}{2011}).
\newblock \urlprefix\url{https://doi.org/10.1088%2F2041-8205%2F743%2F1%2Fl16}.

\bibitem{bedell_chemical_2018}
\bibinfo{author}{Bedell, M.} \emph{et~al.}
\newblock \bibinfo{title}{The {Chemical} {Homogeneity} of {Sun}-like {Stars} in
  the {Solar} {Neighborhood}}.
\newblock \emph{\bibinfo{journal}{ApJ}} \textbf{\bibinfo{volume}{865}},
  \bibinfo{pages}{68} (\bibinfo{year}{2018}).
\newblock \urlprefix\url{https://dx.doi.org/10.3847/1538-4357/aad908}.
\newblock \bibinfo{note}{Publisher: The American Astronomical Society}.

\bibitem{boss1997}
\bibinfo{author}{{Boss}, A.~P.}
\newblock \bibinfo{title}{{Giant planet formation by gravitational
  instability.}}
\newblock \emph{\bibinfo{journal}{Science}} \textbf{\bibinfo{volume}{276}},
  \bibinfo{pages}{1836--1839} (\bibinfo{year}{1997}).

\bibitem{madhusudhan2014a}
\bibinfo{author}{{Madhusudhan}, N.}, \bibinfo{author}{{Amin}, M.~A.} \&
  \bibinfo{author}{{Kennedy}, G.~M.}
\newblock \bibinfo{title}{{Toward Chemical Constraints on Hot Jupiter
  Migration}}.
\newblock \emph{\bibinfo{journal}{\apjl}} \textbf{\bibinfo{volume}{794}},
  \bibinfo{pages}{L12} (\bibinfo{year}{2014}).
\newblock \eprint{1408.3668}.

\bibitem{rauscher_more_2018}
\bibinfo{author}{Rauscher, E.}, \bibinfo{author}{Suri, V.} \&
  \bibinfo{author}{Cowan, N.~B.}
\newblock \bibinfo{title}{A {More} {Informative} {Map}: {Inverting} {Thermal}
  {Orbital} {Phase} and {Eclipse} {Light} {Curves} of {Exoplanets}}.
\newblock \emph{\bibinfo{journal}{The Astronomical Journal}}
  \textbf{\bibinfo{volume}{156}}, \bibinfo{pages}{235} (\bibinfo{year}{2018}).
\newblock \urlprefix\url{https://doi.org/10.3847/1538-3881/aae57f}.
\newblock \bibinfo{note}{Publisher: American Astronomical Society}.

\bibitem{mansfield_eigenspectra_2020}
\bibinfo{author}{Mansfield, M.} \emph{et~al.}
\newblock \bibinfo{title}{Eigenspectra: a framework for identifying spectra
  from {3D} eclipse mapping}.
\newblock \emph{\bibinfo{journal}{Monthly Notices of the Royal Astronomical
  Society}} \textbf{\bibinfo{volume}{499}}, \bibinfo{pages}{5151--5162}
  (\bibinfo{year}{2020}).
\newblock \urlprefix\url{https://doi.org/10.1093/mnras/staa3179}.

\bibitem{challener_theresa_2022}
\bibinfo{author}{Challener, R.~C.} \& \bibinfo{author}{Rauscher, E.}
\newblock \bibinfo{title}{{ThERESA}: {Three}-dimensional {Eclipse} {Mapping}
  with {Application} to {Synthetic} {JWST} {Data}}.
\newblock \emph{\bibinfo{journal}{The Astronomical Journal}}
  \textbf{\bibinfo{volume}{163}}, \bibinfo{pages}{117} (\bibinfo{year}{2022}).
\newblock \urlprefix\url{https://doi.org/10.3847/1538-3881/ac4885}.
\newblock \bibinfo{note}{Publisher: American Astronomical Society}.

\bibitem{arcangeli_climate_2019}
\bibinfo{author}{Arcangeli, J.} \emph{et~al.}
\newblock \bibinfo{title}{Climate of an ultra hot {Jupiter}: {Spectroscopic}
  phase curve of {WASP}-18b with {HST}/{WFC3}}.
\newblock \emph{\bibinfo{journal}{Astronomy \& Astrophysics}}
  \textbf{\bibinfo{volume}{625}}, \bibinfo{pages}{A136} (\bibinfo{year}{2019}).
\newblock \urlprefix\url{https://www.aanda.org/10.1051/0004-6361/201834891}.

\bibitem{komacek_DN1}
\bibinfo{author}{{Komacek}, T.~D.} \& \bibinfo{author}{{Showman}, A.~P.}
\newblock \bibinfo{title}{{Atmospheric Circulation of Hot Jupiters:
  Dayside-Nightside Temperature Differences}}.
\newblock \emph{\bibinfo{journal}{\apj}} \textbf{\bibinfo{volume}{821}},
  \bibinfo{pages}{16} (\bibinfo{year}{2016}).
\newblock \eprint{1601.00069}.

\bibitem{komacek_atmospheric_2017}
\bibinfo{author}{Komacek, T.~D.}, \bibinfo{author}{Showman, A.~P.} \&
  \bibinfo{author}{Tan, X.}
\newblock \bibinfo{title}{Atmospheric {Circulation} of {Hot} {Jupiters}:
  {Dayside}-{Nightside} {Temperature} {Differences}. {II}. {Comparison} with
  {Observations}}.
\newblock \emph{\bibinfo{journal}{The Astrophysical Journal}}
  \textbf{\bibinfo{volume}{835}}, \bibinfo{pages}{198} (\bibinfo{year}{2017}).
\newblock
  \urlprefix\url{https://ui.adsabs.harvard.edu/abs/2017ApJ...835..198K}.
\newblock \bibinfo{note}{ADS Bibcode: 2017ApJ...835..198K}.

\bibitem{perna_magnetic_2010}
\bibinfo{author}{Perna, R.}, \bibinfo{author}{Menou, K.} \&
  \bibinfo{author}{Rauscher, E.}
\newblock \bibinfo{title}{{MAGNETIC} {DRAG} {ON} {HOT} {JUPITER} {ATMOSPHERIC}
  {WINDS}}.
\newblock \emph{\bibinfo{journal}{The Astrophysical Journal}}
  \textbf{\bibinfo{volume}{719}}, \bibinfo{pages}{1421--1426}
  (\bibinfo{year}{2010}).
\newblock \urlprefix\url{https://doi.org/10.1088/0004-637x/719/2/1421}.
\newblock \bibinfo{note}{Publisher: American Astronomical Society}.

\bibitem{batygin_magnetically_2013}
\bibinfo{author}{Batygin, K.}, \bibinfo{author}{Stanley, S.} \&
  \bibinfo{author}{Stevenson, D.~J.}
\newblock \bibinfo{title}{{MAGNETICALLY} {CONTROLLED} {CIRCULATION} {ON} {HOT}
  {EXTRASOLAR} {PLANETS}}.
\newblock \emph{\bibinfo{journal}{The Astrophysical Journal}}
  \textbf{\bibinfo{volume}{776}}, \bibinfo{pages}{53} (\bibinfo{year}{2013}).
\newblock \urlprefix\url{https://doi.org/10.1088/0004-637x/776/1/53}.
\newblock \bibinfo{note}{Publisher: American Astronomical Society}.

\bibitem{Menou:2012fu}
\bibinfo{author}{Menou, K.}
\newblock \bibinfo{title}{Magnetic scaling laws for the atmospheres of hot
  giant exoplanets}.
\newblock \emph{\bibinfo{journal}{The Astrophysical Journal}}
  \textbf{\bibinfo{volume}{745}}, \bibinfo{pages}{138} (\bibinfo{year}{2012}).

\bibitem{beltz_exploring_2021}
\bibinfo{author}{Beltz, H.}, \bibinfo{author}{Rauscher, E.},
  \bibinfo{author}{Roman, M.~T.} \& \bibinfo{author}{Guilliat, A.}
\newblock \bibinfo{title}{Exploring the {Effects} of {Active} {Magnetic} {Drag}
  in a {General} {Circulation} {Model} of the {Ultrahot} {Jupiter} {WASP}-76b}.
\newblock \emph{\bibinfo{journal}{The Astronomical Journal}}
  \textbf{\bibinfo{volume}{163}}, \bibinfo{pages}{35} (\bibinfo{year}{2021}).
\newblock \urlprefix\url{https://doi.org/10.3847/1538-3881/ac3746}.
\newblock \bibinfo{note}{Publisher: American Astronomical Society}.

\bibitem{Rogers:2017}
\bibinfo{author}{Rogers, T.}
\newblock \bibinfo{title}{Constraints on the magnetic field strength of
  {HAT-P-7b} and other hot giant exoplanets}.
\newblock \emph{\bibinfo{journal}{Nature Astronomy}}
  \textbf{\bibinfo{volume}{1}}, \bibinfo{pages}{131} (\bibinfo{year}{2017}).

\bibitem{hindle_shallow-water_2019}
\bibinfo{author}{Hindle, A.~W.}, \bibinfo{author}{Bushby, P.~J.} \&
  \bibinfo{author}{Rogers, T.~M.}
\newblock \bibinfo{title}{Shallow-water {Magnetohydrodynamics} for {Westward}
  {Hotspots} on {Hot} {Jupiters}}.
\newblock \emph{\bibinfo{journal}{The Astrophysical Journal}}
  \textbf{\bibinfo{volume}{872}}, \bibinfo{pages}{L27} (\bibinfo{year}{2019}).
\newblock \urlprefix\url{https://doi.org/10.3847/2041-8213/ab05dd}.
\newblock \bibinfo{note}{Publisher: American Astronomical Society}.

\bibitem{hindle_observational_2021}
\bibinfo{author}{Hindle, A.~W.}, \bibinfo{author}{Bushby, P.~J.} \&
  \bibinfo{author}{Rogers, T.~M.}
\newblock \bibinfo{title}{Observational {Consequences} of {Shallow}-water
  {Magnetohydrodynamics} on {Hot} {Jupiters}}.
\newblock \emph{\bibinfo{journal}{The Astrophysical Journal Letters}}
  \textbf{\bibinfo{volume}{916}}, \bibinfo{pages}{L8} (\bibinfo{year}{2021}).
\newblock \urlprefix\url{https://doi.org/10.3847/2041-8213/ac0fec}.
\newblock \bibinfo{note}{Publisher: American Astronomical Society}.

\bibitem{feinstein_early_2023}
\bibinfo{author}{Feinstein, A.~D.} \emph{et~al.}
\newblock \bibinfo{title}{Early {Release} {Science} of the exoplanet {WASP}-39b
  with {JWST} {NIRISS}}.
\newblock \emph{\bibinfo{journal}{Nature}}  (\bibinfo{year}{2023}).
\newblock \urlprefix\url{https://doi.org/10.1038/s41586-022-05674-1}.

\bibitem{bell_eureka_2022}
\bibinfo{author}{Bell, T.~J.} \emph{et~al.}
\newblock \bibinfo{title}{Eureka!: {An} {End}-to-{End} {Pipeline} for {JWST}
  {Time}-{Series} {Observations}} (\bibinfo{year}{2022}).
\newblock \urlprefix\url{http://arxiv.org/abs/2207.03585}.
\newblock \bibinfo{note}{ArXiv:2207.03585 [astro-ph]}.

\bibitem{scikit-image}
\bibinfo{author}{van~der Walt, S.} \emph{et~al.}
\newblock \bibinfo{title}{scikit-image: image processing in {P}ython}.
\newblock \emph{\bibinfo{journal}{PeerJ}} \textbf{\bibinfo{volume}{2}},
  \bibinfo{pages}{e453} (\bibinfo{year}{2014}).
\newblock \urlprefix\url{https://doi.org/10.7717/peerj.453}.

\bibitem{vandokkum01}
\bibinfo{author}{{van Dokkum}, P.~G.}
\newblock \bibinfo{title}{{Cosmic-Ray Rejection by Laplacian Edge Detection}}.
\newblock \emph{\bibinfo{journal}{\pasp}} \textbf{\bibinfo{volume}{113}},
  \bibinfo{pages}{1420--1427} (\bibinfo{year}{2001}).
\newblock \eprint{astro-ph/0108003}.

\bibitem{ccdproc}
\bibinfo{author}{Craig, M.} \emph{et~al.}
\newblock \bibinfo{title}{astropy/ccdproc: v1.3.0.post1}
  (\bibinfo{year}{2017}).
\newblock \urlprefix\url{https://doi.org/10.5281/zenodo.1069648}.

\bibitem{horne86}
\bibinfo{author}{{Horne}, K.}
\newblock \bibinfo{title}{{An optimal extraction algorithm for CCD
  spectroscopy.}}
\newblock \emph{\bibinfo{journal}{\pasp}} \textbf{\bibinfo{volume}{98}},
  \bibinfo{pages}{609--617} (\bibinfo{year}{1986}).

\bibitem{darveau-bernier_atoca_2022}
\bibinfo{author}{{Darveau-Bernier}, A.} \emph{et~al.}
\newblock \bibinfo{title}{{ATOCA: an Algorithm to Treat Order Contamination.
  Application to the NIRISS SOSS Mode}}.
\newblock \emph{\bibinfo{journal}{\pasp}} \textbf{\bibinfo{volume}{134}},
  \bibinfo{pages}{094502} (\bibinfo{year}{2022}).
\newblock \eprint{2207.05199}.

\bibitem{radica_applesoss_2022}
\bibinfo{author}{{Radica}, M.} \emph{et~al.}
\newblock \bibinfo{title}{{APPLESOSS: A Producer of ProfiLEs for SOSS.
  Application to the NIRISS SOSS Mode}}.
\newblock \emph{\bibinfo{journal}{\pasp}} \textbf{\bibinfo{volume}{134}},
  \bibinfo{pages}{104502} (\bibinfo{year}{2022}).
\newblock \eprint{2207.05136}.

\bibitem{kreidberg_batman_2015}
\bibinfo{author}{Kreidberg, L.}
\newblock \bibinfo{title}{batman : {BAsic} {Transit} {Model} {cAlculatioN} in
  {Python}}.
\newblock \emph{\bibinfo{journal}{Publications of the Astronomical Society of
  the Pacific}} \textbf{\bibinfo{volume}{127}}, \bibinfo{pages}{1161--1165}
  (\bibinfo{year}{2015}).
\newblock \urlprefix\url{http://iopscience.iop.org/article/10.1086/683602}.

\bibitem{cowan_inverting_2008}
\bibinfo{author}{Cowan, N.~B.} \& \bibinfo{author}{Agol, E.}
\newblock \bibinfo{title}{Inverting {Phase} {Functions} to {Map} {Exoplanets}}.
\newblock \emph{\bibinfo{journal}{The Astrophysical Journal}}
  \textbf{\bibinfo{volume}{678}}, \bibinfo{pages}{L129} (\bibinfo{year}{2008}).
\newblock
  \urlprefix\url{https://ui.adsabs.harvard.edu/abs/2008ApJ...678L.129C}.
\newblock \bibinfo{note}{ADS Bibcode: 2008ApJ...678L.129C}.

\bibitem{raftery_bayesian_1995}
\bibinfo{author}{Raftery, A.~E.}
\newblock \bibinfo{title}{Bayesian {Model} {Selection} in {Social} {Research}}.
\newblock \emph{\bibinfo{journal}{Sociological Methodology}}
  \textbf{\bibinfo{volume}{25}}, \bibinfo{pages}{111--163}
  (\bibinfo{year}{1995}).
\newblock \urlprefix\url{https://www.jstor.org/stable/271063}.
\newblock \bibinfo{note}{Publisher: [American Sociological Association, Wiley,
  Sage Publications, Inc.]}.

\bibitem{1974aikake}
\bibinfo{author}{{Akaike}, H.}
\newblock \bibinfo{title}{{A New Look at the Statistical Model
  Identification}}.
\newblock \emph{\bibinfo{journal}{IEEE Transactions on Automatic Control}}
  \textbf{\bibinfo{volume}{19}}, \bibinfo{pages}{716--723}
  (\bibinfo{year}{1974}).

\bibitem{leconte_distorted_2011}
\bibinfo{author}{Leconte, J.}, \bibinfo{author}{Lai, D.} \&
  \bibinfo{author}{Chabrier, G.}
\newblock \bibinfo{title}{Distorted, nonspherical transiting planets: impact on
  the transit depth and on the radius determination}.
\newblock \emph{\bibinfo{journal}{Astronomy \& Astrophysics}}
  \textbf{\bibinfo{volume}{528}}, \bibinfo{pages}{A41} (\bibinfo{year}{2011}).
\newblock
  \urlprefix\url{https://www.aanda.org/articles/aa/abs/2011/04/aa15811-10/aa15811-10.html}.
\newblock \bibinfo{note}{Publisher: EDP Sciences}.

\bibitem{shporer_tess_2019}
\bibinfo{author}{Shporer, A.} \emph{et~al.}
\newblock \bibinfo{title}{\textit{{TESS}} {Full} {Orbital} {Phase} {Curve} of
  the {WASP}-18b {System}}.
\newblock \emph{\bibinfo{journal}{The Astronomical Journal}}
  \textbf{\bibinfo{volume}{157}}, \bibinfo{pages}{178} (\bibinfo{year}{2019}).
\newblock
  \urlprefix\url{https://iopscience.iop.org/article/10.3847/1538-3881/ab0f96}.

\bibitem{blazek_constraints_2022}
\bibinfo{author}{Blažek, M.} \emph{et~al.}
\newblock \bibinfo{title}{Constraints on {TESS} albedos for five hot
  {Jupiters}}.
\newblock \emph{\bibinfo{journal}{Monthly Notices of the Royal Astronomical
  Society}} \textbf{\bibinfo{volume}{513}}, \bibinfo{pages}{3444--3457}
  (\bibinfo{year}{2022}).
\newblock \urlprefix\url{https://doi.org/10.1093/mnras/stac992}.

\bibitem{han_exoplanet_2014}
\bibinfo{author}{Han, E.} \emph{et~al.}
\newblock \bibinfo{title}{Exoplanet {Orbit} {Database}. {II}. {Updates} to
  {Exoplanets}.org}.
\newblock \emph{\bibinfo{journal}{Publications of the Astronomical Society of
  the Pacific}} \textbf{\bibinfo{volume}{126}}, \bibinfo{pages}{827}
  (\bibinfo{year}{2014}).
\newblock
  \urlprefix\url{https://ui.adsabs.harvard.edu/abs/2014PASP..126..827H}.
\newblock \bibinfo{note}{ADS Bibcode: 2014PASP..126..827H}.

\bibitem{fetherolf22}
\bibinfo{author}{{Fetherolf}, T.} \emph{et~al.}
\newblock \bibinfo{title}{{Variability Catalog of Stars Observed During the
  TESS Prime Mission}}.
\newblock \emph{\bibinfo{journal}{arXiv e-prints}}
  \bibinfo{pages}{arXiv:2208.11721} (\bibinfo{year}{2022}).
\newblock \eprint{2208.11721}.

\bibitem{miller_detectability_2012}
\bibinfo{author}{Miller, B.~P.}, \bibinfo{author}{Gallo, E.},
  \bibinfo{author}{Wright, J.~T.} \& \bibinfo{author}{Dupree, A.~K.}
\newblock \bibinfo{title}{{ON} {THE} {DETECTABILITY} {OF} {STAR}-{PLANET}
  {INTERACTION}}.
\newblock \emph{\bibinfo{journal}{The Astrophysical Journal}}
  \textbf{\bibinfo{volume}{754}}, \bibinfo{pages}{137} (\bibinfo{year}{2012}).
\newblock \urlprefix\url{https://doi.org/10.1088/0004-637x/754/2/137}.
\newblock \bibinfo{note}{Publisher: American Astronomical Society}.

\bibitem{pillitteri_no_2014}
\bibinfo{author}{Pillitteri, I.}, \bibinfo{author}{Wolk, S.~J.},
  \bibinfo{author}{Sciortino, S.} \& \bibinfo{author}{Antoci, V.}
\newblock \bibinfo{title}{No {X}-rays from {WASP}-18 - {Implications} for its
  age, activity, and the influence of its massive hot {Jupiter}}.
\newblock \emph{\bibinfo{journal}{Astronomy \& Astrophysics}}
  \textbf{\bibinfo{volume}{567}}, \bibinfo{pages}{A128} (\bibinfo{year}{2014}).
\newblock
  \urlprefix\url{https://www.aanda.org/articles/aa/abs/2014/07/aa23579-14/aa23579-14.html}.
\newblock \bibinfo{note}{Publisher: EDP Sciences}.

\bibitem{shkolnik_signatures_2018}
\bibinfo{author}{Shkolnik, E.~L.} \& \bibinfo{author}{Llama, J.}
\newblock \bibinfo{title}{Signatures of {Star}-planet interactions}.
\newblock \bibinfo{pages}{1737--1753} (\bibinfo{year}{2018}).
\newblock \urlprefix\url{http://arxiv.org/abs/1712.02814}.
\newblock \bibinfo{note}{ArXiv:1712.02814 [astro-ph]}.

\bibitem{scikit-learn}
\bibinfo{author}{Pedregosa, F.} \emph{et~al.}
\newblock \bibinfo{title}{Scikit-learn: Machine learning in {P}ython}.
\newblock \emph{\bibinfo{journal}{Journal of Machine Learning Research}}
  \textbf{\bibinfo{volume}{12}}, \bibinfo{pages}{2825--2830}
  (\bibinfo{year}{2011}).

\bibitem{foreman-mackey_emcee_2013}
\bibinfo{author}{Foreman-Mackey, D.}, \bibinfo{author}{Hogg, D.~W.},
  \bibinfo{author}{Lang, D.} \& \bibinfo{author}{Goodman, J.}
\newblock \bibinfo{title}{emcee : {The} {MCMC} {Hammer}}.
\newblock \emph{\bibinfo{journal}{Publications of the Astronomical Society of
  the Pacific}} \textbf{\bibinfo{volume}{125}}, \bibinfo{pages}{306--312}
  (\bibinfo{year}{2013}).
\newblock \urlprefix\url{http://iopscience.iop.org/article/10.1086/670067}.

\bibitem{wong_tess_2020}
\bibinfo{author}{{Wong}, I.} \emph{et~al.}
\newblock \bibinfo{title}{{Systematic Phase Curve Study of Known Transiting
  Systems from Year One of the TESS Mission}}.
\newblock \emph{\bibinfo{journal}{\aj}} \textbf{\bibinfo{volume}{160}},
  \bibinfo{pages}{155} (\bibinfo{year}{2020}).
\newblock \eprint{2003.06407}.

\bibitem{louden_spiderman_2018}
\bibinfo{author}{Louden, T.} \& \bibinfo{author}{Kreidberg, L.}
\newblock \bibinfo{title}{{SPIDERMAN}: an open-source code to model phase
  curves and secondary eclipses}.
\newblock \emph{\bibinfo{journal}{Monthly Notices of the Royal Astronomical
  Society}} \textbf{\bibinfo{volume}{477}}, \bibinfo{pages}{2613--2627}
  (\bibinfo{year}{2018}).
\newblock \urlprefix\url{https://doi.org/10.1093/mnras/sty558}.

\bibitem{luger_starry_2019}
\bibinfo{author}{Luger, R.} \emph{et~al.}
\newblock \bibinfo{title}{starry: {Analytic} {Occultation} {Light} {Curves}}.
\newblock \emph{\bibinfo{journal}{The Astronomical Journal}}
  \textbf{\bibinfo{volume}{157}}, \bibinfo{pages}{64} (\bibinfo{year}{2019}).
\newblock
  \urlprefix\url{https://ui.adsabs.harvard.edu/abs/2019AJ....157...64L}.
\newblock \bibinfo{note}{ADS Bibcode: 2019AJ....157...64L}.

\bibitem{Gelman1992}
\bibinfo{author}{{Gelman}, A.} \& \bibinfo{author}{{Rubin}, D.~B.}
\newblock \bibinfo{title}{{Inference from Iterative Simulation Using Multiple
  Sequences}}.
\newblock \emph{\bibinfo{journal}{Statistical Science}}
  \textbf{\bibinfo{volume}{7}}, \bibinfo{pages}{457--472}
  (\bibinfo{year}{1992}).

\bibitem{Luger2021}
\bibinfo{author}{{Luger}, R.}, \bibinfo{author}{{Foreman-Mackey}, D.},
  \bibinfo{author}{{Hedges}, C.} \& \bibinfo{author}{{Hogg}, D.~W.}
\newblock \bibinfo{title}{{Mapping Stellar Surfaces. I. Degeneracies in the
  Rotational Light-curve Problem}}.
\newblock \emph{\bibinfo{journal}{\aj}} \textbf{\bibinfo{volume}{162}},
  \bibinfo{pages}{123} (\bibinfo{year}{2021}).
\newblock \eprint{2102.00007}.

\bibitem{winn2011}
\bibinfo{author}{Winn, J.~N.}
\newblock \bibinfo{title}{Transits and occultations} (\bibinfo{year}{2010}).
\newblock \urlprefix\url{https://arxiv.org/abs/1001.2010}.

\bibitem{Bonomo_2017}
\bibinfo{author}{Bonomo, A.~S.} \emph{et~al.}
\newblock \bibinfo{title}{The {GAPS} programme with {HARPS}-n at {TNG}}.
\newblock \emph{\bibinfo{journal}{Astronomy {\&} Astrophysics}}
  \textbf{\bibinfo{volume}{602}}, \bibinfo{pages}{A107} (\bibinfo{year}{2017}).
\newblock \urlprefix\url{https://doi.org/10.1051%2F0004-6361%2F201629882}.

\bibitem{Piskorz2018}
\bibinfo{author}{{Piskorz}, D.} \emph{et~al.}
\newblock \bibinfo{title}{{Ground- and Space-based Detection of the Thermal
  Emission Spectrum of the Transiting Hot Jupiter KELT-2Ab}}.
\newblock \emph{\bibinfo{journal}{\aj}} \textbf{\bibinfo{volume}{156}},
  \bibinfo{pages}{133} (\bibinfo{year}{2018}).
\newblock \eprint{1809.05615}.

\bibitem{Glidic2022}
\bibinfo{author}{{Glidic}, K.} \emph{et~al.}
\newblock \bibinfo{title}{{Atmospheric Characterization of Hot Jupiter CoRoT-1
  b Using the Wide Field Camera 3 on the Hubble Space Telescope}}.
\newblock \emph{\bibinfo{journal}{\aj}} \textbf{\bibinfo{volume}{164}},
  \bibinfo{pages}{19} (\bibinfo{year}{2022}).
\newblock \eprint{2206.03517}.

\bibitem{Toon1989}
\bibinfo{author}{{Toon}, O.~B.}, \bibinfo{author}{{McKay}, C.~P.},
  \bibinfo{author}{{Ackerman}, T.~P.} \& \bibinfo{author}{{Santhanam}, K.}
\newblock \bibinfo{title}{{Rapid calculation of radiative heating rates and
  photodissociation rates in inhomogeneous multiple scattering atmospheres}}.
\newblock \emph{\bibinfo{journal}{\jgr}} \textbf{\bibinfo{volume}{94}},
  \bibinfo{pages}{16287--16301} (\bibinfo{year}{1989}).

\bibitem{McKay1989}
\bibinfo{author}{{McKay}, C.~P.}, \bibinfo{author}{{Pollack}, J.~B.} \&
  \bibinfo{author}{{Courtin}, R.}
\newblock \bibinfo{title}{{The thermal structure of Titan's atmosphere}}.
\newblock \emph{\bibinfo{journal}{\icarus}} \textbf{\bibinfo{volume}{80}},
  \bibinfo{pages}{23--53} (\bibinfo{year}{1989}).

\bibitem{Gordon1994}
\bibinfo{author}{Gordon, S.} \& \bibinfo{author}{Mcbride, B.~J.}
\newblock \bibinfo{title}{Computer program for calculation of complex chemical
  equilibrium compositions and applications. part 1: Analysis}.
\newblock \bibinfo{type}{Tech. Rep.} \bibinfo{number}{19950013764},
  \bibinfo{institution}{NASA Lewis Research Center} (\bibinfo{year}{1994}).

\bibitem{Amundsen2017}
\bibinfo{author}{{Amundsen}, D.~S.}, \bibinfo{author}{{Tremblin}, P.},
  \bibinfo{author}{{Manners}, J.}, \bibinfo{author}{{Baraffe}, I.} \&
  \bibinfo{author}{{Mayne}, N.~J.}
\newblock \bibinfo{title}{{Treatment of overlapping gaseous absorption with the
  correlated-k method in hot Jupiter and brown dwarf atmosphere models}}.
\newblock \emph{\bibinfo{journal}{\aap}} \textbf{\bibinfo{volume}{598}},
  \bibinfo{pages}{A97} (\bibinfo{year}{2017}).
\newblock \eprint{1610.01389}.

\bibitem{LoddersPalme2009}
\bibinfo{author}{{Lodders}, K.} \& \bibinfo{author}{{Palme}, H.}
\newblock \bibinfo{title}{{Solar System Elemental Abundances in 2009}}.
\newblock \emph{\bibinfo{journal}{Meteoritics and Planetary Science
  Supplement}} \textbf{\bibinfo{volume}{72}}, \bibinfo{pages}{5154}
  (\bibinfo{year}{2009}).

\bibitem{Taylor2020}
\bibinfo{author}{{Taylor}, J.} \emph{et~al.}
\newblock \bibinfo{title}{{Understanding and mitigating biases when studying
  inhomogeneous emission spectra with JWST}}.
\newblock \emph{\bibinfo{journal}{\mnras}} \textbf{\bibinfo{volume}{493}},
  \bibinfo{pages}{4342--4354} (\bibinfo{year}{2020}).
\newblock \eprint{2002.00773}.

\bibitem{Buchner2014}
\bibinfo{author}{{Buchner}, J.} \emph{et~al.}
\newblock \bibinfo{title}{{X-ray spectral modelling of the AGN obscuring region
  in the CDFS: Bayesian model selection and catalogue}}.
\newblock \emph{\bibinfo{journal}{Astronomy and Astrophysics}}
  \textbf{\bibinfo{volume}{564}}, \bibinfo{pages}{A125} (\bibinfo{year}{2014}).
\newblock \eprint{1402.0004}.

\bibitem{benneke_atmospheric_2012}
\bibinfo{author}{Benneke, B.} \& \bibinfo{author}{Seager, S.}
\newblock \bibinfo{title}{Atmospheric {Retrieval} for {Super}-{Earths}:
  {Uniquely} {Constraining} the {Atmospheric} {Composition} with {Transmission}
  {Spectroscopy}}.
\newblock \emph{\bibinfo{journal}{The Astrophysical Journal}}
  \textbf{\bibinfo{volume}{753}}, \bibinfo{pages}{100} (\bibinfo{year}{2012}).
\newblock
  \urlprefix\url{https://ui.adsabs.harvard.edu/abs/2012ApJ...753..100B}.
\newblock \bibinfo{note}{ADS Bibcode: 2012ApJ...753..100B}.

\bibitem{benneke_how_2013}
\bibinfo{author}{Benneke, B.} \& \bibinfo{author}{Seager, S.}
\newblock \bibinfo{title}{{HOW} {TO} {DISTINGUISH} {BETWEEN} {CLOUDY}
  {MINI}-{NEPTUNES} {AND} {WATER}/{VOLATILE}-{DOMINATED} {SUPER}-{EARTHS}}.
\newblock \emph{\bibinfo{journal}{The Astrophysical Journal}}
  \textbf{\bibinfo{volume}{778}}, \bibinfo{pages}{153} (\bibinfo{year}{2013}).
\newblock \urlprefix\url{https://doi.org/10.1088/0004-637x/778/2/153}.
\newblock \bibinfo{note}{Publisher: American Astronomical Society}.

\bibitem{benneke_strict_2015}
\bibinfo{author}{Benneke, B.}
\newblock \bibinfo{title}{Strict {Upper} {Limits} on the {Carbon}-to-{Oxygen}
  {Ratios} of {Eight} {Hot} {Jupiters} from {Self}-{Consistent} {Atmospheric}
  {Retrieval}} (\bibinfo{year}{2015}).
\newblock \urlprefix\url{http://arxiv.org/abs/1504.07655}.
\newblock \bibinfo{note}{Number: arXiv:1504.07655 arXiv:1504.07655 [astro-ph]}.

\bibitem{Bell1987}
\bibinfo{author}{{Bell}, K.~L.} \& \bibinfo{author}{{Berrington}, K.~A.}
\newblock \bibinfo{title}{{Free-free absorption coefficient of the negative
  hydrogen ion}}.
\newblock \emph{\bibinfo{journal}{Journal of Physics B Atomic Molecular
  Physics}} \textbf{\bibinfo{volume}{20}}, \bibinfo{pages}{801--806}
  (\bibinfo{year}{1987}).

\bibitem{John1988}
\bibinfo{author}{{John}, T.~L.}
\newblock \bibinfo{title}{{Continuous absorption by the negative hydrogen ion
  reconsidered}}.
\newblock \emph{\bibinfo{journal}{\aap}} \textbf{\bibinfo{volume}{193}},
  \bibinfo{pages}{189--192} (\bibinfo{year}{1988}).

\bibitem{vald1995}
\bibinfo{author}{{Piskunov}, N.~E.}, \bibinfo{author}{{Kupka}, F.},
  \bibinfo{author}{{Ryabchikova}, T.~A.}, \bibinfo{author}{{Weiss}, W.~W.} \&
  \bibinfo{author}{{Jeffery}, C.~S.}
\newblock \bibinfo{title}{{VALD: The Vienna Atomic Line Data Base.}}
\newblock \emph{\bibinfo{journal}{\aaps}} \textbf{\bibinfo{volume}{112}},
  \bibinfo{pages}{525} (\bibinfo{year}{1995}).

\bibitem{Burrows_2003}
\bibinfo{author}{Burrows, A.} \& \bibinfo{author}{Volobuyev, M.}
\newblock \bibinfo{title}{Calculations of the far-wing line profiles of sodium
  and potassium in the atmospheres of substellar-mass objects}.
\newblock \emph{\bibinfo{journal}{The Astrophysical Journal}}
  \textbf{\bibinfo{volume}{583}}, \bibinfo{pages}{985--995}
  (\bibinfo{year}{2003}).
\newblock \urlprefix\url{https://doi.org/10.1086%2F345412}.

\bibitem{Polyansky2018}
\bibinfo{author}{{Polyansky}, O.~L.} \emph{et~al.}
\newblock \bibinfo{title}{{ExoMol molecular line lists XXX: a complete
  high-accuracy line list for water}}.
\newblock \emph{\bibinfo{journal}{\mnras}} \textbf{\bibinfo{volume}{480}},
  \bibinfo{pages}{2597--2608} (\bibinfo{year}{2018}).
\newblock \eprint{1807.04529}.

\bibitem{Rothman2010}
\bibinfo{author}{{Rothman}, L.~S.} \emph{et~al.}
\newblock \bibinfo{title}{{HITEMP, the high-temperature molecular spectroscopic
  database}}.
\newblock \emph{\bibinfo{journal}{\jqsrt}} \textbf{\bibinfo{volume}{111}},
  \bibinfo{pages}{2139--2150} (\bibinfo{year}{2010}).

\bibitem{Yurchenko_2014}
\bibinfo{author}{Yurchenko, S.~N.} \& \bibinfo{author}{Tennyson, J.}
\newblock \bibinfo{title}{{ExoMol} line lists {\textendash} {IV}. the
  rotation{\textendash}vibration spectrum of methane up to 1500~k}.
\newblock \emph{\bibinfo{journal}{Monthly Notices of the Royal Astronomical
  Society}} \textbf{\bibinfo{volume}{440}}, \bibinfo{pages}{1649--1661}
  (\bibinfo{year}{2014}).
\newblock \urlprefix\url{https://doi.org/10.1093%2Fmnras%2Fstu326}.

\bibitem{Coles_2019}
\bibinfo{author}{Coles, P.~A.}, \bibinfo{author}{Yurchenko, S.~N.} \&
  \bibinfo{author}{Tennyson, J.}
\newblock \bibinfo{title}{{ExoMol} molecular line lists {\textendash} {XXXV}. a
  rotation-vibration line list for hot ammonia}.
\newblock \emph{\bibinfo{journal}{Monthly Notices of the Royal Astronomical
  Society}} \textbf{\bibinfo{volume}{490}}, \bibinfo{pages}{4638--4647}
  (\bibinfo{year}{2019}).
\newblock \urlprefix\url{https://doi.org/10.1093%2Fmnras%2Fstz2778}.

\bibitem{Barber_2013}
\bibinfo{author}{Barber, R.~J.} \emph{et~al.}
\newblock \bibinfo{title}{{ExoMol} line lists {\textendash} {III}. an improved
  hot rotation-vibration line list for {HCN} and {HNC}}.
\newblock \emph{\bibinfo{journal}{Monthly Notices of the Royal Astronomical
  Society}} \textbf{\bibinfo{volume}{437}}, \bibinfo{pages}{1828--1835}
  (\bibinfo{year}{2013}).
\newblock \urlprefix\url{https://doi.org/10.1093%2Fmnras%2Fstt2011}.

\bibitem{McKemmish2019}
\bibinfo{author}{{McKemmish}, L.~K.} \emph{et~al.}
\newblock \bibinfo{title}{{ExoMol Molecular linelists - XXXIII. The spectrum of
  Titanium Oxide}}.
\newblock \emph{\bibinfo{journal}{\mnras}} \textbf{\bibinfo{volume}{488}},
  \bibinfo{pages}{2836--2854} (\bibinfo{year}{2019}).
\newblock \eprint{1905.04587}.

\bibitem{McKemmish2016}
\bibinfo{author}{{McKemmish}, L.~K.}, \bibinfo{author}{{Yurchenko}, S.~N.} \&
  \bibinfo{author}{{Tennyson}, J.}
\newblock \bibinfo{title}{{ExoMol line lists - XVIII. The high-temperature
  spectrum of VO}}.
\newblock \emph{\bibinfo{journal}{\mnras}} \textbf{\bibinfo{volume}{463}},
  \bibinfo{pages}{771--793} (\bibinfo{year}{2016}).
\newblock \eprint{1609.06120}.

\bibitem{Wende2010}
\bibinfo{author}{{Wende}, S.}, \bibinfo{author}{{Reiners}, A.},
  \bibinfo{author}{{Seifahrt}, A.} \& \bibinfo{author}{{Bernath}, P.~F.}
\newblock \bibinfo{title}{{CRIRES spectroscopy and empirical line-by-line
  identification of FeH molecular absorption in an M dwarf}}.
\newblock \emph{\bibinfo{journal}{Astronomy and Astrophysics}}
  \textbf{\bibinfo{volume}{523}}, \bibinfo{pages}{A58} (\bibinfo{year}{2010}).
\newblock \eprint{1007.4116}.

\bibitem{fastchem2}
\bibinfo{author}{{Stock}, J.~W.}, \bibinfo{author}{{Kitzmann}, D.} \&
  \bibinfo{author}{{Patzer}, A. B.~C.}
\newblock \bibinfo{title}{{FASTCHEM 2 : an improved computer program to
  determine the gas-phase chemical equilibrium composition for arbitrary
  element distributions}}.
\newblock \emph{\bibinfo{journal}{\mnras}} \textbf{\bibinfo{volume}{517}},
  \bibinfo{pages}{4070--4080} (\bibinfo{year}{2022}).
\newblock \eprint{2206.08247}.

\bibitem{pelletier_where_2021}
\bibinfo{author}{Pelletier, S.} \emph{et~al.}
\newblock \bibinfo{title}{Where {Is} the {Water}? {Jupiter}-like {C}/{H}
  {Ratio} but {Strong} {H}$_2${O} {Depletion} {Found} on
  \${\textbackslash}uptau\$ {Boötis} b {Using} {SPIRou}}.
\newblock \emph{\bibinfo{journal}{The Astronomical Journal}}
  \textbf{\bibinfo{volume}{162}}, \bibinfo{pages}{73} (\bibinfo{year}{2021}).
\newblock \urlprefix\url{https://doi.org/10.3847/1538-3881/ac0428}.
\newblock \bibinfo{note}{Publisher: American Astronomical Society}.

\bibitem{rocchetto_exploring_2016}
\bibinfo{author}{Rocchetto, M.}, \bibinfo{author}{Waldmann, I.~P.},
  \bibinfo{author}{Venot, O.}, \bibinfo{author}{Lagage, P.-O.} \&
  \bibinfo{author}{Tinetti, G.}
\newblock \bibinfo{title}{{EXPLORING} {BIASES} {OF} {ATMOSPHERIC} {RETRIEVALS}
  {IN} {SIMULATED}
  \${\textbackslash}less\$i\${\textbackslash}greater\${JWST}\${\textbackslash}less\$/i\${\textbackslash}greater\$
  {TRANSMISSION} {SPECTRA} {OF} {HOT} {JUPITERS}}.
\newblock \emph{\bibinfo{journal}{The Astrophysical Journal}}
  \textbf{\bibinfo{volume}{833}}, \bibinfo{pages}{120} (\bibinfo{year}{2016}).
\newblock \urlprefix\url{https://doi.org/10.3847/1538-4357/833/1/120}.
\newblock \bibinfo{note}{Publisher: American Astronomical Society}.

\bibitem{Piette2022}
\bibinfo{author}{{Piette}, A. A.~A.}, \bibinfo{author}{{Madhusudhan}, N.} \&
  \bibinfo{author}{{Mandell}, A.~M.}
\newblock \bibinfo{title}{{HyDRo: atmospheric retrieval of rocky exoplanets in
  thermal emission}}.
\newblock \emph{\bibinfo{journal}{\mnras}} \textbf{\bibinfo{volume}{511}},
  \bibinfo{pages}{2565--2584} (\bibinfo{year}{2022}).
\newblock \eprint{2112.05059}.

\bibitem{MacDonald2017}
\bibinfo{author}{{MacDonald}, R.~J.} \& \bibinfo{author}{{Madhusudhan}, N.}
\newblock \bibinfo{title}{{HD 209458b in new light: evidence of nitrogen
  chemistry, patchy clouds and sub-solar water}}.
\newblock \emph{\bibinfo{journal}{\mnras}} \textbf{\bibinfo{volume}{469}},
  \bibinfo{pages}{1979--1996} (\bibinfo{year}{2017}).
\newblock \eprint{1701.01113}.

\bibitem{MacDonald2023}
\bibinfo{author}{{MacDonald}, R.~J.}
\newblock \bibinfo{title}{{POSEIDON: A Multidimensional Atmospheric Retrieval
  Code for Exoplanet Spectra}}.
\newblock \emph{\bibinfo{journal}{Journal of Open Source Software}}
  \textbf{\bibinfo{volume}{8}}, \bibinfo{pages}{4873} (\bibinfo{year}{2023}).

\bibitem{Feroz2009}
\bibinfo{author}{{Feroz}, F.}, \bibinfo{author}{{Hobson}, M.~P.} \&
  \bibinfo{author}{{Bridges}, M.}
\newblock \bibinfo{title}{{MULTINEST: an efficient and robust Bayesian
  inference tool for cosmology and particle physics}}.
\newblock \emph{\bibinfo{journal}{\mnras}} \textbf{\bibinfo{volume}{398}},
  \bibinfo{pages}{1601--1614} (\bibinfo{year}{2009}).
\newblock \eprint{0809.3437}.

\bibitem{Skilling2006}
\bibinfo{author}{Skilling, J.}
\newblock \bibinfo{title}{Nested sampling for general bayesian computation}.
\newblock \emph{\bibinfo{journal}{Bayesian Anal.}}
  \textbf{\bibinfo{volume}{1}}, \bibinfo{pages}{833--859}
  (\bibinfo{year}{2006}).
\newblock \urlprefix\url{https://doi.org/10.1214/06-BA127}.

\bibitem{Richard2012}
\bibinfo{author}{Richard, C.} \emph{et~al.}
\newblock \bibinfo{title}{New section of the hitran database: Collision-induced
  absorption (cia)}.
\newblock \emph{\bibinfo{journal}{Journal of Quantitative Spectroscopy and
  Radiative Transfer}} \textbf{\bibinfo{volume}{113}}, \bibinfo{pages}{1276 --
  1285} (\bibinfo{year}{2012}).
\newblock
  \urlprefix\url{http://www.sciencedirect.com/science/article/pii/S0022407311003773}.
\newblock \bibinfo{note}{Three Leaders in Spectroscopy}.

\bibitem{Harris2006}
\bibinfo{author}{{Harris}, G.~J.}, \bibinfo{author}{{Tennyson}, J.},
  \bibinfo{author}{{Kaminsky}, B.~M.}, \bibinfo{author}{{Pavlenko}, Y.~V.} \&
  \bibinfo{author}{{Jones}, H.~R.~A.}
\newblock \bibinfo{title}{{Improved HCN/HNC linelist, model atmospheres and
  synthetic spectra for WZ Cas}}.
\newblock \emph{\bibinfo{journal}{\mnras}} \textbf{\bibinfo{volume}{367}},
  \bibinfo{pages}{400--406} (\bibinfo{year}{2006}).
\newblock \eprint{astro-ph/0512363}.

\bibitem{Dulick2003}
\bibinfo{author}{{Dulick}, M.} \emph{et~al.}
\newblock \bibinfo{title}{{Line Intensities and Molecular Opacities of the FeH
  F $^{4}${\ensuremath{\Delta}}$_{i}$-X $^{4}${\ensuremath{\Delta}}$_{i}$
  Transition}}.
\newblock \emph{\bibinfo{journal}{\apj}} \textbf{\bibinfo{volume}{594}},
  \bibinfo{pages}{651--663} (\bibinfo{year}{2003}).
\newblock \eprint{astro-ph/0305162}.

\bibitem{Burrows2003}
\bibinfo{author}{{Burrows}, A.} \& \bibinfo{author}{{Volobuyev}, M.}
\newblock \bibinfo{title}{{Calculations of the Far-Wing Line Profiles of Sodium
  and Potassium in the Atmospheres of Substellar-Mass Objects}}.
\newblock \emph{\bibinfo{journal}{\apj}} \textbf{\bibinfo{volume}{583}},
  \bibinfo{pages}{985--995} (\bibinfo{year}{2003}).
\newblock \eprint{astro-ph/0210086}.

\bibitem{Gandhi2017}
\bibinfo{author}{{Gandhi}, S.} \& \bibinfo{author}{{Madhusudhan}, N.}
\newblock \bibinfo{title}{{genesis: new self-consistent models of exoplanetary
  spectra}}.
\newblock \emph{\bibinfo{journal}{\mnras}} \textbf{\bibinfo{volume}{472}},
  \bibinfo{pages}{2334--2355} (\bibinfo{year}{2017}).
\newblock \eprint{1706.02302}.

\bibitem{Madhusudhan2009}
\bibinfo{author}{{Madhusudhan}, N.} \& \bibinfo{author}{{Seager}, S.}
\newblock \bibinfo{title}{{A Temperature and Abundance Retrieval Method for
  Exoplanet Atmospheres}}.
\newblock \emph{\bibinfo{journal}{\apj}} \textbf{\bibinfo{volume}{707}},
  \bibinfo{pages}{24--39} (\bibinfo{year}{2009}).
\newblock \eprint{0910.1347}.

\bibitem{trotta2008}
\bibinfo{author}{{Trotta}, R.}
\newblock \bibinfo{title}{{Bayes in the sky: Bayesian inference and model
  selection in cosmology}}.
\newblock \emph{\bibinfo{journal}{Contemporary Physics}}
  \textbf{\bibinfo{volume}{49}}, \bibinfo{pages}{71--104}
  (\bibinfo{year}{2008}).
\newblock \eprint{0803.4089}.

\bibitem{Burrows1999}
\bibinfo{author}{{Burrows}, A.} \& \bibinfo{author}{{Sharp}, C.~M.}
\newblock \bibinfo{title}{{Chemical Equilibrium Abundances in Brown Dwarf and
  Extrasolar Giant Planet Atmospheres}}.
\newblock \emph{\bibinfo{journal}{\apj}} \textbf{\bibinfo{volume}{512}},
  \bibinfo{pages}{843--863} (\bibinfo{year}{1999}).
\newblock \eprint{astro-ph/9807055}.

\bibitem{Madhusudhan2011}
\bibinfo{author}{{Madhusudhan}, N.}, \bibinfo{author}{{Mousis}, O.},
  \bibinfo{author}{{Johnson}, T.~V.} \& \bibinfo{author}{{Lunine}, J.~I.}
\newblock \bibinfo{title}{{Carbon-rich Giant Planets: Atmospheric Chemistry,
  Thermal Inversions, Spectra, and Formation Conditions}}.
\newblock \emph{\bibinfo{journal}{\apj}} \textbf{\bibinfo{volume}{743}},
  \bibinfo{pages}{191} (\bibinfo{year}{2011}).
\newblock \eprint{1109.3183}.

\bibitem{Piette2020a}
\bibinfo{author}{{Piette}, A. A.~A.} \emph{et~al.}
\newblock \bibinfo{title}{{Assessing spectra and thermal inversions due to TiO
  in hot Jupiter atmospheres}}.
\newblock \emph{\bibinfo{journal}{\mnras}} \textbf{\bibinfo{volume}{496}},
  \bibinfo{pages}{3870--3886} (\bibinfo{year}{2020}).
\newblock \eprint{2006.04807}.

\bibitem{Evans2017}
\bibinfo{author}{{Evans}, T.~M.} \emph{et~al.}
\newblock \bibinfo{title}{{An ultrahot gas-giant exoplanet with a
  stratosphere}}.
\newblock \emph{\bibinfo{journal}{\nat}} \textbf{\bibinfo{volume}{548}},
  \bibinfo{pages}{58--61} (\bibinfo{year}{2017}).
\newblock \eprint{1708.01076}.

\bibitem{Fortney2019}
\bibinfo{author}{{Fortney}, J.~J.}, \bibinfo{author}{{Lupu}, R.~E.},
  \bibinfo{author}{{Morley}, C.~V.}, \bibinfo{author}{{Freedman}, R.~S.} \&
  \bibinfo{author}{{Hood}, C.}
\newblock \bibinfo{title}{{Exploring a Photospheric Radius Correction to Model
  Secondary Eclipse Spectra for Transiting Exoplanets}}.
\newblock \emph{\bibinfo{journal}{\apjl}} \textbf{\bibinfo{volume}{880}},
  \bibinfo{pages}{L16} (\bibinfo{year}{2019}).
\newblock \eprint{1904.00025}.

\bibitem{Taylor2022}
\bibinfo{author}{{Taylor}, J.}
\newblock \bibinfo{title}{{Impact of variable photospheric radius on exoplanet
  atmospheric retrievals}}.
\newblock \emph{\bibinfo{journal}{\mnras}} \textbf{\bibinfo{volume}{513}},
  \bibinfo{pages}{L20--L24} (\bibinfo{year}{2022}).
\newblock \eprint{2203.01839}.

\bibitem{MacDonald2022}
\bibinfo{author}{{MacDonald}, R.~J.} \& \bibinfo{author}{{Lewis}, N.~K.}
\newblock \bibinfo{title}{{TRIDENT: A Rapid 3D Radiative-transfer Model for
  Exoplanet Transmission Spectra}}.
\newblock \emph{\bibinfo{journal}{\apj}} \textbf{\bibinfo{volume}{929}},
  \bibinfo{pages}{20} (\bibinfo{year}{2022}).
\newblock \eprint{2111.05862}.

\bibitem{Karman2019}
\bibinfo{author}{{Karman}, T.} \emph{et~al.}
\newblock \bibinfo{title}{{Update of the HITRAN collision-induced absorption
  section}}.
\newblock \emph{\bibinfo{journal}{\icarus}} \textbf{\bibinfo{volume}{328}},
  \bibinfo{pages}{160--175} (\bibinfo{year}{2019}).

\bibitem{Li2015}
\bibinfo{author}{{Li}, G.} \emph{et~al.}
\newblock \bibinfo{title}{{Rovibrational Line Lists for Nine Isotopologues of
  the CO Molecule in the X $^{1}${\ensuremath{\Sigma}}$^{+}$ Ground Electronic
  State}}.
\newblock \emph{\bibinfo{journal}{The Astrophysical Journal Supplement Series}}
  \textbf{\bibinfo{volume}{216}}, \bibinfo{pages}{15} (\bibinfo{year}{2015}).

\bibitem{Tashkun2011}
\bibinfo{author}{{Tashkun}, S.~A.} \& \bibinfo{author}{{Perevalov}, V.~I.}
\newblock \bibinfo{title}{{CDSD-4000: High-resolution, high-temperature carbon
  dioxide spectroscopic databank}}.
\newblock \emph{\bibinfo{journal}{Journal of Quantitative Spectroscopy and
  Radiative Transfer}} \textbf{\bibinfo{volume}{112}},
  \bibinfo{pages}{1403--1410} (\bibinfo{year}{2011}).

\bibitem{Barber2014}
\bibinfo{author}{{Barber}, R.~J.} \emph{et~al.}
\newblock \bibinfo{title}{{ExoMol line lists - III. An improved hot
  rotation-vibration line list for HCN and HNC}}.
\newblock \emph{\bibinfo{journal}{MNRAS}} \textbf{\bibinfo{volume}{437}},
  \bibinfo{pages}{1828--1835} (\bibinfo{year}{2014}).
\newblock \eprint{1311.1328}.

\bibitem{Brooke2016}
\bibinfo{author}{{Brooke}, J. S.~A.} \emph{et~al.}
\newblock \bibinfo{title}{{Line strengths of rovibrational and rotational
  transitions in the X$^{2}$ {\ensuremath{\Pi}} ground state of OH}}.
\newblock \emph{\bibinfo{journal}{Journal of Quantitative Spectroscopy and
  Radiative Transfer}} \textbf{\bibinfo{volume}{168}},
  \bibinfo{pages}{142--157} (\bibinfo{year}{2016}).

\bibitem{Ryabchikova2015}
\bibinfo{author}{{Ryabchikova}, T.} \emph{et~al.}
\newblock \bibinfo{title}{{A major upgrade of the VALD database}}.
\newblock \emph{\bibinfo{journal}{Physica Scripta}}
  \textbf{\bibinfo{volume}{90}}, \bibinfo{pages}{054005}
  (\bibinfo{year}{2015}).

\bibitem{malik2017helios}
\bibinfo{author}{{Malik}, M.} \emph{et~al.}
\newblock \bibinfo{title}{{HELIOS: An Open-source, GPU-accelerated Radiative
  Transfer Code for Self-consistent Exoplanetary Atmospheres}}.
\newblock \emph{\bibinfo{journal}{\aj}} \textbf{\bibinfo{volume}{153}},
  \bibinfo{pages}{56} (\bibinfo{year}{2017}).
\newblock \eprint{1606.05474}.

\bibitem{malik2019helios}
\bibinfo{author}{{Malik}, M.} \emph{et~al.}
\newblock \bibinfo{title}{{Self-luminous and Irradiated Exoplanetary
  Atmospheres Explored with HELIOS}}.
\newblock \emph{\bibinfo{journal}{\aj}} \textbf{\bibinfo{volume}{157}},
  \bibinfo{pages}{170} (\bibinfo{year}{2019}).
\newblock \eprint{1903.06794}.

\bibitem{tsai2017vulcan}
\bibinfo{author}{{Tsai}, S.-M.} \emph{et~al.}
\newblock \bibinfo{title}{{VULCAN: An Open-source, Validated Chemical Kinetics
  Python Code for Exoplanetary Atmospheres}}.
\newblock \emph{\bibinfo{journal}{\apjs}} \textbf{\bibinfo{volume}{228}},
  \bibinfo{pages}{20} (\bibinfo{year}{2017}).
\newblock \eprint{1607.00409}.

\bibitem{Tsai_2021}
\bibinfo{author}{Tsai, S.-M.} \emph{et~al.}
\newblock \bibinfo{title}{A comparative study of atmospheric chemistry with
  {VULCAN}}.
\newblock \emph{\bibinfo{journal}{The Astrophysical Journal}}
  \textbf{\bibinfo{volume}{923}}, \bibinfo{pages}{264} (\bibinfo{year}{2021}).
\newblock \urlprefix\url{https://doi.org/10.3847%2F1538-4357%2Fac29bc}.

\bibitem{Zhang_2019}
\bibinfo{author}{Zhang, M.}, \bibinfo{author}{Chachan, Y.},
  \bibinfo{author}{Kempton, E. M.-R.} \& \bibinfo{author}{Knutson, H.~A.}
\newblock \bibinfo{title}{Forward modeling and retrievals with {PLATON}, a fast
  open-source tool}.
\newblock \emph{\bibinfo{journal}{Publications of the Astronomical Society of
  the Pacific}} \textbf{\bibinfo{volume}{131}}, \bibinfo{pages}{034501}
  (\bibinfo{year}{2019}).
\newblock \urlprefix\url{https://doi.org/10.1088%2F1538-3873%2Faaf5ad}.

\bibitem{grimm2015helios}
\bibinfo{author}{Grimm, S.~L.} \& \bibinfo{author}{Heng, K.}
\newblock \bibinfo{title}{Helios-k: An ultrafast, open-source opacity
  calculator for radiative transfer}.
\newblock \emph{\bibinfo{journal}{The Astrophysical Journal}}
  \textbf{\bibinfo{volume}{808}}, \bibinfo{pages}{182} (\bibinfo{year}{2015}).

\bibitem{grimm2021helios}
\bibinfo{author}{Grimm, S.~L.} \emph{et~al.}
\newblock \bibinfo{title}{Helios-k 2.0 opacity calculator and open-source
  opacity database for exoplanetary atmospheres}.
\newblock \emph{\bibinfo{journal}{The Astrophysical Journal Supplement Series}}
  \textbf{\bibinfo{volume}{253}}, \bibinfo{pages}{30} (\bibinfo{year}{2021}).

\bibitem{parmentier20133d}
\bibinfo{author}{Parmentier, V.}, \bibinfo{author}{Showman, A.~P.} \&
  \bibinfo{author}{Lian, Y.}
\newblock \bibinfo{title}{3d mixing in hot jupiters atmospheres-i. application
  to the day/night cold trap in hd 209458b}.
\newblock \emph{\bibinfo{journal}{Astronomy \& Astrophysics}}
  \textbf{\bibinfo{volume}{558}}, \bibinfo{pages}{A91} (\bibinfo{year}{2013}).

\bibitem{smith1998estimation}
\bibinfo{author}{Smith, M.~D.}
\newblock \bibinfo{title}{Estimation of a length scale to use with the quench
  level approximation for obtaining chemical abundances}.
\newblock \emph{\bibinfo{journal}{Icarus}} \textbf{\bibinfo{volume}{132}},
  \bibinfo{pages}{176--184} (\bibinfo{year}{1998}).

\bibitem{lewis2010atmospheric}
\bibinfo{author}{Lewis, N.~K.} \emph{et~al.}
\newblock \bibinfo{title}{Atmospheric circulation of eccentric hot neptune
  gj436b}.
\newblock \emph{\bibinfo{journal}{The Astrophysical Journal}}
  \textbf{\bibinfo{volume}{720}}, \bibinfo{pages}{344} (\bibinfo{year}{2010}).

\bibitem{rugheimer2013spectral}
\bibinfo{author}{{Rugheimer}, S.}, \bibinfo{author}{{Kaltenegger}, L.},
  \bibinfo{author}{{Zsom}, A.}, \bibinfo{author}{{Segura}, A.} \&
  \bibinfo{author}{{Sasselov}, D.}
\newblock \bibinfo{title}{{Spectral Fingerprints of Earth-like Planets Around
  FGK Stars}}.
\newblock \emph{\bibinfo{journal}{Astrobiology}} \textbf{\bibinfo{volume}{13}},
  \bibinfo{pages}{251--269} (\bibinfo{year}{2013}).
\newblock \eprint{1212.2638}.

\bibitem{kurucz1979model}
\bibinfo{author}{Kurucz, R.~L.}
\newblock \bibinfo{title}{Model atmospheres for g, f, a, b, and o stars}.
\newblock \emph{\bibinfo{journal}{The Astrophysical Journal Supplement Series}}
  \textbf{\bibinfo{volume}{40}}, \bibinfo{pages}{1--340}
  (\bibinfo{year}{1979}).

\bibitem{Showman2009}
\bibinfo{author}{Showman, A.~P.} \emph{et~al.}
\newblock \bibinfo{title}{Atmospheric {Circulation} of {Hot} {Jupiters}:
  {Coupled} {Radiative}-{Dynamical} {General} {Circulation} {Model}
  {Simulations} of {HD} 189733b and {HD} 209458b}.
\newblock \emph{\bibinfo{journal}{{\textbackslash}apj}}
  \textbf{\bibinfo{volume}{699}}, \bibinfo{pages}{564--584}
  (\bibinfo{year}{2009}).
\newblock \bibinfo{note}{\_eprint: 0809.2089}.

\bibitem{Adcroft2004}
\bibinfo{author}{Adcroft, A.}, \bibinfo{author}{Campin, J.-M.},
  \bibinfo{author}{Hill, C.} \& \bibinfo{author}{Marshall, J.}
\newblock \bibinfo{title}{Implementation of an {Atmosphere}-{Ocean} {General}
  {Circulation} {Model} on the {Expanded} {Spherical} {Cube}}.
\newblock \emph{\bibinfo{journal}{Monthly Weather Review}}
  \textbf{\bibinfo{volume}{132}}, \bibinfo{pages}{2845--2863}
  (\bibinfo{year}{2004}).
\newblock \urlprefix\url{http://dx.doi.org/10.1175/MWR2823.1}.
\newblock \bibinfo{note}{\_eprint: http://dx.doi.org/10.1175/MWR2823.1}.

\bibitem{Marley1999}
\bibinfo{author}{Marley, M.~S.} \& \bibinfo{author}{McKay, C.~P.}
\newblock \bibinfo{title}{Thermal {Structure} of {Uranus}' {Atmosphere}}.
\newblock \emph{\bibinfo{journal}{ıcarus}} \textbf{\bibinfo{volume}{138}},
  \bibinfo{pages}{268--286} (\bibinfo{year}{1999}).

\bibitem{Freedman2014}
\bibinfo{author}{Freedman, R.~S.} \emph{et~al.}
\newblock \bibinfo{title}{Gaseous {Mean} {Opacities} for {Giant} {Planet} and
  {Ultracool} {Dwarf} {Atmospheres} over a {Range} of {Metallicities} and
  {Temperatures}}.
\newblock \emph{\bibinfo{journal}{{\textbackslash}apjs}}
  \textbf{\bibinfo{volume}{214}}, \bibinfo{pages}{25} (\bibinfo{year}{2014}).
\newblock \bibinfo{note}{\_eprint: 1409.0026}.

\bibitem{Lodders2002}
\bibinfo{author}{Lodders, K.} \& \bibinfo{author}{Fegley, B.}
\newblock \bibinfo{title}{Atmospheric {Chemistry} in {Giant} {Planets}, {Brown}
  {Dwarfs}, and {Low}-{Mass} {Dwarf} {Stars}. {I}. {Carbon}, {Nitrogen}, and
  {Oxygen}}.
\newblock \emph{\bibinfo{journal}{ıcarus}} \textbf{\bibinfo{volume}{155}},
  \bibinfo{pages}{393--424} (\bibinfo{year}{2002}).

\bibitem{Visscher2006}
\bibinfo{author}{Visscher, C.}, \bibinfo{author}{Lodders, K.} \&
  \bibinfo{author}{Fegley, B., Jr.}
\newblock \bibinfo{title}{Atmospheric {Chemistry} in {Giant} {Planets}, {Brown}
  {Dwarfs}, and {Low}-{Mass} {Dwarf} {Stars}. {II}. {Sulfur} and {Phosphorus}}.
\newblock \emph{\bibinfo{journal}{{\textbackslash}apj}}
  \textbf{\bibinfo{volume}{648}}, \bibinfo{pages}{1181--1195}
  (\bibinfo{year}{2006}).
\newblock \bibinfo{note}{\_eprint: arXiv:astro-ph/0511136}.

\bibitem{Lee2021picketfence}
\bibinfo{author}{{Lee}, E. K.~H.} \emph{et~al.}
\newblock \bibinfo{title}{{Simulating gas giant exoplanet atmospheres with
  EXO-FMS: comparing semigrey, picket fence, and correlated-k
  radiative-transfer schemes}}.
\newblock \emph{\bibinfo{journal}{\mnras}} \textbf{\bibinfo{volume}{506}},
  \bibinfo{pages}{2695--2711} (\bibinfo{year}{2021}).
\newblock \eprint{2106.11664}.

\bibitem{Malsky2022}
\bibinfo{author}{{Malsky}, I.}
\newblock \bibinfo{title}{A direct comparison between the use of double gray
  and multiwavelength radiative transfer in a general circulation model with
  and without radiatively active clouds}.

\bibitem{RauscherMenou2013}
\bibinfo{author}{Rauscher, E.} \& \bibinfo{author}{Menou, K.}
\newblock \bibinfo{title}{Three-dimensional atmospheric circulation models of
  hd 189733b and hd 209458b with consistent magnetic drag and ohmic
  dissipation}.
\newblock \emph{\bibinfo{journal}{The Astrophysical Journal}}
  \textbf{\bibinfo{volume}{764}}, \bibinfo{pages}{103} (\bibinfo{year}{2013}).

\bibitem{beltz_magnetic_2022}
\bibinfo{author}{Beltz, H.} \emph{et~al.}
\newblock \bibinfo{title}{Magnetic {Drag} and 3-{D} {Effects} in {Theoretical}
  {High}-{Resolution} {Emission} {Spectra} of {Ultrahot} {Jupiters}: the {Case}
  of {WASP}-76b} (\bibinfo{year}{2022}).
\newblock \urlprefix\url{http://arxiv.org/abs/2204.12996}.
\newblock \bibinfo{note}{ArXiv:2204.12996 [astro-ph]}.

\bibitem{2013maxted}
\bibinfo{author}{{Maxted}, P.~F.~L.} \emph{et~al.}
\newblock \bibinfo{title}{{Spitzer 3.6 and 4.5 {\ensuremath{\mu}}m full-orbit
  light curves of WASP-18}}.
\newblock \emph{\bibinfo{journal}{\mnras}} \textbf{\bibinfo{volume}{428}},
  \bibinfo{pages}{2645--2660} (\bibinfo{year}{2013}).
\newblock \eprint{1210.5585}.

\bibitem{wakeford_high-temperature_2016}
\bibinfo{author}{Wakeford, H.~R.} \emph{et~al.}
\newblock \bibinfo{title}{High-temperature condensate clouds in super-hot
  {Jupiter} atmospheres}.
\newblock \emph{\bibinfo{journal}{Monthly Notices of the Royal Astronomical
  Society}} \textbf{\bibinfo{volume}{464}}, \bibinfo{pages}{4247--4254}
  (\bibinfo{year}{2016}).

\end{thebibliography}

\end{document}